\documentclass[a4paper,11pt]{article}
\usepackage{jheppub} 
\usepackage{slashed, mathrsfs,float}
\usepackage{bm}

\usepackage{graphicx}
\usepackage{subfigure}
\newcommand{\MColumn}[2]{
	\begin{pmatrix}
		#1 \\ #2 \end{pmatrix}
}

\newcommand{\vecr}{\bold{r}}

\newcommand{\vecq}{\bold{q}}

\newcommand{\mel}[3]{\langle{#2}|{#1}|{#3}\rangle}

\def\beq{\begin{equation}}
\def\eeq{\end{equation}}

\usepackage[normalem]{ulem}

\title{Neutrino force at all length scales}
\author[a]{Mitrajyoti Ghosh,}
\emailAdd{mghosh2@fsu.edu}
\author[b]{Yuval Grossman,}
\emailAdd{yg73@cornell.edu}
\author[b]{Chinhsan Sieng,}
\emailAdd{cs2284@cornell.edu}
\author[b]{Bingrong Yu}
\emailAdd{bingrong.yu@cornell.edu}

\affiliation[a]{Department of Physics, Florida State University, Tallahassee, FL 32306-4350, USA}
\affiliation[b]{Department of Physics, LEPP, Cornell University, Ithaca, NY 14853, USA}

\abstract{
The Standard Model predicts a long-range force mediated by a pair of neutrinos, known as ``the neutrino force". It scales as $G_F^2/r^5$, where $G_F$ is the Fermi constant. However, as $r \lesssim \sqrt{G_F}$, the four-Fermi theory breaks down and the neutrino force no longer has the $1/r^5$ scaling. For the first time, we derive a complete expression for the neutrino force that is valid at all distances. For $r \gg \sqrt{G_F}$, the result reduces to the known $G_F^2/r^5$; for $r \ll \sqrt{G_F}$, it scales as $1/r$. We explore the implications of this result for atomic parity violation (APV) experiments. 
A key feature of the neutrino force is that it is a long-range effect compared to the atomic length scale. Thus, in general, it cannot be simply treated as a correction to the tree-level $Z$-exchange diagram without considering the atomic wavefunctions.
We calculate the effects in muonium and positronium, finding that the neutrino force contributes about 4\% and 16\%, respectively, compared to the leading $ Z$ exchange.
This indicates a significant impact on APV, with important implications for detecting the neutrino force and measuring the weak mixing angle in APV experiments.}

\begin{document}
\maketitle
\flushbottom

\section{Introduction}

The Standard Model (SM) predicts a long-range force between two fermions a distance $r$ apart due to the exchange of a pair of neutrinos, which goes as (neglecting the neutrino mass)~\cite{Feinberg, Feinberg:1989ps,Hsu:1992tg} 
\begin{align}
\label{eq:Veff}
V_{\rm eff}(r)\sim \frac{G_F^2}{r^5}\;,
\end{align}
where $G_F$ is the Fermi constant. The neutrino force is purely a loop effect and is sometimes called a quantum force, as opposed to a classical force which is mediated by a tree-level boson exchange. In the last two decades, various aspects of the neutrino force have been investigated, such as the effect of neutrino masses and lepton flavor mixing \cite{Grifols:1996fk,Lusignoli:2010gw,LeThien:2019lxh,Costantino:2020bei,Segarra:2020rah}, background corrections~\cite{Horowitz:1993kw, Ferrer:1998ju, Ferrer:1999ad,Ghosh:2022nzo,Blas:2022ovz,VanTilburg:2024tst,Ghosh:2024qai,Barbosa:2024tty}, and astrophysical and cosmological implications~\cite{Fischbach:1996qf,Smirnov:1996vj,Abada:1996nx,Kachelriess:1997cr,Kiers:1997ty,Abada:1998ti,Arafune:1998ft,Orlofsky:2021mmy,Coy:2022cpt}. One noteworthy feature of the neutrino force is that its exact form is sensitive to the absolute neutrino mass as well as the nature of its mass: Dirac or Majorana~\cite{Grifols:1996fk,Segarra:2020rah,Costantino:2020bei,Ghosh:2022nzo}.

\begin{figure}[t]
\centering
\includegraphics[scale=0.8]{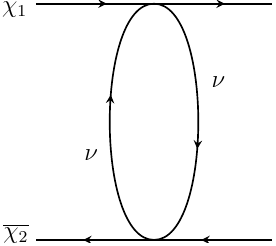}
\caption{\label{fig:4-fermi} The two-neutrino mediated force in the framework of four-Fermi effective theory, which scales as $G_F^2/r^5$. The four-Fermi effective vertex is only valid at long distances when $r \gg \sqrt{G_F}$.}
\end{figure}

One can get Eq.~(\ref{eq:Veff}) from the argument of dimensional analysis. This result, however, is based on the four-Fermi approximation (see Fig.~\ref{fig:4-fermi}), and is only valid when $r \gtrsim \sqrt{G_F}$. As the two fermions approach each other, the four-Fermi approximation is not valid at $r \lesssim \sqrt{G_F}$, and Eq.~(\ref{eq:Veff}) no longer applies.

\begin{figure}[t]
    \centering
    \subfigure[Self-energy Diagram]{
        \includegraphics[scale = 0.8]{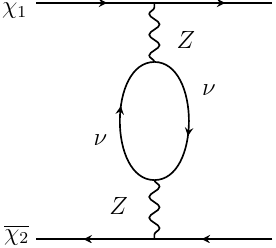}
        \label{subfig:self-energy}
    }
    \hspace{1cm}
    \subfigure[Penguin Diagram (${\rm PG_1}$)]{
        \includegraphics[scale = 0.8]{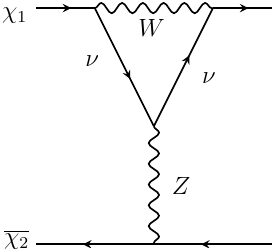}
        \label{subfig:penguin1}
    } \\
    \subfigure[Penguin Diagram (${\rm PG_2}$)]{
        \includegraphics[scale = 0.8]{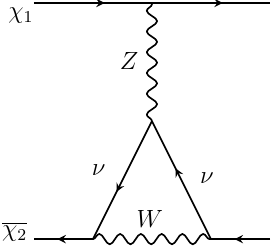} 
        \label{subfig:penguin2}
    }
    \hspace{1cm}
     \subfigure[Box Diagram]{
        \includegraphics[scale = 0.8]{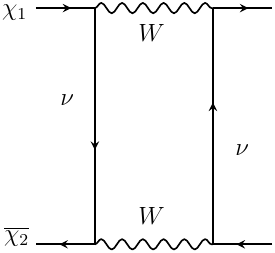}
        \label{subfig:box}
    }
    \caption{\label{fig:beyond-4-fermi}The three possible ways to open up the four-Fermi effective vertex at short distances ($r \lesssim \sqrt{G_F}$) for the Standard Model neutrino force: (a) self-energy diagram, (b)+(c) penguin diagram, and (d) box diagram.}
\end{figure}

To describe the neutrino force at short distances correctly, one has to open up the four-Fermi effective vertex and calculate the amplitude in the ultraviolet theory.
Although the $G_F^2/r^5$ long-range form has been known for more than fifty years~\cite{Feinberg}, the complete short-range behavior of the SM neutrino force is still lacking in the literature. Some partial results were derived recently~\cite{Xu:2021daf,Dzuba:2022vrv,Munro-Laylim:2022fsv}. In Ref.~\cite{Xu:2021daf}, the short-range neutrino force was studied in the beyond-Standard-Model (BSM) scenario, where neutrino is assumed to couple to some new light scalar. Refs.~\cite{Dzuba:2022vrv,Munro-Laylim:2022fsv} studied the SM neutrino force at short distances, but only the self-energy diagram (see Fig.~\ref{fig:beyond-4-fermi}) was included. 

In this work, we go beyond the four-Fermi approximation and perform a complete calculation of the SM neutrino force in the renormalizable electroweak theory. There are three kinds of diagrams which may contribute to the SM neutrino force at short distances: self-energy diagrams, penguin diagrams, and  box diagrams (see Fig.~\ref{fig:beyond-4-fermi}). As we will show, the final result is gauge invariant as it should, and is applicable at all length scales. 

Such an investigation is not only of theoretical interest but is also driven by experimental motivation. The neutrino force in the SM is too small to be measured by current experiments due to the $G_F^2$ suppression and the $1/r^5$ dependence~\cite{Kapner:2006si, Adelberger:2006dh, Chen:2014oda, Vasilakis:2008yn, Terrano:2015sna}. Several approaches have been proposed to address this.

First, neutrino backgrounds, such as the cosmic neutrino background or solar neutrino flux, can significantly enhance the neutrino force~\cite{Horowitz:1993kw, Ferrer:1998ju, Ferrer:1999ad,Ghosh:2022nzo,Blas:2022ovz,VanTilburg:2024tst,Ghosh:2024qai}. Specifically, with a well-directed neutrino beam, the background-induced force scales as $V_{\rm bkg} \sim G_F^2 \Phi E_\nu/r$, where $\Phi$ is the flux and $E_\nu$ is the average energy~\cite{Ghosh:2022nzo}. This results in a notable enhancement at macroscopic scales compared to the vacuum force in Eq.~(\ref{eq:Veff}).
Second, if neutrinos have BSM interactions with dark matter, the neutrino-mediated loop force can influence dark matter self-interactions, affecting small-scale structure formation and the dark matter annihilation cross section~\cite{Orlofsky:2021mmy,Coy:2022cpt}. Alternatively, one can substitute neutrinos with other light particles, leading to similar long-range quantum forces that aid in dark matter searches~\cite{Brax:2017xho,Fichet:2017bng,Costantino:2019ixl,Banks:2020gpu,Brax:2022wrt,Bauer:2023czj}.
In this work, we will focus solely on the SM neutrino force induced by vacuum fluctuations, excluding effects from neutrino backgrounds or non-standard interactions.

Beyond macroscopic probes, the search for the neutrino force extends into atomic and subatomic scales~\cite{Dzuba:2017cas, Ghosh:2019dmi}. One promising phenomenon that probes this force is atomic parity violation (APV) (see e.g. \cite{Safronova:2017xyt,Wieman:2019vik,Arcadi:2019uif} for a review). The neutrino force violates parity and contributes to the forbidden electric-dipole transition element in the atom. As a result, it can be probed by looking at the parity-violating atomic transition matrix element. Ref.~\cite{Ghosh:2019dmi} calculated the neutrino-induced APV effects for the hydrogen
atom in the four-Fermi effective theory. However, the calculation in \cite{Ghosh:2019dmi} is only valid for $\ell \geq 2$ (where $\ell$ is the orbital angular momentum), the result of which is too small to be probed. For smaller $\ell$, the calculations of the atomic matrix element do not converge, indicating the failure of four-Fermi approximation. In APV experiments, so far, measurements were done for $s$-states (i.e., $\ell =0$). The wave function of such states is concentrated at $r=0$, a region where the four-Fermi approximation breaks down. Therefore, in order to perform a self-consistent calculation of the APV effects caused by neutrinos for $\ell = 0$, it is necessary to know the short-range behavior of the neutrino force. This calls for a calculation of APV effects using a form for neutrino force that is valid at all length scales.

In this work, after a complete calculation of the full-range neutrino force including all the relevant diagrams in Fig.~\ref{fig:beyond-4-fermi}, we use our results to study APV effects in muonium ($\mu^+ e^-$) and positronium ($e^+ e^-$). Both are bound states, so the distance can decrease all the way to $r=0$ without spoiling the non-relativistic (NR) approximation. In contrast with a typical atomic nucleus, $\mu^+$ and $e^+$ are point-like particles: the lack of hadronic structure of such atoms not only enables a very precise calculation of their energy levels but also APV observables as well. Moreover, there are noticeably many ongoing and planned experiments such as laser spectroscopy that studies Lamb shifts in muonium and positronium~\cite{Crivelli:2018vfe, Cassidy:2018tgq, Ohayon:2021dec, Mu-MASS:2021uou, Janka:2021xxr, Adkins:2022omi, Sheldon:2023eic}. In this work we concentrate on these systems, but our results are applicable to atoms like hydrogen and Cesium as well.

The rest of the paper is organized as followed.
In Sec.~\ref{sec:four-fermi}, we briefly review the neutrino force in the framework of the four-Fermi effective theory and explain the limitations of the result.
In Sec.~\ref{sec:full-theory}, we present the calculation of the neutrino force in the full electroweak theory, including all the relevant diagrams. The implications of our results to experiments are discussed in the next three sections. In Sec.~\ref{sec:APV}, we introduce the basic idea to use APV as an experimental probe of the neutrino force. In Sec.~\ref{sec:APVneutrino}, we calculate the effects of neutrino forces on APV observables in muonium and positronium. We also highlight the important impact of the neutrino force on extracting the weak mixing angle from APV experiments. Other quantum forces in the SM as well as their APV effects are calculated in Sec.~\ref{sec:APVother} for completeness. We comment on and compare our results with existing literature in Sec.~\ref{sec:litcomp} and conclude in Sec.~\ref{sec:conclusion}. Some technical details of our computations are provided in the three appendices.

\section{The neutrino force in the four-Fermi effective theory}\label{sec:four-fermi}

In this section, we give a brief review of the neutrino forces in the four-Fermi effective theory. Below the electroweak scale $1/\sqrt{G_F}\approx 300~{\rm GeV}$, the interaction between neutrinos and a fermion $\chi$ can be described by the effective Lagrangian \cite{Fermi:1934sk}:
\beq 
\mathcal{L}_{\rm eff} = -\frac{G_F}{\sqrt{2}} \left[ \overline{\chi} \gamma^\mu (g_V^\chi - g_A^\chi \gamma^5)\chi \right] \left[ \overline{\nu}\gamma_\mu (1- \gamma^5) \nu \right], \label{eq:Leff}
\eeq
where $g_V^\chi$ and $g_A^\chi$ are the vector and axial-vector effective couplings between $\chi$ and the neutrinos. Given this Lagrangian, an elastic scattering between two fermions $\chi_1$ and $\chi_2$ can be mediated by a pair of neutrinos, see Fig \ref{fig:4-fermi}. Due to the smallness of the neutrino masses, this interaction is long-ranged. In fact, for neutrino masses $m_\nu \sim 0.1$ eV, the range of the corresponding force is $1/m_\nu \sim 10^{-4}$ cm, which is larger, by about 12 orders of magnitude, than the range of the weak force, $1/m_Z\sim 10^{-16}~{\rm cm}$, where $m_Z$ is the mass of $Z$ boson. 

The known leading-order result is given by~\cite{Feinberg, Feinberg:1989ps,Hsu:1992tg}:
\begin{align}
    \label{eq:Veff2}
    V_{\rm eff}(r) = \frac{g_V^{\chi_1} g_V^{\chi_2} }{4 \pi^3}\frac{G_F^2}{r^5}\;,
\end{align}
where $r$ is the separation between $\chi_1$ and $\chi_2$.
We note the following points regarding Eq.~(\ref{eq:Veff2}):
\begin{enumerate}
\item 
The neutrino mass has been neglected, which is valid as long as $r\ll 1/m_\nu$.
\item 
We work in leading-order in the velocities of the fermions.
\item
The dependence of the force on fermion spins is not included. For a complete consideration of the spin-dependence effects, see Ref.~\cite{Ghosh:2024qai}.
\item
Note that if an antiparticle is involved, then there is an additional minus sign in Eq.~(\ref{eq:Veff2}).
\item 
Depending on the sign of $g_V^\chi$, as well as the particle or antiparticle nature of the fermions, the resulting force can be either attractive or repulsive.
\end{enumerate}

The four-Fermi approximation breaks down for  $r \lesssim \sqrt{G_F}$, a region where the momentum transfer $|{\bf q}|\sim 1/r$ becomes larger than the electroweak scale. Therefore, the $1/r^5$ force in Eq.~(\ref{eq:Veff2}) is not applicable all the way down to $r=0$. In fact, a general inverse-power potential $V\sim 1/r^n$ with $n>2$ is called \emph{singular} at the origin, in the sense that the Schr\"{o}dinger equation does not have a series solution near $r=0$~\cite{Frank:1971xx}. Moreover, for an attractive singular potential, it can be shown that the solution oscillates infinitely rapidly close to the origin, corresponding to an unphysical result without a unique bound-state spectrum~\cite{Frank:1971xx}, see also Refs.~\cite{Coy:2022cpt,Bedaque:2009ri,Chen:2009ch,Chaffey:2021tmj}. 

This can be understood intuitively by an argument of Landau and Lifshitz~\cite{landau2013quantum}:
As $r$ approaches zero, the momentum scales as $1/r$ and  the kinetic energy of the particle increases as $1/r^2$. However, if the potential decreases with a greater power than two, it will dominate over the kinetic energy. This causes the total energy to become unbounded from below, rendering the theory invalid. Therefore, for the consistency of the theory one can expect that the SM neutrino force should not be more singular than $1/r^2$ at short distances.

\section{The neutrino force beyond the four-Fermi approximation}\label{sec:full-theory}

As argued in the previous section, we require a form of the neutrino force at $r \lesssim \sqrt{G_F}$. In this section, we perform a complete calculation of the neutrino force in the renormalizable electroweak theory.

\subsection{Formalism}\label{subsec:formalism}
The relevant SM Lagrangian is given by 
\begin{align}
    \label{eq:Lfull}
   \mathcal{L}  & \supset  \frac{g}{4 c_W} \left[ \overline{\nu_\alpha} \gamma^\mu \left(1 -  \gamma^5\right) \nu_\alpha  +   \overline{\ell_\alpha} \gamma^\mu \left(c^{\alpha}_V - c^{\alpha}_A \gamma^5\right)  \ell_\alpha + \overline{q_j^{}} \gamma^\mu \left(c^{q_j}_V - c^{q_j}_A \gamma^5\right)  q_j^{} \right] Z_\mu \nonumber\\
   &+ \frac{g}{2 \sqrt {2}}\overline{\ell_\alpha} \gamma^\mu\left(1 - \gamma^5\right)  \nu_\alpha W_\mu^{-} +{\rm h.c.} \;,
\end{align}
where $\alpha$ and $j$ sum over the lepton and quark flavors, respectively, $g\approx 0.65$ is the gauge coupling of the ${\rm SU}(2)_{\rm L}$ group, $c_V$ and $c_A$ are the vector and axial-vector couplings of leptons/quarks to the $Z$ boson, given by:
\begin{align}
    \label{eq:fundcoupling}
  \text{charged-leptons} &: \quad c_V^{\alpha} =  4s_W^2 -1\;,\quad
    c_A^{\alpha} = -1\;,\nonumber\\
   \text{up-type quarks} &: \quad c_V^{u} = 1-\frac{8}{3}s_W^2\;,\quad c_A^{u} = 1 \;, \nonumber\\
   \text{down-type quarks} &: \quad c_V^{d} = \frac{4}{3}s_W^2-1\;,\quad c_A^{d} = -1\;,
\end{align}
with $c_W \equiv \cos \theta_W$, $s_W \equiv \sin \theta_W$, and $\theta_W$ is the weak mixing angle. 

In this work we consider massless neutrinos, that is we neglect the neutrino mass and lepton flavor mixing. This is justified as we consider length scales that are much smaller than the inverse of neutrino mass. (We will briefly discuss the effect of the neutrino mass in Sec. \ref{sec:numass} and at the end of Sec.~\ref{sec:conclusion}.)

Given Eq.~(\ref{eq:Lfull}), there are three different ways in the SM to open up the four-Fermi effective vertex that results in the neutrino forces valid at all length scales: self-energy diagram 
(SE), penguin diagrams (${\rm PG_1}$, ${\rm PG_2}$) and box diagram (box), see Fig.~\ref{fig:beyond-4-fermi}.

The full-range neutrino force is determined by the amplitude of the elastic scattering $\chi_1 (p_1) + \overline{\chi_2} (p_2) \to \chi_1 (p'_1) + \overline{\chi_2}(p'_2) $  in Fig.~\ref{fig:beyond-4-fermi}, which can be factored into the following general form:
\begin{align}
    \label{eq:form}
    i \mathcal{M} &= [\Bar{u}(p'_1)\gamma^\mu(G^{\chi_1}_V - G^{\chi_1}_A \gamma_5) u(p_1)][\Bar{v}(p_2)\gamma^\nu(G^{\chi_2}_V - G^{\chi_2}_A \gamma_5) v(p'_2)] I_{\mu \nu}\;, \nonumber\\
    &\equiv \Sigma_{\chi_1}^\mu \Sigma_{\chi_2}^\nu I_{\mu \nu} \;,
\end{align}
where $p_i$ and $p_i'$ are the incoming and outgoing momentum of the external fermions, $u$ and $v$ denote the spinor wavefunctions of the fermions and anti-fermions, 
$G_V^{\chi}$ and $G_A^{\chi}$ are the effective vector and axial-vector couplings given in the Tab.~\ref{table:couplings}, and $I_{\mu \nu}$ is the loop integral specific to the diagram considered. 
Furthermore, let $m_i$ be the mass of fermion $\chi_i$. 

\renewcommand\arraystretch{1.1}
\begin{table}[t]
\centering
\begin{tabular}{|c | c | c | c | c|} 
 \hline
 Effective Couplings & SE & $\rm{PG_1}$ & $\rm{PG_2}$ & box  \\ [0.5ex] 
 \hline
 $G_V^{\chi_1}$ & $c_V^{\chi_1}$ & $\delta_{\ell, \chi_1}$ & $c_V^{\chi_1}$ & $\delta_{\ell, \chi_1}$ $\delta_{\chi_1, \chi_2}$ \\ 
 \hline
 $G_A^{\chi_1}$ & $c_A^{\chi_1}$ & $\delta_{\ell, \chi_1}$ & $c_A^{\chi_1}$ & $\delta_{\ell, \chi_1}$ $\delta_{\chi_1, \chi_2}$ \\
  \hline
 $G_V^{\chi_2}$ & $c_V^{\chi_2}$ & $c_V^{\chi_2}$ & $\delta_{\ell, \chi_2}$  & $\delta_{\ell, \chi_1}$ $\delta_{\chi_1, \chi_2}$ \\ 
 \hline
 $G_A^{\chi_2}$ & $c_A^{\chi_2}$ & $c_A^{\chi_2}$ & $\delta_{\ell, \chi_2}$ & $\delta_{\ell, \chi_1}$ $\delta_{\chi_1, \chi_2}$ \\ 
 \hline
\end{tabular}
\caption{Effective couplings $G_V^\chi$ and $G_A^\chi$ for neutrino self-energy diagram (SE), penguin diagrams ($\rm{PG}_1$, $\rm{PG}_2$), and box diagram (box), where the fundamental couplings $c_V^\chi$ and $c_A^\chi$ are given in Eq.~(\ref{eq:fundcoupling}). Note that $\ell$ denotes charged-lepton, and $\delta_{\ell, \chi} =1$ if $\chi = \ell$ and $\delta_{\ell, \chi} = 0$, otherwise.
}
\label{table:couplings}
\end{table}
\renewcommand\arraystretch{1}
The neutrino force is given by the Fourier transform of the amplitude in the NR limit:
\begin{align}
    \label{eq:Fourertransform}
    V({\bf r}) = -\int \frac{{\rm d}^3 {\bf q}}{\left(2\pi\right)^3} e^{i \bf q \cdot {\bf r}} {\cal A}_{\rm NR}({\bf q})\;,
\end{align}
where $\mathcal{A}_\text{NR} \equiv \mathcal{M} / (4m_1 m_2)$ denotes the normalized amplitude in the NR limit and $q=p_1-p_1'$ is the momentum transfer. In the NR limit, we have $q\approx (0,{\bf q})$.

The integral in Eq.~(\ref{eq:Fourertransform}) can be computed using the dispersion technique. See Ref.~\cite{Feinberg:1989ps} for a detailed review of the dispersion technique and its application to calculate long-range forces. (The original formalism developed in Ref.~\cite{Feinberg:1989ps} assumed that there was no additional pole on the real $t$-axis apart from the branch cut. In Appendix~\ref{app:dispersion}, we generalize the dispersion formalism in the presence of additional poles, which is relevant for the calculation of the neutrino force from self-energy and penguin diagrams.)

Using the dispersion formalism, the spin-independent potential is given by
\begin{align}\label{eq:dispersion}
    V_{\rm SI}(r) &= \frac{1}{8 \pi^2 r} \int_0^\infty {\rm d} t\ \text{Disc}\left[i \mathcal{A}_{\text{NR}} (t)\right] e^{-\sqrt{t} r}\;,
\end{align}
where 
\begin{align}
\text{Disc}\left[i \mathcal{A}_{\text{NR}} (t)\right] \equiv i \mathcal{A}_{\text{NR}} (t+i\epsilon)-i \mathcal{A}_{\text{NR}} (t-i\epsilon)  =-2\,{\rm Im}\left[\mathcal{A}_{\text{NR}} (t)\right] \quad (\text{with $\epsilon \to 0^+$})   
\end{align}
is the discontinuity of the spin-independent amplitude in the complex $t$-plane. Here, $t$ is the value of $q^2$ after analytic continuation (see Appendix~\ref{app:dispersion} for details).
Although the amplitude ${\cal A}_{\rm NR}$ calculated from quantum field theories may be divergent, the discontinuity part of the one-loop amplitude is always finite according to the optical theorem. Therefore, the result in Eq.~(\ref{eq:dispersion}) is also always finite. As we show below, the parity-violating potentials can be also be obtained from Eq.~(\ref{eq:dispersion}) with some spin-dependent prefactors.

In the most general case, there are three different components of the neutrino forces in Eq.~(\ref{eq:Fourertransform}): the spin-independent (SI) part, the spin-dependent parity-conserving (SD-PC) part, and the spin-dependent parity-violating (PV) part~\cite{Ghosh:2024qai}. Note that the spin-independent part is always parity-conserving and the parity-violating part must be spin-dependent. In this paper, we focus on the SI part and the PV part of the force, where the latter contributes to the APV experiments. We work in the Pauli-Dirac basis in the NR limit, where the wave-functions have the following solution:
\begin{equation}
    u(p) \approx \sqrt{2m_1} \MColumn{\xi_1}{\frac{\bm{\sigma}\cdot \mathbf{p}_1}{2m_1} \xi_1}, \qquad v(p) \approx \sqrt{2m_2} \MColumn{\frac{\bm{\sigma}\cdot \mathbf{p}_2}{2m_2} \xi_2}{\xi_2} \;,
\end{equation}
where ${\bm \sigma}=(\sigma_1,\sigma_2,\sigma_3)$ is the vector of Pauli matrices, $\mathbf{p}_i$ is the three momentum, and $\xi_i$ is the two-component spinor of fermion $\chi_i$.
With these approximations, the wave-function part of the amplitude in Eq.~(\ref{eq:form}) is given by:
\begin{align} \label{spinor}
    \Sigma^\mu_{\chi_1} = 2 m_1\left( G_V^{\chi_1} - G_A^{\chi_1} \left(\frac{\bm{\sigma}_1.\mathbf{p}_1}{m_1} - \frac{\bm{\sigma}_1.\mathbf{q}}{2m_1} \right), -G_A^{\chi_1} \bm{\sigma}_1 + G_V^{\chi_1}\left(\frac{\mathbf{p}_1}{m_1} - \frac{\mathbf{q}}{2 m_1} - { i} \frac{\bm{\sigma}_1 \times \mathbf{q}}{2 m_1} \right) \right) \;, \nonumber \\
    \Sigma^\mu_{\chi_2} = 2m_2 \left( G_V^{\chi_2} - G_A^{\chi_2} \left(\frac{\bm{\sigma}_2.\mathbf{p}_2}{m_2} + \frac{\bm{\sigma}_2.\mathbf{q}}{2m_2} \right), -G_A^{\chi_2} \bm{\sigma}_2 + G_V^{\chi_2}\left(\frac{\mathbf{p}_2}{m_2} + \frac{\mathbf{q}}{2 m_2} + { i} \frac{\bm{\sigma}_2 \times \mathbf{q}}{2 m_2} \right) \right) \;,
\end{align}
with ${\bm \sigma}_i \equiv \xi_i^\dagger \bm{\sigma} \xi_i$. 

We next consider the loop integral in  Eq.~(\ref{eq:form}). At leading order in the NR approximation, it takes the form:
\begin{equation}
    \label{eq:decomp}
    I^{\mu \nu} (q) = g^{\mu \nu} I_0 (q^2) + q^\mu q^\nu I_1 (q^2)\;.
\end{equation}
The $q^{\mu} q^\nu$ term when combined with the wave-function part in Eq.~(\ref{spinor}) and the equation of motion only contributes to the spin-dependent parity-conserving part, and thus we do not study it here. Taking the hermitian conjugate, we can then extract the SI part and PV part of the neutrino force in Eq. (\ref{eq:Fourertransform}) as followed:
\begin{align} 
    V_{\text{SI}}(r) &= G_V^{\chi_1} G_V^{\chi_2} V_0(r)\;, \label{eq:SI} \\    
    V_{\text{PV}}(r) & =  \left[-2  G_V^{\chi_2} G_A^{\chi_1} \left( \frac{\bm{\sigma}_1 \cdot \mathbf{p}}{\mu} \right) - \frac{G_V^{\chi_1} G_A^{\chi_2}}{m_1} (\bm{\sigma}_1 \times \bm{\sigma}_2) \cdot \mathbf{\nabla} \right. \nonumber \\
    & \left. \quad\;\; +2 G_V^{\chi_1} G_A^{\chi_2} \left( \frac{\bm{\sigma}_2 \cdot \mathbf{p}}{\mu} \right) + \frac{G_V^{\chi_2} G_A^{\chi_1}}{m_2} (\bm{\sigma}_2 \times \bm{\sigma}_1) \cdot \mathbf{\nabla} \right] V_0(r)\ \label{eq:PV},
\end{align}
where 
\begin{align}
    V_0(r) & = \frac{1}{8 \pi^2 r} \int_0^\infty {\rm d} t\ \text{Disc}[I_0(t)] e^{-\sqrt{t} r}\;,  \label{eq:V0}
\end{align}
and
\begin{equation}
    \mathbf{p} \equiv \frac{\mathbf{p}_1 m_2 - \mathbf{p}_2 m_1}{m_1+m_2}\;, \quad \mu \equiv \frac{m_1 m_2}{m_1+m_2}\;.
\end{equation}
Note that $V_0(r)$ is independent of the fundamental couplings $c_V^\chi$ or $c_A^\chi$. In the special case of $m_2 \gg m_1$ (such as in muonium or normal atoms), we have $\mathbf{p}\approx \mathbf{p}_1$ and $\mu \approx m_1$; in the case of $m_2 = m_1$ (such as in positronium), we have $\mathbf{p} = 0$ and $\mu = m_1/2$, which implies the helicity terms ($\bm{\sigma}_i \cdot \mathbf{p}$) drop out in the center-of-mass frame.

In the following section, we compute $V_0(r)$ for each diagram. From $V_0(r)$ and the effective couplings in Table.~\ref{table:couplings}, one can find $V_{\text{SI}}(r)$ and $V_{\text{PV}}(r)$ as desired using Eqs.~(\ref{eq:SI}) and (\ref{eq:PV}).

\subsection{Calculation of $V_0(r)$}
\label{subsec:V0}
To get $V_0(r)$, one needs to calculate the discontinuity of the one-loop amplitude of each diagram in Fig.~\ref{fig:beyond-4-fermi} across the branch cut in the complex $t$-plane. The details of the calculation are provided in Appendix~\ref{app:neutrino-force} and the results are both finite and gauge invariant, as expected. After obtaining the discontinuity, we substitute it into Eq.~(\ref{eq:V0}) and obtain $V_0(r)$ for each diagram as followed.

\subsubsection{The self-energy diagram}
For the self-energy diagram in Fig.~\ref{subfig:self-energy}, we obtain 
\begin{align} \label{eq:V0SE}
    V_0^{\rm SE}(r) = \left(\frac{g}{4 c_W}\right)^4 \frac{1}{48 \pi^3 r} \left[e^x\left(2+x\right) {\rm Ei}\left(-x\right)+e^{-x}\left(2-x\right) {\rm Ei}\left(x\right)+2\right],
\end{align}
where 
\begin{align}
x \equiv m_Z r\;, \quad  {\rm Ei} (x) \equiv -\int_{-x}^\infty {\rm d}t \  t^{-1} e^{-t}\;.
\end{align}
In Fig.~\ref{subfig:VSE}, $V_0^{\rm SE}$ is plotted as a function of the distance.

In the long-distance limit, $r\gg 1/m_Z$, the expression in Eq.~(\ref{eq:V0SE}) is reduced to
\begin{align}
    \label{eq:SElong}
    V_0^{\rm SE}(r) &= -\frac{g^4}{512\pi^3 m_W^4 r^5} = -\frac{G_F^2}{16 \pi^3 r^5}\;.
\end{align}
In the short-distance limit $r\ll 1/m_Z$, we have
 \begin{align}
    V_0^{\rm SE}(r) &=  -\left(\frac{g}{4 c_W}\right)^4 \frac{1}{24 \pi^3 r}\left[2\log\left(\frac{1}{m_Z r}\right)-1-2\gamma_{\rm E}\right].
\end{align}
where $\gamma_{\rm E}\approx 0.577$ is the  Euler-Mascheroni constant. 

\subsubsection{The penguin diagrams} 
\label{Penguin}

\begin{figure}[t]
    \centering
    
    \subfigure[Self-energy diagram]{
        \includegraphics[width=0.47\textwidth]{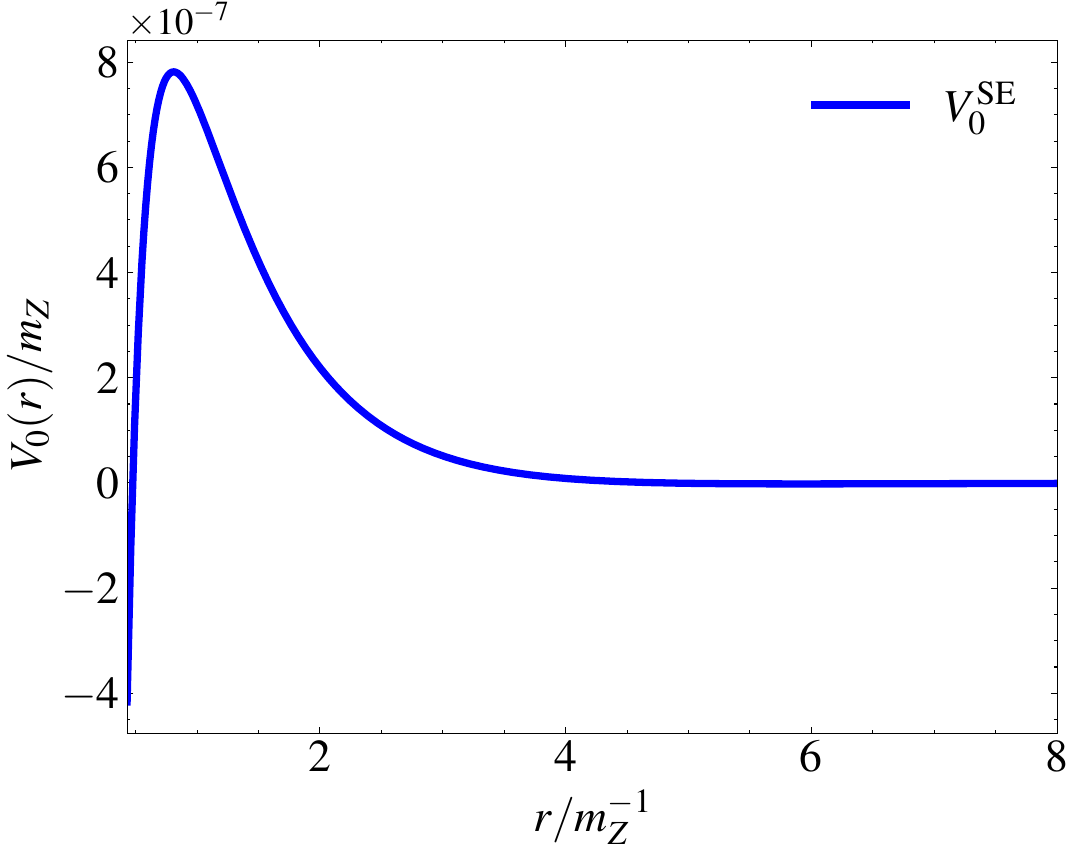} 
        \label{subfig:VSE}
    }
    \hfill
    \subfigure[Penguin diagram]{
        \includegraphics[width=0.47\textwidth]{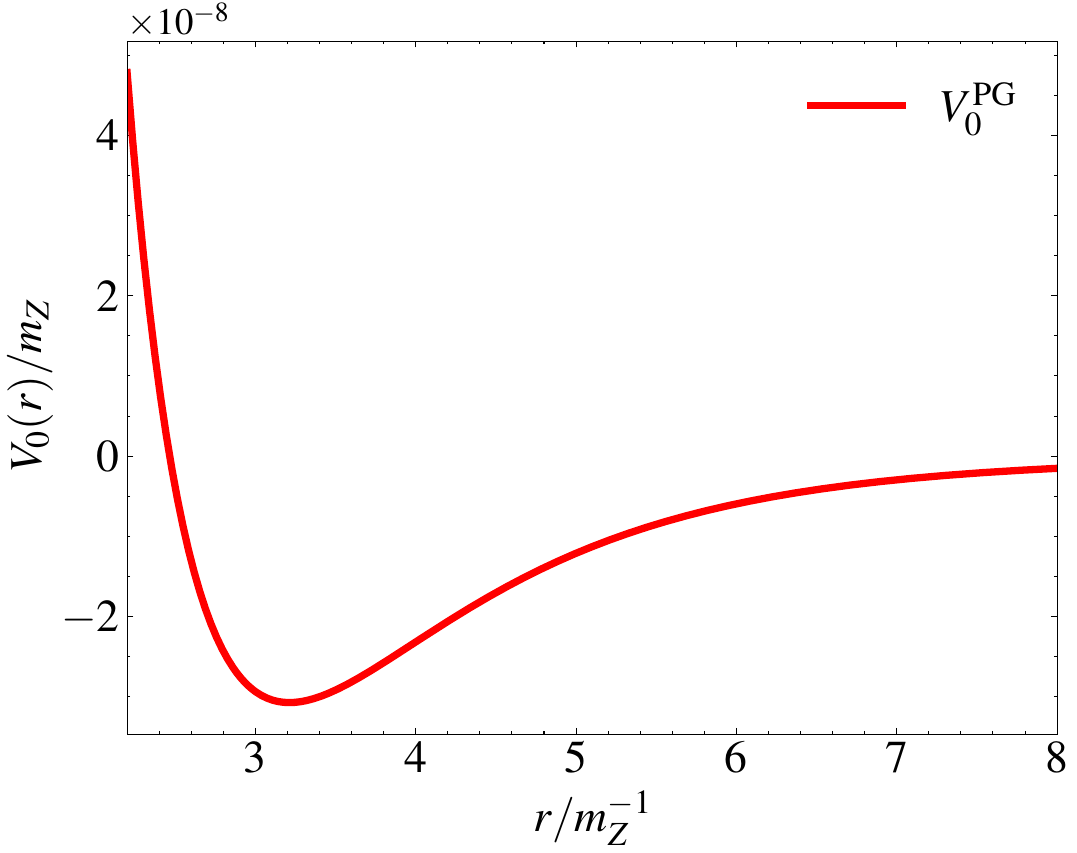} 
        \label{subfig:VPG}
    }
    
    \subfigure[Box diagram]{
        \includegraphics[width=0.47\textwidth]{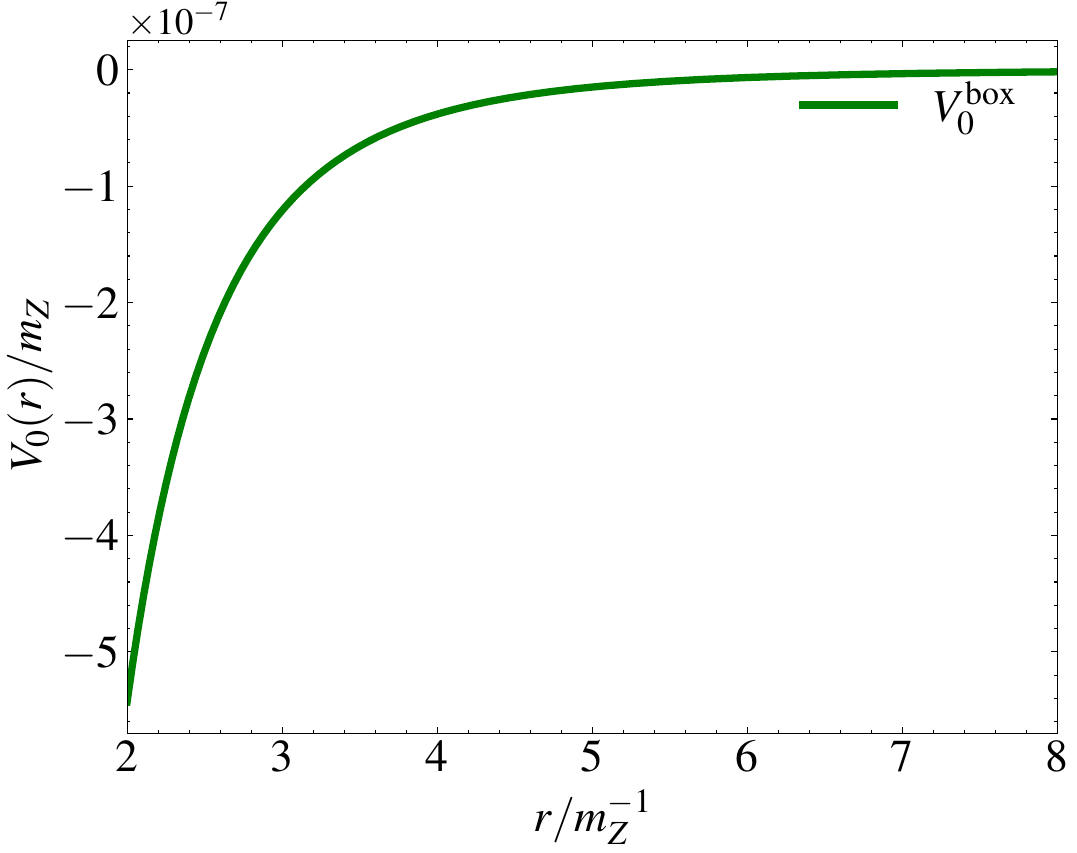} 
        \label{subfig:Vbox}
    }
    
    \caption{$V_0(r)/m_Z$ as a function of the radial distance for the different diagrams that contribute to the neutrino force. Note that the self-energy and penguin potentials change sign while the box potential does not. For a comparison of all three potentials in the same diagram, see Fig.~\ref{fig:V0_all}.}
    \label{fig:V0_separate}
\end{figure}
There are two penguin diagrams of concern, one with a vertex correction at the fermion $\chi_1$, as shown in Fig.~\ref{subfig:penguin1}, and another diagram with a vertex correction at the anti-fermion $\overline{\chi_2}$ as in Fig.~\ref{subfig:penguin2}. The potential $V_0$ for two penguin diagrams has the same structure except for the dependence on the external fermion mass $m_i$. At leading order of $m_i^2/m_Z^2$ we have:
\begin{align} \label{eq:V0PG}
V_0^{\rm PG}(r) = -\frac{g^4}{2048 \pi^3  c_W^2 r} \int_0^\infty {\rm d}t\,e^{-\sqrt{t}\,r}\,\frac{t\left(2m_W^2 + 3t\right) - 2 \left(m_W^2 + t\right)^2 \log \left(1+t/{m_W^2}\right)} {t^2\left(t-m_Z^2\right)} \;,
\end{align} 
where the dependence on the fermion mass drops out.

The $1/(t-m_Z^2)$ factor in Eq.~(\ref{eq:V0PG}) comes from the $Z$ propagator in the penguin diagram, which brings about a singularity for the integral. To obtain a finite and physical result, the pole should be shifted by an infinitesimally small imaginary part according to the Feynman propagator prescription: $t-m_Z^2 \to t-m_Z^2 + i\epsilon$. In the limit of $\epsilon \to 0^+$, a finite imaginary part appears due to the identity ${\rm Im}\left[1/(t-m_Z^2 + i0^+)\right] = -\pi \delta(t-m_Z^2)$. However, as we show in Appendix~\ref{app:dispersion}, this imaginary part is unphysical and is exactly canceled by the imaginary part of the residue of the amplitude at $Z$ pole. As a result, the final expression for the interaction potential is real, as it should be. Therefore, practically, the integral in Eq.~(\ref{eq:V0PG}) should be computed in the following way:
\begin{align} \label{eq:V0PGmod}
V_0^{\rm PG}(r) &= -\frac{g^4}{2048 \pi^3  c_W^2 r}\nonumber\\
& \times {\rm Re}\left[\lim_{\epsilon \to 0^+} \int_0^\infty {\rm d}t\,e^{-\sqrt{t}\,r}\,\frac{t\left(2m_W^2 + 3t\right) - 2 \left(m_W^2 + t\right)^2 \log \left(1+t/{m_W^2}\right)} {t^2\left(t-m_Z^2 + i\epsilon\right)}\right].
\end{align} 
The integral in Eq.~(\ref{eq:V0PGmod}) is well defined and can be calculated numerically at any distance. The result is shown in Fig.~\ref{subfig:VPG}.

The analytical expressions for $V_0^{\rm PG}$ can be found in the asymptotic regions.
In the long-distance limit, $r\gg m_Z^{-1}$, Eq.~(\ref{eq:V0PG}) reduces to the familiar $1/r^5$ form:
\begin{align}
\label{eq:PGlong}
 V_0^{\rm PG}(r) = -\frac{g^4}{256\pi^3 c_W^2 m_Z^2 m_W^2 r^5}
 = -\frac{G_F^2}{8\pi^3 r^5}\;.
\end{align}
In the short-distance limit $r\ll m_Z^{-1}$, using a Mellin transform, we get
\begin{align}
\label{eq:VPGshort}
V_0^{\rm PG}(r) &= \frac{g^4}{2048\pi^3 c_W^2r} \left[ 4 \log ^2\left(e^{\gamma_{\rm E}}
   m_W r\right) + 6 \log \left(e^{\gamma_{\rm E}}m_Z r\right)-\log ^2\left(\frac{m_W^2}{m_Z^2}\right) + \frac{4 \pi ^2}{3} \right],\nonumber\\
&\approx \frac{g^4}{512\pi^3 c_W^2} \frac{1}{r}\log^2(m_W r)\;,
\end{align}
where in the last step we have used the fact that at sufficiently short distances, $\log^2(m_W r)$ dominates over all other terms. 

\subsubsection{The box diagram}\label{Box}
For the box diagram in Fig.~\ref{subfig:box}, at leading order in $m_i^2/m_W^2$ (for $i=1,2$), we find that the result does not depend on the fermion mass:
\begin{align}\label{eq:V0box}
    V_0^{\rm box}(r) = -\frac{g^4}{512\pi^3 r} \int_0^\infty {\rm d}t\,e^{-\sqrt{t}\,r} \frac{m_W^2 t\left(t+2 m_W^2\right) + \left(t^2-2m_W^4\right)\left(t+m_W^2\right)\log\left(1+t/m_W^2\right)}{t^2 \left(t+2m_W^2\right)^2}\;.
\end{align}
Note that this integral is well defined and can be numerically computed at arbitrary distances, as shown in Fig.~\ref{subfig:Vbox}.

At long distances ($r \gg m_W^{-1}$), the result is reduced to
\begin{align}
    \label{eq:boxlong}
    V_0^{\rm box}(r) = -\frac{g^4}{128\pi^3 m_W^4 r^5} = -\frac{G_F^2}{4\pi^3 r^5}\;.
\end{align}
At short distances ($r\ll m_W^{-1}$), the result becomes
\begin{align} 
    V_0^{\rm box}(r) &=-\frac{g^4}{1024 \pi ^3 r} \left[ 4 \log
   ^2\left(e^{ \gamma_{\rm E}} m_W r\right)+ \frac{\pi^2}{2}-1\right]\nonumber \\
    &\approx -\frac{g^4}{256\pi^3 }\frac{\log^2\left(m_W r\right)}{r}\;.
\end{align}

\subsubsection{Combination of all diagrams}
\begin{figure}[t]
    \centering
    \includegraphics[width=\textwidth]{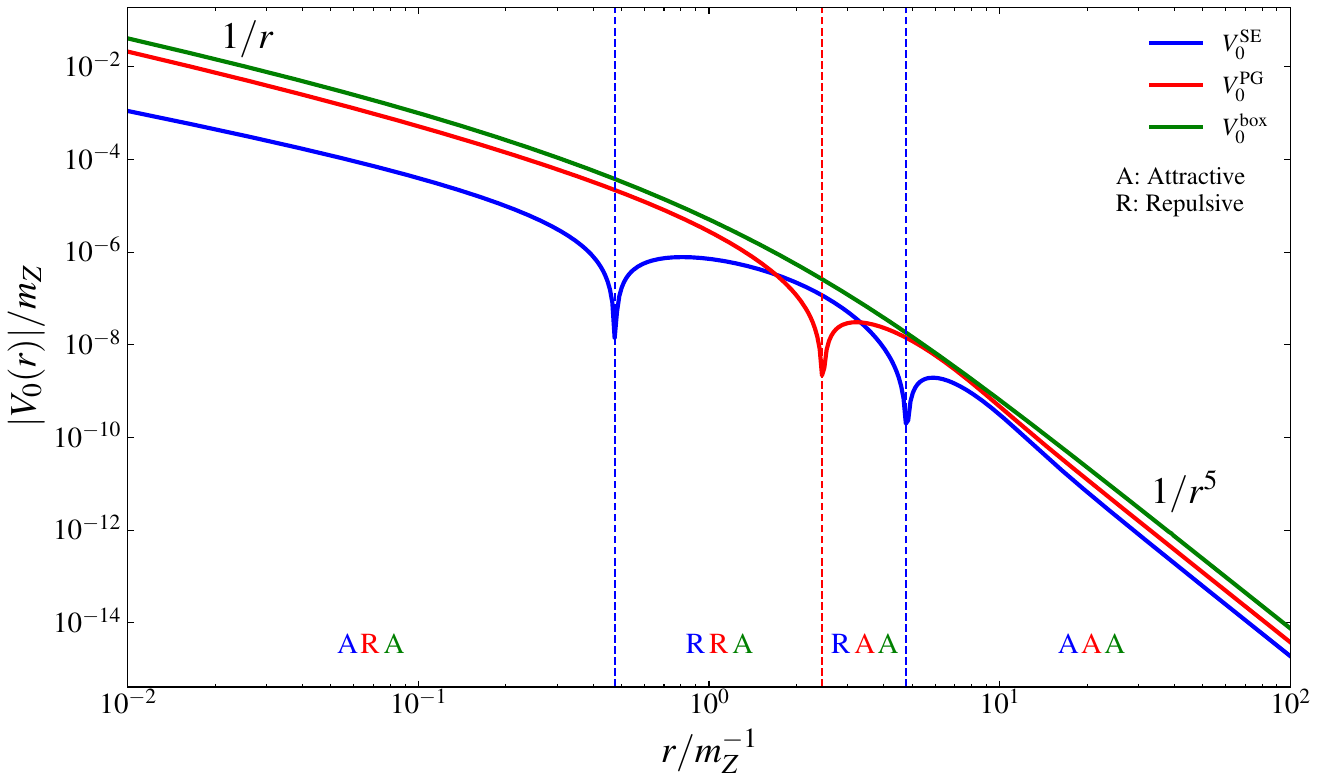} 
    \caption{Magnitudes of the neutrino force from self-energy (SE), penguin (PG), and box diagrams plotted as a function of the radial distance. Labels `A' and `R' represent the sign of the potential that renders it either attractive or repulsive. Note that the plot is made on a log-log scale and so the cusps in the potential occur when the potential changes sign. In reality there are no cusps in the potential.}
    \label{fig:V0_all}
\end{figure}


The neutrino forces given by Eqs.~(\ref{eq:V0SE}), (\ref{eq:V0PG}) and (\ref{eq:V0box}) are calculated in the electroweak theory, so they are valid all the way down to $r=0$. They reduce to $1/r^5$ at long distances as expected, and scale as $1/r$ at short distances when the four-Fermi approximation breaks down. We comment here that the self-energy result in Eq.~(\ref{eq:V0SE}) has the same structure as its BSM counterpart studied in \cite{Xu:2021daf}, where the neutrino is coupled to some light scalar; its long- and short-range limits also agree with \cite{Dzuba:2022vrv,Munro-Laylim:2022fsv}. The results for the neutrino force from penguin and box diagrams in this work are new.

To have a better understanding of the evolution of the neutrino force with distances, we showed $V_0$ for each diagram in Figs.~\ref{fig:V0_separate} and \ref{fig:V0_all}, based on Eqs.~(\ref{eq:V0SE}), (\ref{eq:V0PGmod}) and (\ref{eq:V0box}).

From Fig.~\ref{fig:V0_all}, it can be seen that $V_0$ that results from all the three diagram types scale as $1/r^5$ at long distances while $1/r$ at short distances. However, the penguin and box potentials are much larger than the self-energy potential at short distances due to the $\log^2(m_W r)$ enhancement. In addition, it is interesting to note that $V_0^{\rm SE}$ changes sign twice (corresponding to the cusps in the log-log scale plot in Fig.~\ref{fig:V0_all}),
$V_0^{\rm PG}$ change signs once, while $V_0^{\rm box}$ does not change its sign over the whole range.

As a self-consistent check, one can take the long-range limit ($r\gg m_Z^{-1}$) of the full-range neutrino forces calculated above and obtain Eqs.~(\ref{eq:SElong}), (\ref{eq:PGlong}) and (\ref{eq:boxlong}) for self-energy, penguin and box diagrams. This should be compared with the result predicted in the four-Fermi effective theory. Taking muonium and positronium as an example: The muonium includes SE with three-flavor neutrinos in the loop and two PGs with $\nu_e$ or $\nu_\mu$ in the loop; the positronium includes SE with three-flavor neutrino in the loop, two PGs with $\nu_e$ in the loop, and one box with $\nu_e$ in the loop. 
Using Eq.~(\ref{eq:SI}) and Tab.~\ref{table:couplings}, as well as the couplings in Eq.~(\ref{eq:fundcoupling}), we get the spin-independent neutrino force at long distances:
\begin{align}
    \label{eq:VSImuonium}
    V_{\rm SI}^{e\overline{\mu}}\left(r\gg m_Z^{-1}\right) &= 3 \left(4s_W^2-1\right)^2 V_0^{\rm SE} + 2\left(4s_W^2-1\right) V_0^{\rm PG}\nonumber\\
    &= \frac{G_F^2}{4\pi^3 r^5}\left(1-4s_W^2\right)\left(\frac{1}{4}+3s_W^2\right),\\
    \label{eq:VSIpositronium}
    V_{\rm SI}^{e\overline{e}}\left(r\gg m_Z^{-1}\right) &= 3 \left(4s_W^2-1\right)^2 V_0^{\rm SE} + 2\left(4s_W^2-1\right) V_0^{\rm PG} + V_0^{\rm box}\nonumber\\
    &= -\frac{G_F^2}{4\pi^3 r^5}\left[1-\left(1-4s_W^2\right)\left(\frac{1}{4}+3s_W^2\right)\right].
\end{align}

On the other hand, in the four-Fermi effective theory, the spin-independent neutrino force is given by Eq.~(\ref{eq:Veff2}), where the effective couplings $g_V^\chi$ are determined by matching the electroweak theory onto the four-Fermi effective theory at the tree level: $g_V^e = 1/2+2s_W^2$ (for $\nu_e$ in the loop) and $g_V^e = -1/2+2s_W^2$ (for $\nu_\mu$ or $\nu_\tau$ in the loop);  $g_V^\mu = 1/2+2s_W^2$ (for $\nu_\mu$ in the loop) and $g_V^\mu = -1/2+2s_W^2$ (for $\nu_e$ or $\nu_\tau$ in the loop). Substituting the values of $g_V^\chi$ into Eq.~(\ref{eq:Veff2}), we get (note that an additional minus sign should be added due to the existence of an antiparticle):
\begin{align}
    \label{eq:Veffmuonium}
    V_{\rm eff}^{e\overline{\mu}}(r) &= -\frac{G_F^2}{4\pi^3 r^5}\left[2\left(\frac{1}{2}+2s_W^2\right)\left(-\frac{1}{2}+2s_W^2\right) + \left(-\frac{1}{2}+2s_W^2\right)^2\right]\nonumber\\
    &=\frac{G_F^2}{4\pi^3 r^5}\left(1-4s_W^2\right)\left(\frac{1}{4}+3s_W^2\right),\\
    \label{eq:Veffpositronium}
    V_{\rm eff}^{e\overline{e}}(r) &= -\frac{G_F^2}{4\pi^3 r^5}\left[\left(\frac{1}{2}+2s_W^2\right)^2 + 2\left(-\frac{1}{2}+2s_W^2\right)^2\right]\nonumber\\
    &=-\frac{G_F^2}{4\pi^3 r^5}\left[1-\left(1-4s_W^2\right)\left(\frac{1}{4}+3s_W^2\right)\right].
\end{align}
Comparing Eqs.~(\ref{eq:VSImuonium})-(\ref{eq:VSIpositronium}) with (\ref{eq:Veffmuonium})-(\ref{eq:Veffpositronium}), one can verify that the full-range neutrino force we calculated in the electroweak theory agree with the result predicted from four-Fermi effective theory in the long-range limit.

\subsection{The effect from neutrino mass} \label{sec:numass}
We close this section, with a brief discussion of the effect from finite neutrino mass $m_\nu$ on our result. At long distances $r\gg 1/m_Z$, one can use the four-Fermi approximation, and the neutrino forces including the neutrino mass term are known~\cite{Grifols:1996fk}:
\begin{align}
    V_0^{\rm D} \left(r\gg m_Z^{-1}\right) &= \frac{G_F^2 m_\nu^3}{4\pi^3 r^2}K_3 \left(2m_\nu r\right),\label{eq:VDeff}\\
    V_0^{\rm M} \left(r\gg m_Z^{-1}\right) &= \frac{G_F^2 m_\nu^2}{2\pi^3 r^3}K_2 \left(2m_\nu r\right),\label{eq:VMeff}
\end{align}
where $V_0^{\rm D}$ and $V_0^{\rm M}$ refer to the case of Dirac and Majorana neutrinos, respectively, and $K_3$ and $K_2$ are the modified Bessel functions. Both Eqs.~(\ref{eq:VDeff}) and (\ref{eq:VMeff}) reduce to $1/r^5$ in the limit $m_\nu \to 0$.

The four-Fermi approximation breaks down at short distances below $1/m_Z$. Since $m_\nu \ll m_Z$, it is then sufficient to consider the $r \ll 1/m_\nu$ range. At leading order in the neutrino mass, we obtain:
\begin{align}
V_0^{{\rm D},j} \left(r\ll m_\nu^{-1}\right)  &= V_0^j (r) \left[1+ m_\nu^2 r^2f_{\rm D}^j (m_Z r)\right] + {\cal O}  \left(m_\nu^4 r^4\right) ,\\
V_0^{{\rm M},j} \left(r\ll m_\nu^{-1}\right)  &= V_0^j (r) \left[1+ m_\nu^2 r^2f_{\rm M}^j (m_Z r)\right] + {\cal O}  \left(m_\nu^4 r^4\right) ,
\end{align}
where $V_0^j$ (for $j={\rm SE,PG,box}$) are the potentials in the massless neutrino limit, which have been computed in Eqs.~(\ref{eq:V0SE}), (\ref{eq:V0PG}), (\ref{eq:V0box}). The functions $f_{\rm D}^j$ and $f_{\rm M}^j$ can be calculated using the same strategy in Appendix~\ref{app:neutrino-force}. 

For the self-energy diagram, we are able to compute these functions analytically (the results for the penguin and box diagrams can be calculated numerically), with $x \equiv m_Z r$:
\begin{align}
f_{\rm D}^{\rm SE}(x) = \frac{1}{2}f_{\rm M}^{{\rm SE}}(x) = -\frac{3\left[x e^x {\rm Ei}(-x)-x e^{-x} {\rm Ei}(x)+2\right]}{x^2\left[e^x\left(2+x\right) {\rm Ei}(-x)+e^{-x}\left(2-x\right){\rm Ei}(x)+2\right]}\;.    
\end{align}
The asymptotic behaviors are as follows:
\begin{align}
f_{\rm D}^{\rm SE}\left(x\gg 1\right) = -\frac{1}{2} + {\cal O}\left(\frac{1}{x^2}\right),\;\;
f_{\rm D}^{\rm SE}\left(x\ll 1\right) = -\frac{3}{x^2 \left(1+2\gamma_{\rm E} + 2\log x\right)} +  {\cal O}\left(x^0\right).
\end{align}
Therefore, in the asymptotic range of interest we obtain:
\begin{align}
 V_0^{{\rm D},\,{\rm SE}} \left(m_Z^{-1}\ll r \ll m_\nu^{-1}\right)  &\approx V_0^{\rm SE} (r) \left(1-\frac{1}{2}m_\nu^2 r^2\right),\label{eq:VDfull}\\
  V_0^{{\rm M},\,{\rm SE}} \left(m_Z^{-1}\ll r \ll m_\nu^{-1}\right)  &\approx V_0^{\rm SE} (r) \left(1-m_\nu^2 r^2\right)\label{eq:VMfull}, 
\end{align}
and 
\begin{align}
 V_0^{{\rm D},\,{\rm SE}} \left(r \ll m_Z^{-1}\right) &\approx V_0^{{\rm SE}} (r) \left[1-\frac{m_\nu^2}{m_Z^2} \frac{3}{1+2\gamma_{\rm E}+2\log\left(m_Z r\right)}\right],\\
V_0^{{\rm M},\,{\rm SE}} \left(r \ll m_Z^{-1}\right) &\approx V_0^{{\rm SE}} (r) \left[1-\frac{m_\nu^2}{m_Z^2} \frac{6}{1+2\gamma_{\rm E}+2\log\left(m_Z r\right)}\right], 
\end{align}
where $V_0^{\rm SE}$ is given by Eq.~(\ref{eq:V0SE}). Note the result in Eqs.~(\ref{eq:VDfull})-(\ref{eq:VMfull}) agrees with that derived from four-Fermi approximation (\ref{eq:VDeff})-(\ref{eq:VMeff}) in the corresponding range.  

To conclude, while the neutrino mass affects the neutrino force at short distances $r\lesssim 1/m_Z$, its effect is highly suppressed by $m_\nu^2/ m_Z^2$. It is therefore valid to ignore the neutrino mass for the phenomenological purpose in the remainder of this paper.

\section{Probing the neutrino force: atomic parity violation} \label{sec:APV}

Several potential probes of the neutrino force at atomic/subatomic scales have been studied, such as the spectroscopy method in \cite{Dzuba:2017cas}. The parity-conserving part of the force, specifically the spin-spin coupling term, causes energy shifts between singlet and triplet states in the $s$-wave, which is a correction to the hyperfine splitting. However, it is challenging to isolate the neutrino contributions to these energy differences as higher-order Quantum Electrodynamics (QED) effects can produce energy shifts of comparable magnitude. Since hyperfine splitting is a parity-conserving phenomenon, QED effects always dominate this process. Nevertheless, the contribution of electroweak effects to hyperfine splitting is not new to the literature, see also, for example \cite{Asaka:2018qfg}. 

Given that the neutrino force is so weak, in this work we concentrate on atomic parity violation (APV) as a probe of this force. Since QED strictly conserves parity and neutrinos can only interact weakly, such parity-violating process provides a chance to distinguish the neutrinos contributions from higher-order QED effects.

\subsection{Review of APV}
Parity is violated by electroweak interactions, and the neutrino force, essentially being a correction to weak boson exchanges, also exhibits parity violation. APV refers to the various effects of parity-violating interactions in atomic systems, particularly in the transition rates of electrons between states of opposite parities. In the absence of such interactions, the eigenstates of an atomic system are also eigenstates of parity. Thus atomic transitions obey to selection rules that arise due to parity conservation. An observable such as atomic birefringence (optical rotation of incident light on atoms) is a direct consequence of transitions that violate such selection rules (for detailed reviews and early literature on APV,  see \cite{Bouchiat:1974kt, Bouchiat:1976vg,Bouchiat:1977zp,Bouchiat:1980gp,Bouchiat:1983uf,Bouchiat:1984wz,Bouchiat:1984yh,Bouchiat:1986rm,Bouchiat:1993ek,Bouchiat:1997mj,Bouchiat:2001tba, Bouchiat:2011fs, Guena:2004sq,Guena:2005uj, Johnson:2003ba, Ghosh:2019dmi,Safronova:2017xyt,Wieman:2019vik,Arcadi:2019uif}).

The leading-order parity-violating effect in atoms is due to the tree-level $Z$ exchange. Its form is the same as the one for the neutrino force in Eq.~(\ref{eq:PV}), with $V_0$ replaced by $V_0^Z$ given as:
\beq 
V_0^Z (r) = {g^2 \over 64 \pi c_W^2} {e^{-m_Z r} \over r}\;. \eeq
Note that, for any antiparticle involved as external fermion, a minus sign needs to be added. This ${\cal O}(G_F)$ leading effect was observed in Cesium in 1997 with the experimental error of 0.35\%~\cite{Wood:1997zq}.
Beyond the leading $Z$-effect, radiative corrections from the electroweak sector to APV have been studied long ago~\cite{Marciano:1978ed,Marciano:1982mm,Marciano:1983ss}. 

As previously mentioned, one observable of APV is the optical rotation of light. By measuring the rotation angle of the polarization plane of incident light, one can extract experimentally the value of the nuclear weak charge $Q_W$, defined by:
\beq 
Q_W  \equiv \left(1 - 4 s_W^2+\kappa_1\right){\cal Z}- \left(1+\kappa_2\right){\cal N}\;, \label{eq:qwdef} 
\eeq 
where ${\cal Z}$ is the atomic number, ${\cal N}$ is the neutron number, and $\kappa_1$ and $\kappa_2$ denote the radiative corrections to the weak charge. For example, at one-loop level, we have $\kappa_1, \kappa_2 \sim {\cal O} (\alpha)$, where $\alpha$ is the fine structure constant; their precise forms are predicted by calculating the parity-violating amplitudes at one-loop level and then taking the limit $q^2 \to 0$ (zero momentum transfer)~\cite{Marciano:1978ed,Marciano:1982mm,Marciano:1983ss}. From the measured weak charge and using Eq.~(\ref{eq:qwdef}) one can extract the weak mixing angle $s_W^2$. Comparing the weak mixing angle extracted from APV experiments with the value predicted by the SM serves as a precision test of the SM and can also be used to constrain new physics beyond SM (such as models with extra $Z$ bosons)~\cite{Safronova:2017xyt,Wieman:2019vik,Arcadi:2019uif}.

However, previous studies have neglected the effects of neutrinos on APV. For instance, Refs.~\cite{Marciano:1978ed,Marciano:1982mm,Marciano:1983ss} calculated the one-loop corrections to APV in the SM up to ${\cal O}(\alpha G_F)$. The effects of two-neutrino exchange, which are of the order ${\cal O}(G_F^2)$, were ignored because they were assumed to be too small. The calculations in \cite{Marciano:1978ed,Marciano:1982mm,Marciano:1983ss} were performed only at the amplitude level in the limit of $q^2\to 0$ without taking into account the radial dependence of the atomic wavefunctions. This method can only capture the APV effect when the atomic wavefunctions are concentrated around the origin and is not generically applicable when there exist long-range parity-violating potentials.

To improve upon the existing results, one needs to take into account the full distribution of the parity-violating neutrino force in the atomic system
and convolve it with the atomic wavefunctions. This is precisely described by the accumulated charge defined in Eq.~(\ref{eq:Qc}) and plotted in Fig.~\ref{fig:Q}. For $s$-states ($\ell = 0$), since the atomic wavefunctions are roughly constant around the origin, this convolution is essentially the integral of the neutrino force itself over the atomic length scale; for $\ell \geq 1$ states, the radial dependence of the atomic wavefunctions plays a nontrivial role in the convolution. Phenomenologically, the current most promising observable for APV is the $s$-state parity-violating matrix elements in heavy atoms~\cite{Wood:1997zq}. 
As we will show below, after correctly taking into account the short-range behavior of the neutrino force, the generic contribution from two-neutrino exchange to the atomic parity-violating matrix elements for $s$-states is of order $g^2/(16\pi^2)\sim 0.3\%$ when compared with the leading $Z$ effect. Therefore, the neutrino contribution to APV is at the same order as the ${\cal O}(\alpha G_F)$ contributions computed in \cite{Marciano:1978ed,Marciano:1982mm,Marciano:1983ss}, and should be included for a complete calculation of the one-loop radiative corrections to the parity-violating matrix elements in atoms. Remarkably, the contribution from pure neutrinos is comparable to both the current APV experimental error ($\sim$ 0.35\%)~\cite{Wood:1997zq} and the theoretical uncertainty in the atomic many-body systems calculations ($\sim$ 0.45\%)~\cite{Porsev:2009pr,Dzuba:2012kx}.
This makes us optimistic about the use of atomic systems in probing the neutrino force in the not too distant future.

Apart from the general discussions above, there are two effects that can enhance the relative size of the neutrino force contribution to APV compared to the tree-level $Z$-effect: 
\begin{itemize}
\item[($i$)] In simple atoms such as the muonium and positronium studied in this work, as well as the hydrogen studied in Ref.~\cite{Ghosh:2019dmi}, the tree-level contribution to APV is proportional to $1-4 s_W^2$, which is accidentally suppressed because $s_W^2$ is close to 1/4. As a result, the relative contribution from the neutrino force is accidentally enhanced by a factor of $(1-4 s_W^2)^{-1}\sim 20$. 
\item[($ii$)] For states with angular momentum $\ell > 0$, in the non-relativistic approximation, the electron wavefunctions are not concentrated at the origin, making the tree-level contribution highly suppressed by the very short range of the $Z$ force. 
In this case, the relative contribution of the neutrino force is enhanced. 

(In heavy atoms, the relativistic corrections from the Dirac equation, as well as the effects from electron correlations, are non-negligible, which increases the electron density around the origin for $\ell > 0$ states. As a result, the relative enhancement of the neutrino contribution is suppressed.)
\end{itemize}
Note that these two enhancement effects are not realized in current APV experiments, which have been done only in heavy atoms for $s$-states.

\subsection{Atomic energy states in the presence of parity violation}
The eigenstates of an atomic system where the electrons and nuclei interact via QED are states of definite parity. For example, the eigenstates of the $1/r$ potential in the hydrogen atom have definite parity $(-1)^{\ell}$. In the presence of parity-violating interactions 
the QED Hamiltonian of the atom is perturbed by parity-violating terms, denoted by $V^j_{\rm PV}$, where $j$ denotes the specific source that gives rise to such a term, for example, $j$ = $Z$, SE, $\text{PG}_1$, $\text{PG}_2$, box. Such terms cause mixing of states with opposite parities. 
Schematically, we have:
\beq 
| A \rangle \to |A\rangle + C_{AB} |B \rangle + \dots\;, \label{eq:corrcoeff}
\eeq
where $|A\rangle$ and $|B\rangle$ are two eigenstates with opposite parities that mixed due to the parity-violating interactions. Here
\beq C_{AB} = \sum _j C^j_{AB} \label{eq:corrcoeff2} \eeq is the net correction to the eigenstate in the presence of parity-violating forces, and \beq C^j_{AB} = { \langle B | V^j _{\rm PV} | A \rangle \over E_{A}-E_{B}} \label{eq:corrcoeff3} \eeq are the correction due to the parity-violating forces $V_{\rm PV}^j$. 

Note that these expressions are schematic. In reality the eigenstates we must work with are states of definite total angular momentum $\hat{J} = \hat{L}+\hat{S}_1+ \hat{S}_2$, where $\hat{L}$ is the orbital angular momentum and $\hat{S}_i$ are spins of the particles forming the bound state. To get the precision correction of the neutrino force to the eigenstates, one also needs to include the fine structure and hyperfine splitting since the perturbations induced by the forces here are much smaller than these effects~\cite{Ghosh:2019dmi}. However, in the discussion that follows, we are interested in the APV effects caused by neutrinos relative to the leading $Z$ effect, thus using the above approximation is justified.

In this work, we focus on the ratio between the coefficient $C_{AB}$ induced by loop-level parity-violating forces to that of $C_{AB}$ induced by tree-level $Z$-exchange. For a given loop contribution $j$, we define this ratio as 
\beq \label{eq:etadef}
\eta_j \equiv {C^{j}_{AB} \over C^{Z}_{AB}}\;. 
\eeq 

In the next two sections, we will calculate $\eta_{\nu}$ (contributions from neutrino forces, see Figs.~\ref{fig:4-fermi} and \ref{fig:beyond-4-fermi}) and $\eta_{\gamma}$ (defined as contributions from parity-violating diagrams with a photon-exchange, see Figs.~\ref{fig:4-fermi-photon} and \ref{fig:photon_penguins}), respectively. 

\section{APV from neutrino forces}
\label{sec:APVneutrino}
In this section, we use the full-range neutrino forces derived in Sec.~\ref{sec:full-theory} to calculate the APV effects caused by neutrinos in  muonium ($e \overline{\mu}$) and positronium ($e \overline{e}$). Then we discuss the possible generalization of our results to heavy atom systems such as Cesium. We also demonstrate that the neutrino force has an important effect on the extraction of $\sin^2\theta_W$ from APV experiments.

\subsection{Parity-violating forces in muonium and positronium  systems}

First we consider the parity-violating force due to the tree-level $Z$ exchange. Using Eq.~(\ref{eq:PV}) with $G_V^e = G_V^{\mu} = 4 s_W^2 -1, G_A^e = G_A^{\mu} = -1$, for muonium and positronium we obtain:
\begin{align}
    \label{eq:VPV-tree-muonium}
    V_{{\rm PV}}^{e \overline{\mu},Z}(r) &= \left[\frac{2\left(\bm{\sigma}_{\overline{\mu}}-\bm{\sigma}_e\right) \cdot \mathbf{p}_e}{m_e} +\frac{(\bm{\sigma}_{\overline{\mu}} \times \bm{\sigma}_e) \cdot \mathbf{\nabla}}{m_e}  \right] \left(1-4 s_W^2\right) V_0^Z(r)\;,\\
        \label{eq:VPV-tree-positronium}
     V_{{\rm PV}}^{e \overline{e},Z}(r) &= \frac{2\,(\bm{\sigma}_{\overline{e}} \times \bm{\sigma}_e) \cdot \mathbf{\nabla}}{m_e}  \left(1-4 s_W^2\right) V_0^Z(r)\;,
\end{align}
where
\beq V_0^Z (r) = -{g^2 \over 64 \pi c_W^2} {e^{-m_Z r} \over r}\;. \label{eq:V0Z} \eeq 
The minus sign in $V_0^Z$ reflects that the force between the electron and the anti-muon/positron is attractive. Note that the tree-level parity-violating force has a factor of $(1-4 s_W^2) \approx 0.05$, which arises from the vector coupling of the $Z$ to the leptons (see Eq.~(\ref{eq:fundcoupling}) and use $s_W^2 \approx 0.239$~\cite{ParticleDataGroup:2022pth}).
This suppression factor is accidental, but it is important in the subsequent discussion. 

We now turn to the parity-violating neutrino force. For muonium, two types of diagrams in Fig.~\ref{fig:beyond-4-fermi} contribute: the self-energy diagram with three neutrino flavors in the loops and the penguin diagram with only two flavors allowed in the loop $(\nu_e$ and $ \nu_\mu)$. Note that in the SM, flavor eigenstates are identical to mass eigenstates. Using Eq.~(\ref{eq:PV}), the parity-violating neutrino force in muonium system is found to be:
\begin{align}
    V_{\rm PV}^{e \overline{\mu}}(r) &= \left[\frac{2\left(\bm{\sigma}_{\overline{\mu}}-\bm{\sigma}_e\right) \cdot \mathbf{p}_e}{m_e} +\frac{(\bm{\sigma}_{\overline{\mu}} \times \bm{\sigma}_e) \cdot \mathbf{\nabla}}{m_e}  \right] \left[3 \left(1-4s_W^2\right)  V_0^{\rm SE}(r) - \left(2- 4s_W^2\right) V_0^{\rm PG}(r)  \right],\label{eq:VPV-nu-muonium}
\end{align} 
where $V_0^{\rm SE}$ and $V_0^{\rm PG}$ are given by Eqs.~(\ref{eq:V0SE}) and (\ref{eq:V0PG}), respectively. The factors in front of each $V_0$ are determined by the effective couplings $G_VG_A$ in Tab.~\ref{table:couplings}, see Eq.~(\ref{eq:PV}). In particular, there are two penguins diagrams giving the factor $(+1)(-1) + (+1)(4s_W^2-1) = -(2-4s_W^2)$.

In the case of positronium, there exists 
a box diagram with $\nu_e$ in the loop in addition to the two types of diagram that contribute to the muonium system. Therefore, the parity-violating neutrino force in positronium system turns out to be:
\begin{align}
    V_{\rm PV}^{e \overline{e}}(r) =& \frac{2\,(\bm{\sigma}_{\overline{e}} \times \bm{\sigma}_e) \cdot \mathbf{\nabla}}{m_e}  \left[3 \left(1-4s_W^2\right)  V_0^{\rm SE}(r) - \left(2- 4s_W^2\right) V_0^{\rm PG}(r) + V_0^{\rm box}(r) \right],
    \label{eq:VPV-nu-positronium}
\end{align}
where $V_0^{\rm box}$  is given by Eq.~(\ref{eq:V0box}). 

From Eqs.~(\ref{eq:VPV-nu-muonium}) and (\ref{eq:VPV-nu-positronium}), it can be seen that the contribution from self-energy diagram is suppressed by a factor of $(1-4s_W^2)\sim {\cal O}(10^{-1})$ compared with that from penguin and box diagrams. In addition, as shown in Sec.~\ref{subsec:V0} and Fig.~\ref{fig:V0_all}, at short distances $r\ll m_Z^{-1}$, $V_0^{\rm SE}$ itself is much smaller than $V_0^{\rm PG}$ and $V_0^{\rm box}$, as the latter are enhanced by $\log^2(m_W r)$; while at long distances $r\gg m_Z^{-1}$, $V_0^{\rm SE}$ is only a half of $V_0^{\rm PG}$ (or a quarter of $V_0^{\rm box}$). 
Therefore, for muonium, the dominant contribution to APV comes from penguin diagram, while in positronium it comes from both penguin and box diagrams.

\subsection{Calculation in muonium and positronium systems}
Now we focus on the neutrino force contribution to the parity-violating matrix elements in muonium and positronium. More specifically, we want to compute the ratio:

\beq \eta_{\nu} \equiv {C^{\nu}_{AB} \over C^{Z}_{AB}}\;, \label{eq:etanu}\eeq
where $C_{AB}^\nu$ and $C_{AB}^Z$ are the coefficients which characterize the parity-violating transition rates induced by the neutrino force and the $Z$ force, respectively, as defined in Eq.~(\ref{eq:corrcoeff3}).

\subsubsection{$\ell = 0$ state}
So far, the APV measurements have only been done for $s$-states ($\ell = 0$). For concreteness, we consider the transition matrix element between the $s$- and $p$-states, i.e, schematically  $|A\rangle = |1S \rangle$ and $|B\rangle = |2P\rangle$. (We have checked that the numerical value of the ratio $\eta_\nu$ does not change significantly for other small values of the principal quantum number.)
Note that the energy difference in the denominator of Eq.~(\ref{eq:corrcoeff3}) shall cancel in this ratio and we can focus on the matrix elements $\langle 2P | V^j _{\rm PV} | 1S \rangle$ alone.

Based on the full-range neutrino forces calculated in Sec.~\ref{sec:full-theory}, we can split the neutrino contributions into three parts (self-energy, penguin, and box):
\begin{align}
    \eta_\nu = \eta_\nu^{\rm SE} + \eta_\nu^{\rm PG} + \eta_\nu^{\rm box}\;,
\end{align}
where, according to Eqs.~(\ref{eq:VPV-tree-muonium})-(\ref{eq:VPV-tree-positronium}) and (\ref{eq:VPV-nu-muonium})-(\ref{eq:VPV-nu-positronium}),
\begin{align}
    \eta_\nu^{\rm SE} &= 3\frac{\langle 2P|\nabla V_0^{\rm SE}| 1S \rangle}{\langle 2P| \nabla V_0^{Z}| 1S \rangle}\;,\label{eq:etaSE}\\
    \eta_\nu^{\rm PG} &= -\frac{2-4s_W^2}{1-4s_W^2}\frac{\langle 2P| \nabla V_0^{\rm PG}| 1S \rangle}{\langle 2P| \nabla V_0^{Z}| 1S \rangle}\;,\label{eq:etaPG}\\
    \eta_\nu^{\rm box} &= \frac{1}{1-4s_W^2}\frac{\langle 2P| \nabla V_0^{\rm box}| 1S \rangle}{\langle 2P| \nabla V_0^{Z}| 1S \rangle}\;.\label{eq:etabox}
\end{align}

An important observation regarding the $s$-states that simplifies our efforts greatly is that the $s$-state wavefunctions do not change greatly with respect to the parity-violating potential. This, in turn allows the wave functions and their derivatives to be treated as roughly constant in the region where the potential $V_0$ is defined. In this approximation, the spin operator contributions [see Eqs.~(\ref{eq:VPV-tree-muonium})-(\ref{eq:VPV-tree-positronium}) and (\ref{eq:VPV-nu-muonium})-(\ref{eq:VPV-nu-positronium})] to $\eta$ can be canceled in the numerator and denominator, leaving only the radial integrals to be performed. Note that this simplification can only be performed in the case of $s$-states. Neglecting the spin effects for states with higher angular momentum is not advisable. Nevertheless, we provide as an estimate for the case of $p$-states and $d$-states later in the paper [see Eqs.~\eqref{eq:etaP}-\eqref{eq:etaD}]. For a complete treatment of the spin operators for $\ell \geq 2$ states with four-Fermi approximation, see \cite{Ghosh:2019dmi}.

Thus, as discussed in the previous paragraph, to compute the ratio of the matrix elements in Eqs.~(\ref{eq:etaSE})-(\ref{eq:etabox}), we only need to include the radial part of the wavefunctions: $ |1S\rangle \propto e^{-r/a_0}\;,
 |2P\rangle \propto r\,e^{-r/(2a_0)}$, where $a_0$ is the Bohr radius in muonium or positronium.
Using the explicit expressions of $V_0^Z$, $V_0^{\rm SE}$, $V_0^{\rm PG}$ and $V_0^{\rm box}$ in Eqs.~(\ref{eq:V0Z}), (\ref{eq:V0SE}), (\ref{eq:V0PG}) and (\ref{eq:V0box}), respectively, one can arrive at the analytical result for each of the three terms (at the leading order of $m_e^2/m_Z^2$):
\begin{align}
\eta_\nu^{\rm SE} &= -\frac{g^2}{32\pi^2 c_W^2}\;,\label{eq:etaSEanalytical}\\
\eta_\nu^{\rm PG} &= \frac{g^2\left(1-2s_W^2\right)}{96\pi^2\left(1-4s_W^2\right)}\Bigg[2\pi^2\left(1+c_W^2\right)^2-3\left(7+4c_W^2\right)+6\left(3+2c_W^2\right)\log\left(\frac{1}{c_W^2}\right)\nonumber
\\
&\qquad-6\left(1+c_W^2\right)^2\log^2\left(1+\frac{1}{c_W^2}\right)- 12\left(1+c_W^2\right)^2 {\rm Li}_2\left(\frac{c_W^2}{1+c_W^2}\right)
\Bigg],\label{eq:etaPGanalytical}\\
\eta_\nu^{\rm box} &= \frac{g^2\left(4+3\pi^2\right)}{256\pi^2\left(1-4 s_W^2\right)}\;.\label{eq:etaboxanalytical}
\end{align}
To obtain Eqs.~(\ref{eq:etaPGanalytical})-(\ref{eq:etaboxanalytical}) from Eqs.~(\ref{eq:etaPG})-(\ref{eq:etabox}), we have changed the order of integrations --- we first perform the radial part integral $\int {\rm d}r$ and then perform the momentum square integral $\int {\rm d}t$. Using this trick, although the potentials $V_0^{\rm PG}$ and $V_0^{\rm box}$ in Eqs.~(\ref{eq:V0PG}) and (\ref{eq:V0box}) do not have analytical expressions, the atomic matrix elements induced by them can be computed analytically, as shown in Eqs.~(\ref{eq:etaPGanalytical})-(\ref{eq:etaboxanalytical}). We have also verified that they agree very well with the direct result obtained from 2D numerical integrals.

Taking the SM predicted value of the weak mixing angle (in the $\overline{\rm MS}$ scheme, and at zero momentum transfer)~\cite{ParticleDataGroup:2022pth}, $s_W^2 = 0.23863$, one obtains numerically:
\begin{align}
\eta_\nu^{\rm SE} &\approx  -0.2\%\;,\\
\eta_\nu^{\rm PG} &\approx   4\%\;,\\
\eta_\nu^{\rm box} &\approx 12\%\;.
\end{align}
Therefore, we conclude that the contributions of the neutrino force to APV relative to the tree-level effect in muonium and positronium are respectively given by:
\begin{align}
    \eta_\nu^{e\overline{\mu}} &= \eta_\nu^{\rm SE} + \eta_\nu^{\rm PG} \approx 4\%\;,\label{eq:nuetamu}\\
    \eta_\nu^{e\overline{e}} &= \eta_\nu^{\rm SE} + \eta_\nu^{\rm PG}  + \eta_\nu^{\rm box} \approx 16\%\;. \label{eq:nuetae}
\end{align}
\vspace{0.2cm}
A few comments are in order:
\begin{enumerate}
    \item Naively, the neutrino force is weaker compared to the $Z$ force by a factor of:
    \beq {V_{\text{loop}} \over V_{\text{tree}}} \sim  {g^2 \over 16 \pi^2 \left(1- 4  s_W^2 \right)} \sim \mathcal{O}(5 \%)\;,\eeq
    where the $1-4 s_W^2$ comes from the accidental suppression of the parity-violating $Z$ force in muonium and positronium.
    The value of $\eta_{\nu}$ obtained for muonium is 4\%, which is in agreement with our naive expectations, while for positronium there is an additional enhancement from the box diagram (see below), leading to $\eta_{\nu} \approx 16 $\%.
\item APV in positronium is enhanced because the box diagram is  not suppressed by any couplings. In contrast, both the tree-level $Z$ diagram and the self-energy diagram are suppressed by $(1-4s_W^2)\sim {\cal O}(10^{-1})$ due to the $Z$-boson couplings to fermions.  Note that the potential due to the box diagram $V_0^{\rm box}$ maintain the same sign throughout (see Fig.~\ref{fig:V0_all}), unlike the potential of the penguin diagrams, which changes sign in the middle. In addition, the magnitude of $V_0^{\rm box}$ is about twice of $V_0^{\rm PG}$, as calculated in Sec.~\ref{sec:full-theory}. As a result, $\eta_\nu$ for positronium is about four times that of muonium. However, despite its large APV effect, positronium has a very short lifetime, on the order of nanoseconds, making the observation of APV in positronium impractical at this time. 
 \end{enumerate}   

\subsubsection{Higher-$\ell$ states}
For higher-$\ell$ states, in the non-relativistic limit, the atomic wavefunctions are not concentrated at the origin, thus the tree-level $Z$ 
 contribution is heavily suppressed, thereby increasing the ratio $\eta_\nu$. For instance, using the same strategy above, for the self-energy diagram, we obtain:
    \begin{align} 
        \eta_\nu^{\rm SE} (2P\to 3D) &= \frac{ g^2}{16\pi^2 c_W^2} \left[\log\left(\frac{m_Z}{\alpha m_e}\right)+\log\left(\frac{6}{5}\right)-\frac{38}{15}\right]\approx 5\%\;,\label{eq:etaP}\\
        \eta_\nu^{\rm SE} (3D\to 4F) &= \frac{9 g^2}{1372 \pi^2 c_W^2}\left(\frac{m_Z}{\alpha m_e}\right)^2 \approx 2\times 10^{11}\;. \label{eq:etaD}
    \end{align}
To get the $\eta_\nu$ from penguin and box diagrams, we only need to compare the long-range expansion of their potentials with that of the self-energy diagram.
This is because for $\ell \geq 1$ states, the most contributions of the neutrino force to APV matrix elements come from the region $r\gg 1/m_Z$ (see Fig.~\ref{fig:Q}). By comparing Eqs.~(\ref{eq:SElong}), (\ref{eq:PGlong}), (\ref{eq:boxlong}) as well as (\ref{eq:etaSE})-(\ref{eq:etabox}), we can directly arrive at:
\begin{align}
\ell \geq 1:\quad 
\eta_\nu^{\rm PG} &\approx -\frac{4}{3}\left(\frac{1-2s_W^2}{1-4s_W^2}\right)\eta_\nu^{\rm SE}\approx -15\eta_\nu^{\rm SE}\;,\\
\eta_\nu^{\rm box} &\approx \frac{4}{3(1-4 s_W^2)}\,\eta_\nu^{\rm SE}\approx 30 \eta_\nu^{\rm SE}\;.\label{eq:eta-high-l}
\end{align}

Note that $V_0$ from each of the three diagrams have the same sign at long distances (see Fig.~\ref{fig:V0_all}), however there is an additional minus sign from the factor $-(2-4s_W^2)$ with $V_0^{\rm PG}$ in Eqs.~(\ref{eq:VPV-nu-muonium}) and (\ref{eq:VPV-nu-positronium}); this is why the penguin contribution differs a sign from self-energy and box for $\ell \geq 1$ states. 

In conclusion, due to the very short range of the $Z$ force, for $\ell = 1$ state, $\eta_\nu$ is enhanced by a factor of $\log\left(m_Z^2/(\alpha^2 m_e^2)\right) \approx 30$ compared with the $\ell = 0$ state; while for $\ell \geq 2$ states, $\eta_\nu$ is enhanced by $\left(m_Z^2/(\alpha^2 m_e^2)\right)^{\ell -1}\gg 1$ and the neutrino force contribution is much larger than the tree-level $Z$ contribution. The result for $\ell \geq 2$ agrees with that in \cite{Ghosh:2019dmi} which uses the four-Fermi approximation. 

One should note, however, that the current experiments are still not able to measure APV for $\ell > 0$ states. Moreover, for heavy atoms, the relativistic corrections and electron correlations are significant, which enhances the wavefunctions around the origin, and thus the results in Eqs.~(\ref{eq:etaP})-(\ref{eq:eta-high-l}) might be changed.

\subsubsection{Accumulated charge}
\begin{figure}[t]
    \centering
        \includegraphics[scale = 0.5]{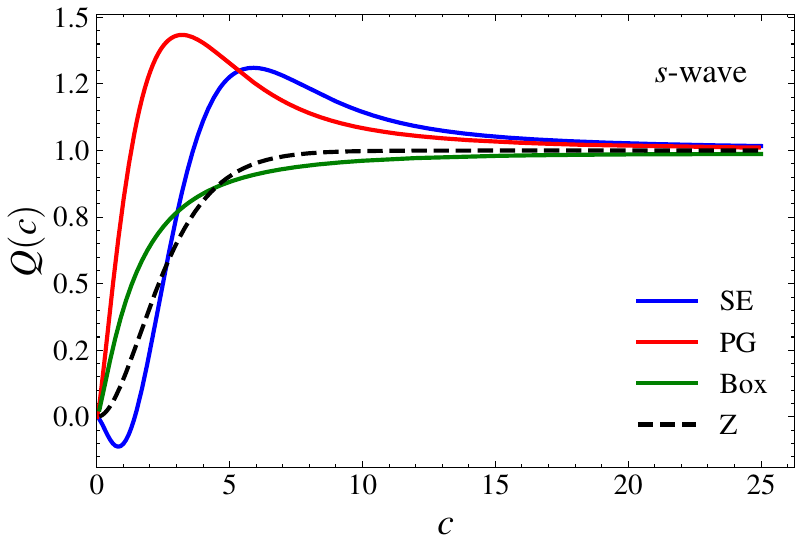}\qquad
        \includegraphics[scale = 0.5]{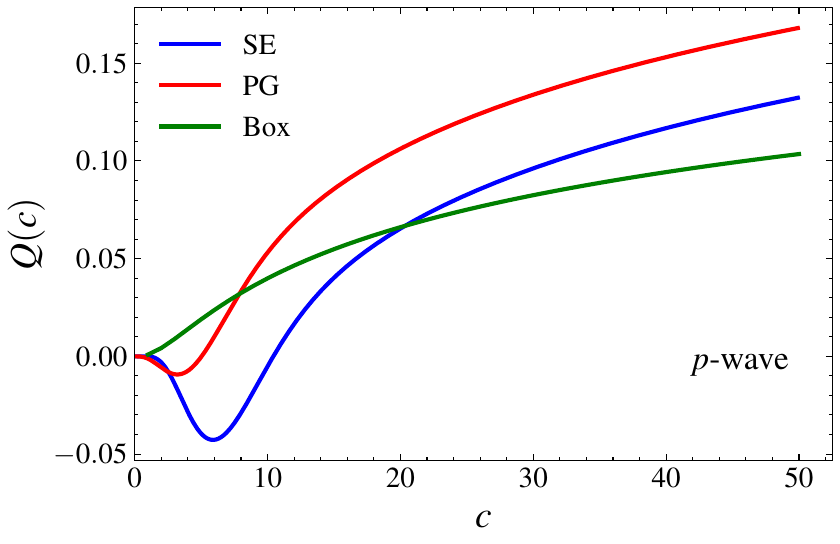}\\
        \includegraphics[scale = 0.5]{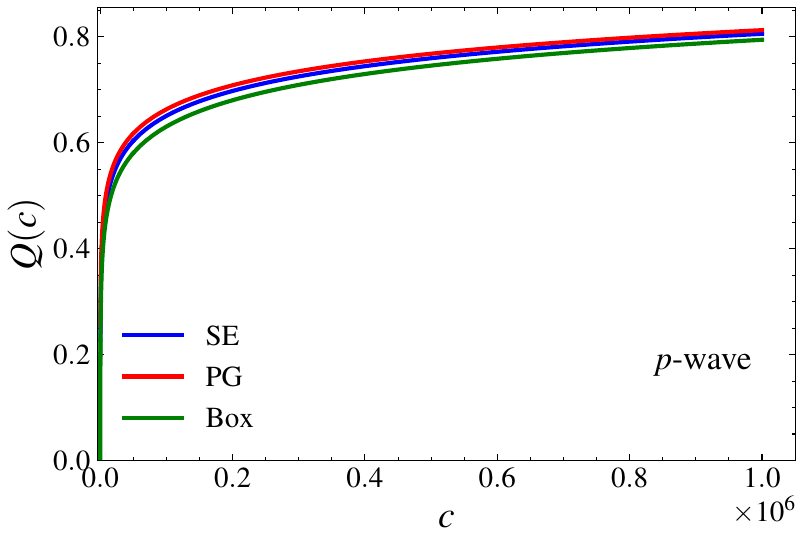}\qquad
        \includegraphics[scale = 0.5]{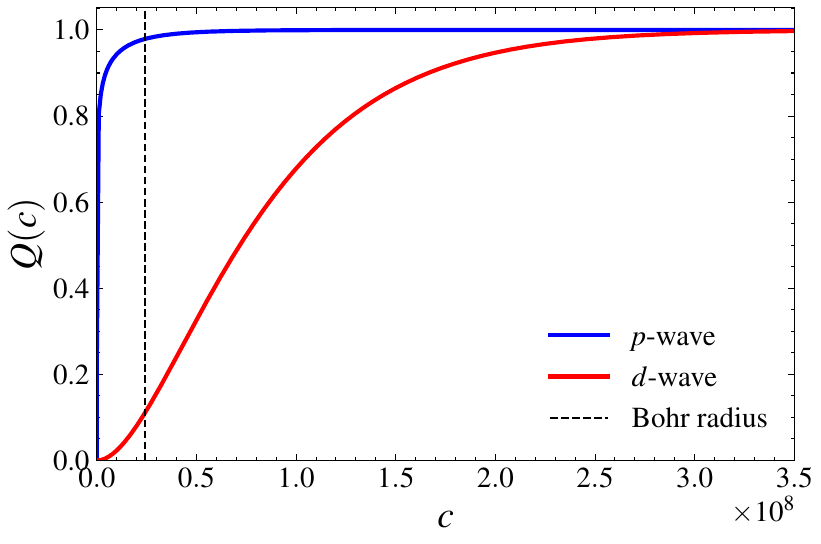}
\caption{\label{fig:Q}Accumulated charges $Q(c)$ for APV matrix elements as a function of the length scale $c\equiv m_Z r$. \emph{Upper left}: $s$-wave $Z$ force and neutrino forces (SE, PG, and Box). \emph{Upper right} and \emph{lower left}: $p$-wave neutrino forces at short and long distances.
\emph{Lower right}: Comparison of $p$-wave and $d$-wave neutrino forces across atomic length scales, where the dashed line corresponds to the Bohr radius.}
\end{figure}
To see the region in which the neutrino force contributes to APV matrix elements most significantly, we define the ``accumulated charge'' of a specific parity-violating potential $V_{\rm PV}^j$:
    \begin{align}
        Q_j^\ell (c) \equiv \frac{\int_0^{c/m_Z} {\rm d}r\,r^2 \psi_{\ell +1}^*V_{\rm PV}^j \psi_\ell^{}} {\int_0^{\infty} {\rm d}r\,r^2 \psi_{\ell +1}^*V_{\rm PV}^j \psi_\ell^{}}\;,\label{eq:Qc}
    \end{align}
where $\psi_\ell$ is the radial wavefunction with angular momentum $\ell$. The accumulated charge characterizes the contribution to APV in the region $r<c/m_Z$ (where $c$ is any positive constant) normalized by the total contribution. For the tree-level $Z$ force with $s$-states, one obtains:
\begin{align}
Q_Z^{\ell = 0}(c) = 1 - \left(1+c+\frac{1}{3}c^2\right)e^{-c}\;.
\end{align}
The exponential suppression tells us 
that the $Z$ contribution to APV mostly comes from the region within $c \lesssim 1$, that is, $r \lesssim 1/m_Z$. At larger distances the change of the accumulated charge gets exponentially suppressed. 

The accumulated charge of the neutrino contribution is nontrivial. For instance, the self-energy diagram gives:
\begin{align}
    Q_\nu^{\ell = 0}(c) &= -\frac{1}{6} c^2 \left[2+e^c\left(c-1\right) {\rm Ei}\left(-c\right) - e^{-c}\left(c+1\right) {\rm Ei}\left(c\right) \right]\nonumber\\
    &\overset{c\gg 1}{=} 1 + \frac{10}{c^2} + {\cal O}\left(\frac{1}{c^4}\right).
\end{align}
As opposed to the $Z$ force that is exponentially suppressed, the neutrino contribution is only quadratically suppressed at large distances. In the upper left panel of Fig.~\ref{fig:Q}, we compare the accumulated charge of the $Z$ force and the neutrino forces for the $s$-state. The bumps of the neutrino curves correspond to the places where the neutrino forces change signs (see Fig.~\ref{fig:V0_all}). This nontrivial feature cannot be captured by using the four-Fermi approximation of the neutrino force. In addition, we see that most of contributions of the neutrino force to the $\ell = 0$ APV matrix element are located in the region $1\lesssim c \lesssim 10$. This is a very short distance compared to typical atomic scales, but is significantly larger than the range of the $Z$ force.

The accumulated charges for higher-$\ell$ states can be computed in a similar way. In particular, in the region $1/m_Z \ll r \ll 1/(\alpha m_e)$, it scales as:
\begin{align}
    Q_\nu^{\ell = 1} (c) &\sim \frac{\log c}{\log\left(\frac{m_Z}{\alpha m_e}\right)}\;,\\
    Q_\nu^{\ell \geq 2} (c) &\sim c^{2(\ell -1)} \left(\frac{\alpha m_e}{m_Z}\right)^{2(\ell -1)}\;.
\end{align}
The accumulated charges for the $p$-wave neutrino forces are shown in the upper right and lower left panels of Fig.~\ref{fig:Q}, which grow logarithmically when $ 1/m_Z \ll r\ll 1/(\alpha m_e)$. As $r$ exceeds the Bohr radius (corresponding to $c=m_Z/(\alpha m_e)\approx 2\times 10^7$), all the accumulated charges tend to unit exponentially.
In the lower right panel of Fig.~\ref{fig:Q}, we compare the accumulated charges of the $p$-state and the $d$-state at large distances. As one can see, for $\ell \geq 1$ states, the contribution from the neutrino force to APV matrix elements has an extensive range up to the atomic length scale.
    
\subsection{Extension to heavy atom systems}
APV experiments using muonium and positronium are not yet realized.
The magnitude of APV from leading $Z$-exchange in these systems, as in the case of hydrogen, is too small to be detected with current experiments. To enhance the sensitivity larger atomic systems with atomic number ${\cal Z}$ are used. They  have a ${\cal Z}^3$-enhancement effect~\cite{Bouchiat:1997mj} for both loop and tree contributions. In particular, the most sensitive experiments were done with Cesium atoms~\cite{Bouchiat:1976vg,Wood:1997zq}.

The ratio $\eta_\nu$ defined in Eq.~(\ref{eq:etanu}) correspond to the corrections to the wavefunctions induced by the neutrino force relative to that from the tree-level $Z$ force. 
These relative corrections are not expected to change significantly with the system size, as increases in size affect both the numerator and denominator of the ratio $\eta_\nu$ similarly. For example, comparing the values of $\eta_{\nu}^{\text{SE}}$ computed in this work for muonium and positronium with the same quantity computed in \cite{Dzuba:2022vrv} for heavy atoms reveal that they are all of the same order of magnitude (see Sec.~\ref{sec:litcomp} for more discussions).  This observation suggests that the effects of many-body corrections in heavy atoms are not significant. However, a detailed atomic many-body calculation, including relativistic corrections from the Dirac equation, is necessary to accurately determine the impact of neutrino forces in such systems. Such analyses, however, are beyond the scope of this work.

In line with the above argument, we estimate $\eta_\nu$ in heavy atoms using the value of $\eta_\nu$ computed previously. Unlike the cases of muonium, positronium, and hydrogen, where there are no neutrons in the nucleus, the accidental tree-level suppression of $1 - 4 s_W^2$ does not exist in heavy atoms. This is because, in the latter case, both the $Z$ force and the neutrino force contributions to APV are proportional to the vector coupling of the neutron to the $Z$ boson, which is not accidentally suppressed like the vector couplings of charged leptons or protons. According to the above analysis, we expect:
\begin{align}
\text{muonium/hydrogen}&: \quad \left|\eta_\nu\right| \sim \frac{g^2}{16\pi^2\left(1-4 s_W^2\right)}\sim 5\%\;,\\
\text{heavy atoms}&: \quad \left|\eta_\nu\right| \sim \frac{g^2}{16\pi^2}\sim 0.3\%\;.\label{eq:nuetaheavy}
\end{align}

Remarkably, $\eta_\nu \sim 0.3\%$ in heavy atoms indicates that the contribution of the neutrino force is comparable to the current experimental error in APV experiments, which is about $0.35\%$~\cite{Wood:1997zq}, and to the theoretical uncertainty of many-body calculations in the cesium atom, estimated to be about $0.45\%$~\cite{Porsev:2009pr,Dzuba:2012kx}. Note that this $\sim 0.3\%$ effect refers to the contribution from two-neutrino exchange only, and there could be other comparable contributions to APV at the one-loop level that are not discussed here. Yet, we emphasize that the neutrino force contribution we computed above is nontrivial, in the sense that it cannot be absorbed into the tree-level vertex, is independent of the renormalization scheme, and was missed in the previous literature~\cite{Marciano:1978ed,Marciano:1982mm,Marciano:1983ss}.

Using the values of $\eta_\nu$ obtained above, in the next subsection we calculate the neutrino effects on the extraction of the weak mixing angle from APV experiments in both muonium/positronium and heavy atom systems.

\subsection{Effects of neutrino forces on the extraction of $\sin^2\theta_W$}
The current goal of APV experiments is to measure weak charges of heavy atoms and from the weak charge, extract the weak mixing angle $\theta_W$, which can be compared with the $\theta_W$ predicted by the SM at zero momentum transfer. Interestingly, there exists a discrepancy between the SM predicted value and the value reported from APV experiments of about 1.5$\sigma$ (see \cite{Dzuba:2002kx, Cadeddu:2018izq} and the references therein). APV measurements report a value of $\sin^2 \theta_W$ that is about 1\% lower than the results obtained from $Z$ pole measurements, with an experimental error of about 0.8\%. While this discrepancy is not statistically significant, it is interesting to explore possible sources of it.
In \cite{Cadeddu:2018izq}, for example, the authors considered the omission of the distribution of the protons and neutrons in the nucleus as a possible explanation for this discrepancy. In the following, we investigate the effects of the neutrino force on the extraction of $\sin^2\theta_W$ from APV experiments.

During the extraction of the weak charge in Cesium atoms, the tree-level $Z$ force is assumed to be the only source of APV. Because of this assumption, the measured value of $\sin^2\theta_W$ will be different from the SM value, since in the SM, there are other contributions to this force. 

With this assumption in place, let the angle measured in this case be $\theta_W'$, which is defined via the following (the true value is denoted by $\theta_W$):
\beq {-g^2 \over 64\pi \cos^2 \theta_W' }Q_W' {e^{-m_Z r}\over r} \equiv {-g^2 \over 64\pi \cos^2 \theta_W}Q_W {e^{-m_Z r}\over r} + \text{loop contributions}\;, \eeq 
where $Q_W ^{(')}$ is the weak charge of the nucleus, given at the tree level by [see Eq.~(\ref{eq:qwdef})]:
\beq 
\label{eq:qwtree}
Q_W ^{(')} = {\cal Z}\left(1 - 4 \sin ^2 \theta _W ^{(')}\right)- {\cal N}\;, 
\eeq 
Radiative corrections slightly modify the coefficients in front of ${\cal Z}$ and ${\cal N}$ in Eq.~(\ref{eq:qwtree}), but that only causes higher-ordered corrections to our following analysis and thus can be neglected.

If the APV is assumed to come from a tree-level $Z$-exchange that gives a potential $V(r) \sim {g^2}Q_W^{'} e^{-m_Z r}/(r\cos^2\theta_W')$, then,

\beq {\cos ^2 \theta_W'  \over Q_W^{'} } C_{AB} = {\cos ^2 \theta_W  \over Q_W} C^Z_{AB} \propto {g^2 \over 64\pi} \mel{{e^{-m_Z r}\over r}}{B}{A}\;, \eeq
where $C_{AB}$ is the total correction received by the $|A\rangle$ state from the $|B\rangle$ state due to all APV-inducing effects, as defined in Eqs.~(\ref{eq:corrcoeff})-(\ref{eq:corrcoeff3}), while $C^Z_{AB}$ is the correction received only due to the tree-level $Z$-exchange.

Now we use Eq.~(\ref{eq:qwtree}) and the relation $ C_{AB} = C^Z_{AB} (1+\eta)$, where $\eta$ is the ratio of loop contributions to the coefficient $C$ relative to the tree contribution, as defined in Eq.~(\ref{eq:etadef}). We eventually obtain, for a general atomic system:

\beq  {\sin^2 \theta_W ' - \sin^2 \theta_W  \over \sin^2 \theta_W} = \frac{\eta \left[{\cal N}-(1-4 \sin^2 \theta_W ) {\cal Z}\right]   \cot^2 \theta_W }{ \left(1+\eta\right) {\cal N}+{\cal Z} \left[3-\eta  \left(1-4 \sin^2 \theta_W\right)\right]}\;. \eeq
Assuming $\eta \ll 1$, one obtains:
\begin{align}
{\sin^2 \theta_W ' - \sin^2 \theta_W  \over \sin^2 \theta_W} \approx \frac{\eta}{{\cal N}+3{\cal Z}}\left[3{\cal N}+\left(4{\cal N}-3{\cal Z}\right)\left(1-4\sin^2\theta_W\right)\right].
\end{align}
Things are different for atoms without neutrons in the nucleus and for heavy atoms:
\begin{itemize}
\item For muonium/positronium/hydrogen, the result is obtained by taking ${\cal N} \to 0$ and ${\cal Z} \to 1$: 
\beq  {\sin^2 \theta_W ' - \sin^2 \theta_W  \over \sin^2 \theta_W} \approx - \left(1-4 \sin^2 \theta_W\right)\eta\;. \label{eq:devmuonium}\eeq

\item For heavy atoms with ${\cal N}\sim {\cal Z}$, on the other hand, one can neglect the term suppressed by $1-4 \sin^2 \theta_W$ and obtain:
\beq  {\sin^2 \theta_W ' - \sin^2 \theta_W  \over \sin^2 \theta_W} \approx \frac{3{\cal N}}{{\cal N}+3{\cal Z}}\,\eta\;.  \label{eq:devheavy}\eeq
\end{itemize}

Now we compute the effects from the neutrino force in different atoms. For muonium and positronium, the $\eta_\nu$ has been calculated in Eqs.~(\ref{eq:nuetamu}) and (\ref{eq:nuetae}), while for heavy atoms we can use Eq.~(\ref{eq:nuetaheavy}) as an estimate:
\begin{align}
\text{muonium}&: \quad 
\left( {\sin^2 \theta_W ' - \sin^2 \theta_W  \over \sin^2 \theta_W} \right)_\nu \approx -0.2\%\;,\\
\text{positronium}&: \quad 
\left( {\sin^2 \theta_W ' - \sin^2 \theta_W  \over \sin^2 \theta_W} \right)_\nu \approx -0.7\%\;,\\
\text{heavy atoms}&: \quad
\left( {\sin^2 \theta_W ' - \sin^2 \theta_W  \over \sin^2 \theta_W} \right)_\nu \approx \frac{3{\cal N}}{{\cal N}+3{\cal Z}}  \left(-\frac{g^2}{16\pi^2} \right)\overset{\text{$^{133}$Cs}}{\approx} -0.3\%\;.
\end{align}

It is interesting to note that the deviation in the measurement of $\sin^2 \theta_W$ caused by the neutrino force is the same order of magnitude when measured in muonium versus when it is measured in heavy atoms like Cesium. This is because the accidental enhancement factor $(1-4 \sin^2 \theta_W)^{-1}$ to $\eta_\nu$ in muonium is canceled by the same factor in Eq.~(\ref{eq:devmuonium}), making the deviation insensitive to the true value of $\sin^2 \theta_W$, just as the case of heavy atoms. Currently, the discrepancy in measurement of $\sin^2\theta_W$ between SM predicted value and the APV measurement in Cesium is about ${\cal O}(1\%)$, and in that regard $0.3\%$ is definitely not negligible. This is another indication that the contribution to APV from neutrino effects is important, even though for most of the literature it has been ignored.

\section{APV from other SM fermions}
\label{sec:APVother}
The neutrino loops are not the only contributor to APV. In this section, we look at diagrams of other SM fermions at one-loop level that contribute to APV.  The fundamental difference is the fact that all other fermions have masses that are large compared to the inverse Bohr radius.

\subsection{Other quantum forces at one-loop level}
\begin{figure}[t]
    \centering
        \includegraphics[scale = 0.9]{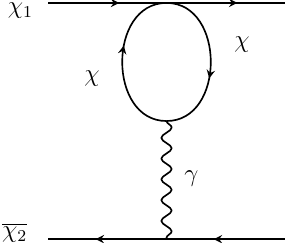}\qquad\qquad
        \includegraphics[scale = 0.9]{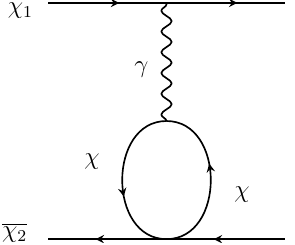}
\caption{\label{fig:4-fermi-photon}Four-Fermi effective photon penguin diagrams, where $\chi$ in the loop denotes Standard Model fermions other than neutrinos.}
\end{figure}

To obtain the effect due to other fermions, we replace the neutrino loop in the self-energy diagram Fig.~\ref{subfig:self-energy} with any SM charged fermion $\chi$. The result is the same as $V_0^{\rm SE}$ for $r\ll m_\chi^{-1}$. When $r\gtrsim m_\chi^{-1}$, the contribution will get exponentially suppressed by $e^{-2m_\chi r}$. As a result, due to the shorter range, the contribution of $\chi$ to the parity-violating matrix element cannot exceed that of the neutrino $V_0^{\rm SE}$.
Therefore, we expect that the contribution to APV from such kind of diagrams is subdominant compared to the neutrino force.

\begin{figure}[t]
    \centering
    \subfigure[Self-energy ($\gamma\text{SE}_1$)]{
        \includegraphics[scale = 0.8]{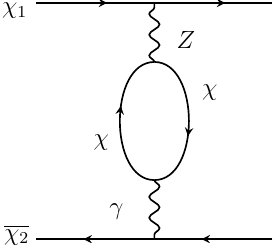}
        \label{subfig:SE-gamma1}
    }
    \hspace{1cm}
    \subfigure[Self-energy ($\gamma\text{SE}_2$)]{
        \includegraphics[scale = 0.8]{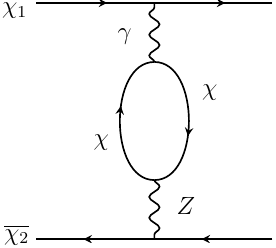}
        \label{subfig:SE-gamma2}
    } \\
    \subfigure[Penguin ($\gamma\text{PG}_1$)]{
        \includegraphics[scale = 0.8]{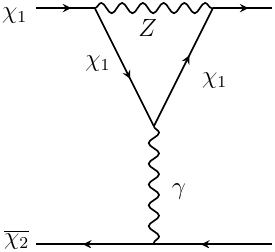}
        \label{subfig:penguin-gamma1}
    }
    \hspace{1cm}
    \subfigure[Penguin ($\gamma\text{PG}_2$)]{
        \includegraphics[scale = 0.8]{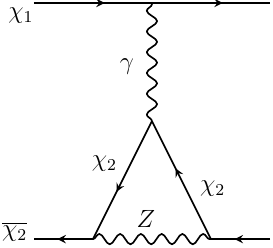}
        \label{subfig:penguin-gamma2}
    }
    \caption{Two possibilities to open up the four-Fermi effective vertex in the photon penguin diagrams in Fig.~\ref{fig:4-fermi-photon} at short distances ($r \lesssim \sqrt{G_F}$): (a)+(b) self-energy diagram and (c)+(d) penguin diagram.}
    \label{fig:photon_penguins}
\end{figure}

\begin{figure}[t]
    \centering
    \includegraphics[width=\textwidth]{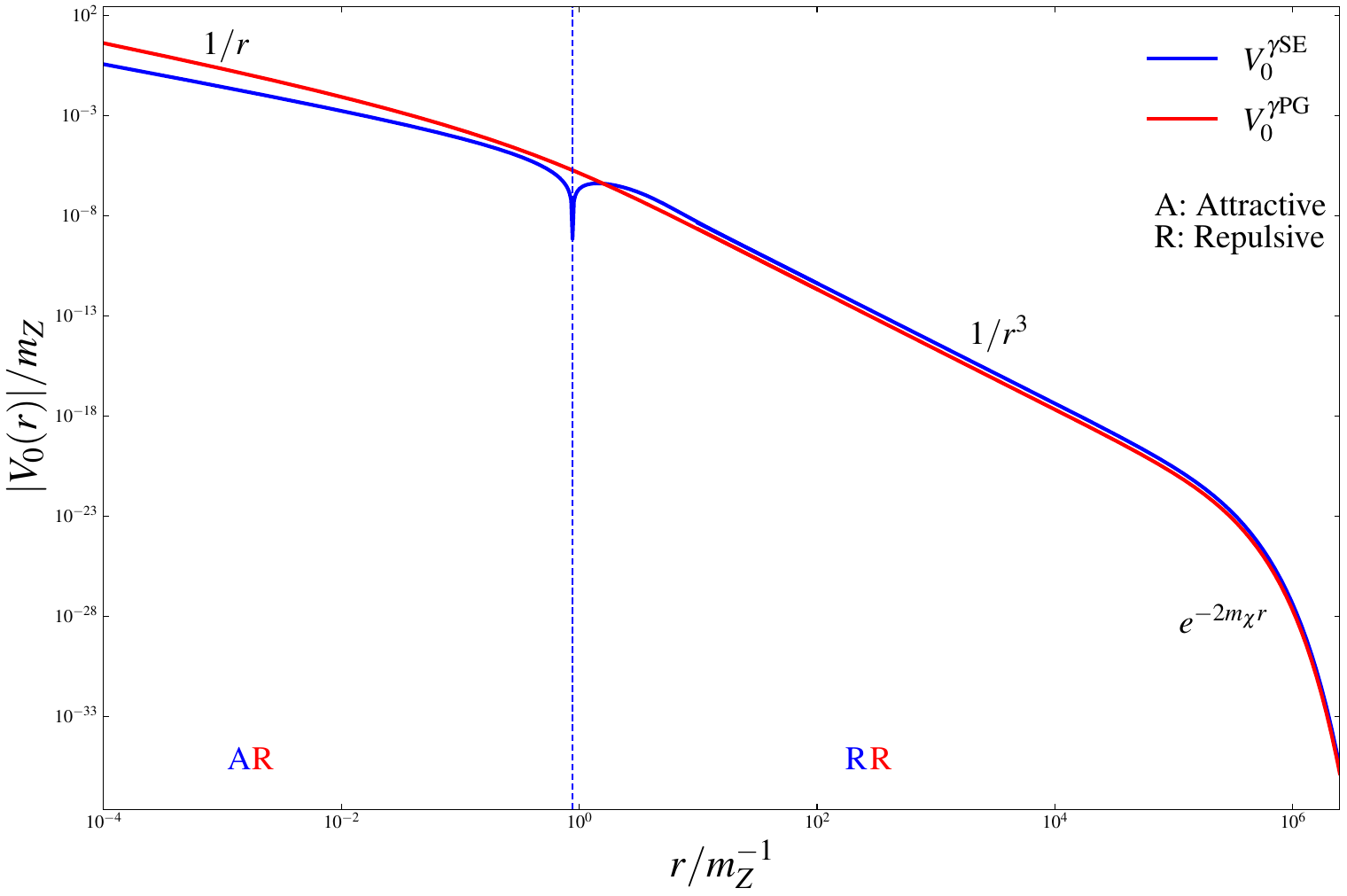} 
    \caption{Magnitudes of the quantum force mediated by a pair of non-neutrino fermions $\chi$ (we have chosen $\chi$ to be electron here for illustration) from different diagrams in Fig.~\ref{fig:photon_penguins} plotted as a function of the radial distance. The conventions are the same as in Fig.~\ref{fig:V0_all}.}
    \label{fig:V0Gamma_all}
\end{figure}

The more significant contribution from other fermions comes from the ``photon penguin diagrams'' in Figs. \ref{fig:4-fermi-photon} and \ref{fig:photon_penguins}, where one $Z$ propagator in the neutrino case is replaced with a photon. As a result, these diagrams contribute a force that is suppressed by ${\cal O}(\alpha G_F)$, compared to the ${\cal O}(G_F^2)$ dependence of the neutrino force. 
The forces due to these diagrams are computed in Appendix~\ref{app:fermion-force}. 
For the parity-violating force that we are interested in, the result can be split into two regions: (i) $r\ll m_\chi^{-1}$ and (ii) $r\gg m_Z^{-1}$. With the exception of the top quark all the other quarks have $m_\chi \ll m_Z$, the combination of the results in both regions is able to cover the whole range in the atomic system, as shown in Fig.~\ref{fig:V0Gamma_all}.

\begin{figure}[t]
    \centering
    \includegraphics[width=\textwidth]{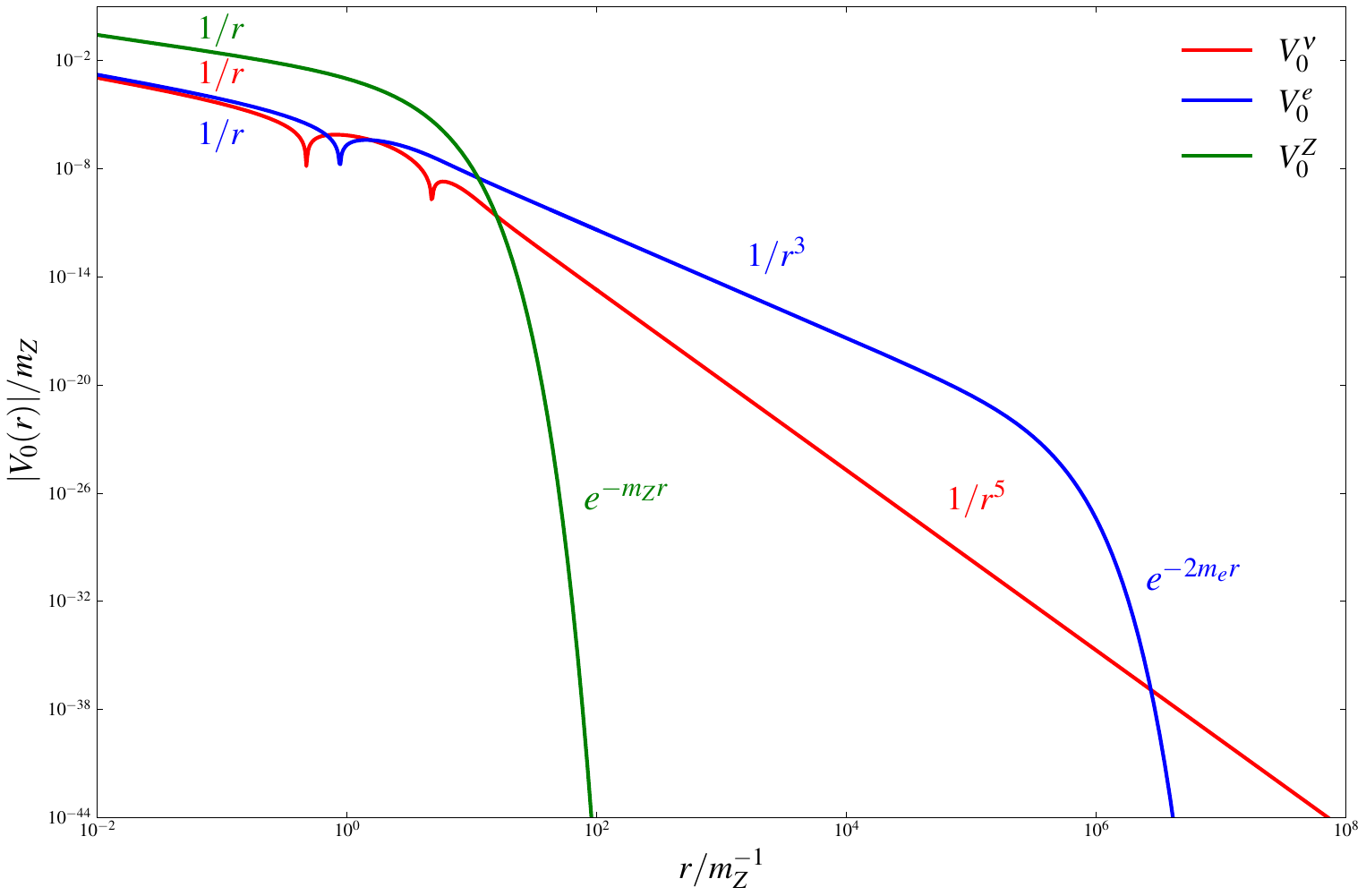} 
    \caption{The comparison of the magnitude of three typical parity-violating forces within atomic length scales: The tree-level $Z$ force $V_0^Z$, the two-neutrino force $V_0^\nu$, and the two-electron force with one photon propagator $V_0^e$. For the purposes of illustration, we have taken $V_0^\nu = V_0^{\rm SE}$ and $V_0^e = V_0^{\gamma\rm SE}$. The cusps correspond to the places where the potentials change sign.}
    \label{fig:VCompare_all}
\end{figure}

In region (i), one can neglect the fermion mass and we have a full theory with two kinds of diagrams, namely self-energy ($\gamma$SE) and penguin ($\gamma$PG), as shown in Fig.~\ref{fig:photon_penguins}. Note that in the self-energy diagram, $\chi$ can be charged-leptons or quarks; while in the penguin diagram, $\chi$ can only be the external fermion.
The parity-violating forces in muonium and positronium are given by:
\begin{align}
    V_{\rm PV}^{\gamma e \overline{\mu}}(r) &= \left[\frac{2\left(\bm{\sigma}_{\overline{\mu}}-\bm{\sigma}_e\right) \cdot \mathbf{p}_e}{m_e}+\frac{(\bm{\sigma}_{\overline{\mu}} \times \bm{\sigma}_e) \cdot \mathbf{\nabla}}{m_e}  \right] \left[\sum_\chi Q_\chi c_V^\chi V_0^{ \gamma \rm SE}(r) +  2\left(1-4s_W^2 \right)V_0^{ \gamma \rm PG}(r)\right],
    \label{eq:VPV-other-muonium}\\
    V_{\rm PV}^{\gamma e \overline{e}}(r) &= \frac{2(\bm{\sigma}_{\overline{e}} \times \bm{\sigma}_e) \cdot \mathbf{\nabla}}{m_e} \left[\sum_\chi Q_\chi c_V^\chi V_0^{ \gamma \rm SE}(r) + 2\left(1-4s_W^2 \right)V_0^{ \gamma \rm PG}(r)\right],
\end{align}
where $\chi$ sums over all SM charged-fermions (each quark contribution has a factor of $N_C=3$ due to three color degrees of freedom), and $Q_\chi$ is the electric charge of $\chi$ (in unit of $e$). Compared with the penguin and box diagrams mediated by neutrinos in Eqs.~(\ref{eq:VPV-nu-muonium}) and (\ref{eq:VPV-nu-positronium}), the penguin diagram mediated by other fermions is generically suppressed by a factor of $(1-4 s_W^2)\sim {\cal O}(10^{-1})$, which comes from the $Z$-boson couplings to the charged-leptons in the loop.
The full expressions of $V_0^{\gamma {\rm SE}}$ and $V_0^{\gamma{\rm PG}}$ are given by Eqs.~(\ref{eq:V0gammaSE}) and (\ref{eq:V0gammaPG}), respectively. In particular, we have
\begin{align}
    &r\ll m_Z^{-1}:\quad V_0^{\gamma {\rm SE}}\sim \frac{\alpha g^2}{r}\log(m_Z r)\;,\quad
     V_0^{\gamma {\rm PG}}\sim \frac{\alpha g^2}{r}\log^2(m_Z r)\;,\\
     & m_Z^{-1} \ll r\ll m_\chi^{-1}:\quad  V_0^{\gamma {\rm SE}}\sim V_0^{\gamma {\rm PG}} \sim \frac{\alpha G_F}{r^3}\;.
\end{align}

On the other hand, in region (ii), one can integrate out the gauge bosons and use the four-Fermi effective theory. The parity-violating potential in this case is given by a different form as shown:
\begin{align}
    V_{\rm PV}^{\gamma e \overline{\mu}}(r) &= \left[\frac{2\left(\bm{\sigma}_{\overline{\mu}}-\bm{\sigma}_e\right) \cdot \mathbf{p}_e}{m_e} +\frac{\left(\bm{\sigma}_{\overline{\mu}} \times \bm{\sigma}_e\right) \cdot \mathbf{\nabla}}{m_e}  \right]\sum_{\chi\neq e,\mu,t} 2 Q_\chi c_V^\chi V_0^{\rm \gamma \chi}(r) \nonumber\\
&\quad+\left[\frac{+2\bm{\sigma}_{\overline{\mu}} \cdot \mathbf{p}_e}{m_e}+\frac{\left(\bm{\sigma}_{\overline{\mu}} \times \bm{\sigma}_e\right) \cdot \mathbf{\nabla}}{m_e}\right]
    6\left(1-4s_W^2 \right)V_0^{\gamma \mu}(r)\nonumber\\
&\quad+\left[\frac{-2\bm{\sigma}_{e} \cdot \mathbf{p}_e}{m_e}+\frac{\left(\bm{\sigma}_{\overline{\mu}} \times \bm{\sigma}_e\right) \cdot \mathbf{\nabla}}{m_\mu}\right]
    6\left(1-4s_W^2 \right)V_0^{\gamma e}(r)\;,
\end{align}
where in the first line $\chi$ sums over all SM charged-fermions other than the electron, the muon and the top quark,\footnote{Since the top quark is even heavier than $Z$ boson, at $r > m_Z^{-1}$, the top quark contribution is exponentially suppressed by $e^{-2 m_t r}$ (where $m_t$ is the top quark mass) so it can be neglected.} and
\begin{align}
V_{\rm PV}^{\gamma e \overline{e}}(r) =   \frac{2\left(\bm{\sigma}_{\overline{e}} \times \bm{\sigma}_e\right) \cdot \mathbf{\nabla}}{m_e}\left[ \sum_{\chi\neq e,t} 2 Q_\chi c_V^\chi V_0^{\rm \gamma \chi}(r) + 4\left(1-4s_W^2\right)V_0^{\gamma e}(r) \right].
\end{align}
The full expression of $V_0^{\gamma \chi}$ including the fermion mass is given by Eq.~(\ref{eq:V0gammachi}). For the asymptotic behaviors, we have:
\begin{align}
     &m_Z^{-1} \ll r \ll m_\chi^{-1}, \quad V_0^{\gamma \chi} \sim \frac{\alpha G_F}{r^3}\;,\\
    & r \gg m_\chi^{-1}, \quad V_0^{\gamma \chi} \sim  \frac{\alpha G_F \sqrt{m_\chi}}{r^{5/2}}\,e^{-2 m_\chi r}\;.
\end{align}

In Fig.~\ref{fig:VCompare_all}, we compare the magnitude of all the three kinds of parity-violating forces in atoms that we have calculated: The ${\cal O}(G_F)$ tree-level $Z$ force, the ${\cal O}(G_F^2)$ two-neutrino exchange force, and the ${\cal O}(\alpha G_F)$ two-electron exchange force with one photon propagator. Below $m_Z^{-1}$, the $Z$ force is dominant, while the neutrino force and electron force are at the same order. Above $m_Z^{-1}$, the $Z$ force dies off quickly as $e^{-m_Z r}$, while the neutrino and the electron force scale as $1/r^5$ and $1/r^3$, respectively. So the electron force dominates in the region $m_Z^{-1}<r<m_e^{-1}$. When $r>m_e^{-1}$, the electron force exponentially decreases as $e^{-2m_e r}$, and the neutrino force becomes dominant. Only the neutrino force is long range within the whole atomic length scale (characterized by the Bohr radius $a_0 = 1/(\alpha m_e)$). Above Bohr radius, although the neutrino force still scales as $1/r^5$, the atomic wavefunctions are exponentially suppressed, making the contribution to the parity-violating matrix elements negligible.

\subsection{Contribution to APV matrix elements}
Now we estimate the contribution from photon penguin diagrams to APV, by convoluting the potential calculated above with the atomic wavefunctions. More specifically, we want to calculate:
\begin{align}
\eta_{\gamma} \equiv {C^{\gamma}_{AB} \over C^{Z}_{AB}}\equiv \eta_{\gamma}^{\text{SE}} + \eta_{\gamma}^{\text{PG}}\;,
\end{align}
where $\eta_\gamma^{\rm SE}$ and $\eta_\gamma^{\rm PG}$ correspond to the contribution from self-energy $\gamma {\rm SE}$ (Figs.~\ref{subfig:SE-gamma1} and \ref{subfig:SE-gamma2}) and penguin $\gamma {\rm PG}$ (Figs.~\ref{subfig:penguin-gamma1} and \ref{subfig:penguin-gamma2}), respectively. Similar to the case of the neutrino force, we calculate $\eta_\gamma$ by taking $|A\rangle =|1S\rangle$ and $|B\rangle = |2P \rangle$ as an illustration.

The self-energy contribution comes from all SM charged fermions in the loop except the top quark. Using the same strategy as the neutrino case, we obtain:
\begin{align}
    \eta_\gamma^{\rm SE} = -\frac{2\alpha}{3\pi\left(1-4 s_W^2\right)} \sum_{\chi\neq t} Q_\chi c_V^\chi N_C^{\chi}\left[\log\left(\frac{m_Z}{m_\chi}\right)+\gamma_{\rm E}-\frac{4}{3}\right],
\end{align}
where $Q_\chi$ is the electric charge, $c_V^\chi$ is the vector coupling to $Z$ boson given in Eq.~(\ref{eq:fundcoupling}), and $N_C^\chi$ is the color factor ($N_C^\chi = 1$ for charged leptons and $N_C^\chi=3$ for quarks). Substituting the corresponding couplings, we get
\begin{align}
 \eta_\gamma^{\rm SE} &= -\frac{2\alpha}{3\pi}\Bigg\{\sum_{\chi=e,\mu,\tau}\left[\log\left(\frac{m_Z}{m_\chi}\right)+\gamma_{\rm E}-\frac{4}{3}\right]+ \frac{2\left(1-\frac{8}{3}s_W^2\right)}{1-4s_W^2}\sum_{\chi=u,c}\left[\log\left(\frac{m_Z}{m_\chi}\right)+\gamma_{\rm E}-\frac{4}{3}\right]
\nonumber\\
 &\qquad\qquad+\frac{1-\frac{4}{3}s_W^2}{1-4s_W^2}\sum_{\chi=d,s,b}\left[\log\left(\frac{m_Z}{m_\chi}\right)+\gamma_{\rm E}-\frac{4}{3}\right] \Bigg\}\nonumber\\
 &\approx -3\% -23\% - 29\%\nonumber\\
 &\approx -55\%\;,\label{eq:etaSEgamma}
\end{align}
where in the second equality we have explicitly split the contributions into charged leptons, up-type quarks, and down-type quarks.

On the other hand, the penguin contribution can only come from electron or muon in the loop. We obtain (for $\chi=e,\mu$): 
\begin{align}
    \eta_\gamma^{{\rm PG},\chi} \approx -\frac{2\alpha}{3\pi}\log\left(\frac{m_Z}{m_\chi}\right),
\end{align}
which is numerically about $-2\%$ for $\chi=e$ and $-1\%$ for $\chi=\mu$. 

The total loop contributions from photon penguin diagrams to APV relative to the tree-level contribution in muonium and positronium are given by
\begin{align}
    \eta_\gamma^{e\overline{\mu}} \approx \eta_\gamma^{\rm SE} +  \eta_\gamma^{{\rm PG},\mu} \approx -56\%\;,\\
    \eta_\gamma^{e\overline{e}} \approx \eta_\gamma^{\rm SE} +  \eta_\gamma^{{\rm PG},e} \approx -57\%\;.
\end{align}

\vspace{0.2cm}
A few comments are in order:
\begin{enumerate}
    \item The suppression by a factor of $\alpha$ is expected since the photon penguin diagrams have two extra photon vertices compared to the $Z$-exchange diagram, thereby contributing an additional factor of $\alpha$.

    \item The photon penguin contribution is larger than the neutrino contribution for two main reasons: (a) The force due to photon penguin diagrams decreases as $1/r^3$ in the range $m_Z^{-1}<r<m_\chi^{-1}$, dominating over the $1/r^5$ neutrino force (see Fig.~\ref{fig:VCompare_all}). From $m_\chi^{-1}$ to Bohr radius, although the neutrino force becomes dominant, its magnitude is too small to contribute sizably to the parity-violating matrix elements. (b) The main contribution of the photon penguin diagrams comes from the self-energy diagrams ($\gamma {\rm SE}$) with quarks in the loop. This is because the vector couplings of quarks to $Z$ boson are not accidentally suppressed, as opposed to the $(1-4s_W^2)$ suppression in the tree-level $Z$ diagram and other photon penguin diagrams with charged-leptons in the loop. In addition, each of the five quarks contributes three color degrees of freedom, further enhancing the value of $\eta_\gamma$.
\item 
Note that the contribution to APV from photon penguin diagrams has already been included as the ${\cal O}(\alpha G_F)$ radiative corrections in \cite{Marciano:1978ed,Marciano:1982mm,Marciano:1983ss}, although the result therein was only given at the amplitude level. 

\item  For completeness, we also estimate the effects from photon penguin diagrams on the extraction of the weak mixing angle using the value of $\eta_\gamma$ computed above:
\begin{align}
\text{muonium/positronium}&: \left( {\sin^2 \theta_W ' - \sin^2 \theta_W  \over \sin^2 \theta_W} \right)_\gamma \approx 3\%\;,\\
\text{Cesium}&: \left( {\sin^2 \theta_W ' - \sin^2 \theta_W  \over \sin^2 \theta_W} \right)_\gamma \approx 2\%\;.
\end{align}
These contributions have been included as the radiative corrections to the weak charge in \cite{Marciano:1978ed,Marciano:1982mm, Marciano:1983ss} and thus should not be regarded as something new.
\end{enumerate}

\section{Comparison with existing literature}\label{sec:litcomp}

In this section, we comment on the previous relevant work and compare them with our results.

The one-loop radiative corrections of APV in the electroweak theory up to ${\cal O}(\alpha G_F)$ were calculated in Refs.~\cite{Marciano:1978ed,Marciano:1982mm, Marciano:1983ss}. However, the neutrino contribution was neglected in these works. As we have demonstrated above, the neutrino force, due to its long-range nature, significantly contributes to APV. The effect of the neutrino force on APV is comparable to that of other one-loop diagrams and should be included. In particular, in muonium and positronium, the relative contribution from the neutrino force is further enhanced by $(1 - 4s_W^2)^{-1}$ due to the accidental suppression of the tree-level $Z$ force.

The neutrino force contribution to APV in heavy atoms was first studied in Ref.~\cite{Dzuba:2022vrv}. However,  only self-energy diagrams were included and the  contribution from penguin diagrams were neglected (note that the box diagram is not present in the case of heavy atoms). Our result of the contribution from self-energy diagram seems to agree with that in~\cite{Dzuba:2022vrv}. In Ref.~\cite{Dzuba:2022vrv}, the authors report $\eta_{\nu} \approx -0.72\%$ in heavy atoms, but this number includes contributions from self-energy diagrams with all possible SM fermions except top quark in the loop, not just the neutrinos. By extracting pure neutrino contributions one can get $-0.17\%$, which is very close to what we obtain $\eta_\nu^{\rm SE}\approx -0.2\%$ in muonium and positronium systems.
This indicates that the relative neutrino force contribution to APV may not change significantly with the atomic size.
In this work we calculate the full-range potentials and the APV effects from neutrino penguin and box diagrams for the first time. We show that in muonium the penguin effect $\eta_\nu^{\rm PG}$ dominates over that of self-energy by a factor of  $\eta_\nu^{\rm PG}/\eta_\nu^{\rm SE}\sim (1-4 s_W^2)^{-1}\sim {\cal O}(10)$. Therefore, it is reasonable that in muonium we obtain $\eta_\nu = \eta_\nu^{\rm SE}+\eta_\nu^{\rm PG} \approx \eta_\nu^{\rm PG}$. In positronium, the box diagram gives the largest contribution since it does not involve any $Z$-boson couplings that can suppress it, and the potential from box diagram $V_0^{\rm box}$ keeps the same sign through the whole range.

Next, we compare the contribution to APV from photon penguin diagrams, which is parametrized by  $\eta_{\gamma} = \eta_{\gamma}^{\text{SE}} + \eta_{\gamma}^{\text{PG}}$. The first term was calculated in Ref.~\cite{Dzuba:2022vrv} for heavy atoms, which found that $\eta_{\gamma}^{\text{SE}} \approx 2 \%$, with the contribution from the electron $\eta_{\gamma}^{\text{SE},e} \approx 0.1 \%$.  Our calculation in muonium and positronium yields $\eta_{\gamma}^{\text{SE},e} \approx -2 \%$. At the same time we find the total $\eta_{\gamma}^{\text{SE}} \approx -55 \%$, with quarks giving a large contribution of $-52\%$, as seen in Eq.~(\ref{eq:etaSEgamma}). The difference is attributed to the different vector couplings of the nucleus to the $Z$ boson. For heavy nucleus, the vector coupling is proportional to the weak charge $Q_W\approx -{\cal N} + (1-4 s_W^2){\cal Z} \approx -{\cal N}$, whereas in muonium and positronium it is $1-4s_W^2$. As a result, the relative loop/tree 
coupling in heavy atoms is roughly ${\cal Z/(-N)}$, while that in muonium and positronium is $1/(1-4s_W^2)$. This explains why the magnitude of $\eta_\gamma^{\rm SE}$ in muonium and positronium is about 20 times larger than that in heavy atoms. The second term $\eta_\gamma^{\rm PG}$ was not included in \cite{Dzuba:2022vrv} and we find it only contributes approximately $-1\%$ ($-2\%$) in muonium (positronium). This is because the $\gamma {\rm PG}$ diagrams can only have charged-leptons in the loop, whose vector couplings to $Z$ boson are more suppressed than quarks. Both $\gamma {\rm SE}$ and $\gamma {\rm PG}$ diagrams have been calculated long time ago~\cite{Marciano:1978ed,Marciano:1982mm, Marciano:1983ss} as radiative corrections to the weak charge. However, the results in \cite{Marciano:1978ed,Marciano:1982mm, Marciano:1983ss} were only given at the amplitude level in the $q^2 \to 0$ limit; the authors did not perform the Fourier transform and convolve with the atomic wave functions to get the parity-violating matrix elements.

Apart from the two-neutrino exchange diagrams and the photon penguin diagrams, there are also other diagrams that contribute solely to short-distance forces and therefore to APV at short distances, notably diagrams with multiple gauge boson exchange. We do not compute them here, as such radiative corrections have been computed in~\cite{Marciano:1978ed,Marciano:1982mm, Marciano:1983ss}. As those diagrams only mediate short-range forces at $r<m_Z^{-1}$, their contribution to APV matrix elements can be well described by the parity-violating amplitude in the limit of zero momentum transfer, which was included in~\cite{Marciano:1978ed,Marciano:1982mm, Marciano:1983ss}.

Finally, we would like to comment on the zero-momentum transfer limit that is commonly used in the literature to calculate APV. At the one-loop amplitude level, compared with ${\cal O}(\alpha G_F)$, the ${\cal O} (G_F^2)$ contribution from two-neutrino exchange is suppressed by $q^2/m_Z^2$, and therefore may be naively neglected in the limit of $q^2 \to 0$. Taking the limit of zero momentum transfer is equivalent to setting $m_Z^2 \to \infty$. This results in taking the $\delta$-function approximation to the $Z$ force: $V_Z \sim G_F m_Z^2 e^{-m_Z r}/r \sim G_F \delta^3(r)$. This is usually how the radiative corrections to the weak charge are obtained~\cite{Marciano:1978ed,Marciano:1982mm, Marciano:1983ss}: the theoretical prediction of the weak charge (\ref{eq:qwdef}) is given by the parity-violating Feynman amplitude in the $q^2\to 0$ limit. However, this is a good approximation only when the atomic wavefunctions do not change rapidly around the origin and when the potential decays quickly at $r > 1/m_Z$. In the more general case, the weak charge cannot capture the APV effect at $r>1/m_Z$, because it does not contain information about atomic wavefunctions away from the origin. When there exist long-range parity-violating interactions, the more general method is to transform the amplitude into position space and then convolve it with the atomic wavefunctions. As explicitly shown in Fig.~\ref{fig:Q}, for the $s$-state, the range of the neutrino contribution to APV is concentrated at very short distances compared to the atomic length scale (though still longer than the $Z$-force range), and the $\delta$-function approximation still holds, while for $\ell \geq1$ states, the neutrino contribution to APV has an extensive range across the entire atomic length scale.
Quantitatively, our calculation implies that for $s$-states, the generic ${\cal O}(G_F^2)$ neutrino contribution to APV is actually of order $g^2/(16\pi^2)$ [without including the accidental enhancement factor $1/(1-4s_W^2)$] compared with the leading ${\cal O}(G_F)$ contribution from $Z$-exchange. We thus conclude that it is of the same order as the ${\cal O}(\alpha G_F)$ terms calculated in \cite{Marciano:1978ed,Marciano:1982mm, Marciano:1983ss}. Therefore, the ${\cal O}(G_F^2)$ neutrino contribution to APV cannot be neglected at the one-loop level.

\section{Conclusions} \label{sec:conclusion}

In this work, we studied the neutrino force beyond the four-Fermi approximation. The previous computation of the two-neutrino exchange force within the four-Fermi effective theory, which yields the $1/r^5$ dependence~\cite{Feinberg, Feinberg:1989ps,Hsu:1992tg}, is not valid at short distances $r \lesssim 1/m_Z$, as the heavy gauge bosons cannot be integrated out at such length scales. We conducted a complete calculation of the neutrino force in the Standard Model by including all relevant ultraviolet diagrams. We thus derived a formula for the Standard Model neutrino force that is valid across all length scales, summarized in Eqs.~(\ref{eq:V0SE}), (\ref{eq:V0PG}), (\ref{eq:V0box}) and Fig.~\ref{fig:V0_all}.
At short distances, the force exhibits a $\log^2(m_W r)/r$ dependence, while at long distances, the familiar $1/r^5$ dependence is recovered as expected. 

Apart from its long range, another distinguishing feature of the neutrino force is that it violates parity. This provides a promising way to probe this force. Parity-violating transitions are forbidden by the dominant electromagnetic interactions, so parity-violating effects can thus distinguish the neutrino force from higher-order QED effects. Using the full-range neutrino force, we computed the corrections from neutrinos to the parity-violating matrix elements of $s$-wave atomic states using Eq.~(\ref{eq:PV}). These corrections were not included in past works~\cite{Marciano:1978ed,Marciano:1982mm, Marciano:1983ss}. We found that 
the neutrino force contributes sizably to APV.  In muonium, due to the accidental suppression of the leading $Z$-exchange, the relative contribution to APV is of order $\eta_\nu\sim g^2/(16\pi^2 (1-4 s_W^2))\sim 5\%$. In positronium, the additional box diagram enhances $\eta_\nu$ to 16\%. In heavy atoms such as Cesium, the accidental enhancement is absent since the weak charge is dominated by the vector couplings of neutrons, making $\eta_\nu \sim g^2/(16\pi^2)\sim 0.3\%$. This is comparable to the current APV experimental error of 0.35\% in Cesium~\cite{Wood:1997zq}.

We also estimated the effects of neutrino forces on the extraction of the weak mixing angle from APV experiments. We found that ignoring the neutrino force would lead to a deviation in the extracted value of $\sin^2 \theta_W$ by about 0.2\% (0.7\%) in muonium (positronium) and 0.3\% in heavy atoms. This number is, again, not negligible when compared with the current experimental sensitivity.

This result underscores the importance of taking the limits in the correct order. Indeed, for distances much larger than $1/m_Z$, the neutrino force is negligible compared to other one-loop effects, such as the photon penguin, with a ratio of order $G_F^2/r^5$ versus $G_F \alpha/r^3$. However, for $r \sim 1/m_Z$, the full electroweak theory must be used, and the ratio between these diagrams is of the order of the coupling constants. Since the relevant scale for calculating APV for $s$-states is approximately $1/m_Z$, the neutrino corrections become as significant as other one-loop diagrams.

Before concluding, we would like to briefly mention the effect of neutrino masses on APV. Although we have neglected neutrino masses and lepton flavor mixings throughout this work, it is theoretically interesting to look at the possibility of probing the neutrino mass nature (Dirac or Majorana) by the detection of the neutrino force. Since the exact form of the neutrino force is sensitive to the neutrino absolute mass and the nature of the mass~\cite{Grifols:1996fk,Segarra:2020rah,Costantino:2020bei,Ghosh:2022nzo}, those dependence will also enter the APV matrix elements. However, at the atomic length scale (characterized by the Bohr radius $a_0$), the neutrino mass effect on APV is additionally suppressed by a factor of $m_\nu^2 a_0^2\sim m_\nu^2/(\alpha^2 m_e^2) \sim 10^{-9}$ for a neutrino mass $m_\nu \sim 0.1~{\rm eV}$, hence cannot be practically probed in the current APV experiments. (The fact that the neutrino mass effect is proportional to $m_\nu^2$ rather than $m_\nu$ indicates that the neutrino force is a pure quantum effect, i.e., it is mediated by two neutrino propagators.)
On the other hand, if one can probe the parity-violating effect caused by the neutrino force at a length scale of $m_\nu^{-1} \sim {\cal O}(\mu{\rm m})$ (which is the typical size of a cell or some large molecules), then it would provide an interesting possibility to determine the nature of the neutrino mass, because at such scales, the magnitude of the neutrino force as well as its parity-violating effect will be different by about ${\cal O}(1)$ if the neutrinos were Majorana fermions as opposed to Dirac.

To conclude, our results highlight the need for a more detailed calculation of APV
in Cesium, incorporating the neutrino force accurately across all length scales, along with a comprehensive consideration of atomic many-body effects and relativistic corrections. This calculation is necessary for two reasons: first, to enhance the likelihood of detecting the neutrino force in such systems, and second, to extract $\sin^2 \theta_W$ more precisely from APV experiments as a test of the Standard Model.

\acknowledgments
We would like to thank Zackaria Chacko, Anson Hook, Yushan Su, Raman Sundrum, and Ken Van Tilburg for helpful discussions and feedback.
MG acknowledges support from the US Department of Energy grant DE-SC0010102. YG is supported in part by the NSF grant PHY- 2309456.
\appendix

\section{A gauge-invariant calculation of the renormalizable neutrino force}

\label{app:neutrino-force}
In this appendix, we provide details of the calculation of the neutrino force in electroweak theory, which is valid at all length scales. We perform the calculation in $R_\xi$ gauge and explicitly show that the final result is gauge invariant. The gauge-boson propagator reads
\begin{align}
    i G_{\mu \nu}^X(q) = \frac{-i}{q^2 - m_X^2} \left(g_{\mu \nu} - (1- \xi_X)\frac{q_\mu q_\nu}{q^2 - \xi_X m_X^2}  \right),
\end{align}
where $\xi_X$ is the gauge parameter and $X=W$ or $Z$. Since the external fermions $\chi_i$ are much lighter than the gauge bosons, one can always expand the result as a series of $m_i^2/m_X^2$ and only keep the leading term. In this case, the contributions from diagrams involving Goldstone bosons can be neglected, as they are all suppressed by the ratio of fermion masses to the heavy boson masses. Therefore, we only need to consider the three kinds of diagrams (self-energy, penguin, and box) in Fig.~\ref{fig:beyond-4-fermi}.

The neutrino force $V_0$ can be calculated using the dispersion technique in Eq.~(\ref{eq:V0}). It is proportional to the discontinuity of the loop integral $I_0$ across its branch cut in the complex $t$-plane, as defined in Eqs.~(\ref{eq:form}) and (\ref{eq:decomp}). In the following, we compute $I_0$ and $V_0$ for each diagram in Fig.~\ref{fig:beyond-4-fermi}, where the one-loop Feynman integrals are evaluated with the help of the publicly available Package-X~\cite{Patel:2015tea,Patel:2016fam}.

\subsection{The self-energy diagram} 
For the self-energy diagram in Fig.~\ref{subfig:self-energy}, the loop integral is given by:
\begin{align}    
I_{\mu \nu}(q) &= \left(\frac{g}{4 c_W}\right)^ 4 G^Z_{\mu \rho}(q) G^Z_{\nu \sigma}(q)\int \frac{{\rm d}^4 k}{(2 \pi)^4} \text{Tr}
    \left[ \gamma^\rho (1 - \gamma^5) \frac{1}{\slashed{k}+\slashed{q}}\gamma^\sigma (1-  \gamma^5) \frac{1}{\slashed{k}} \right].
\end{align}
Notice that the minus sign from the neutrino loop has been canceled by the minus sign from the anti-commutativity between fermion and anti-fermion fields. The one-loop integral can be worked out analytically in terms of the Passarino-Veltman functions \cite{Passarino:1978jh}:
\begin{align}
I_{\mu \nu}(q) &= \left( \frac{g}{4 c_W}\right)^4 \frac{i}{ 2 \pi^2 \left(q^2 - m_Z^2\right)^2
   \left(q^2- \xi_Z m_Z^2  \right)^2} \nonumber\\
   &\times \left(q_{\mu } q_{\nu } \left( 2 (\xi_Z -1)
   \left(2 m_Z^2 \xi_Z -(\xi_Z +1) q^2\right)  B_{00}\left(q^2;0,0\right) \right.\right.\nonumber\\
   & \left.\left.+\left(2
   m_Z^4 \xi_Z ^2-2 m_Z^2 \xi_Z  (\xi_Z +1) q^2+\left(\xi_Z ^2+1\right) q^4 \right)\left(B_{11}\left(q^2;0,0\right)+B_1\left(q^2;0,0\right)\right)\right) \right. \nonumber  \\
   &- \left. g_{\mu \nu} \left(q^2- \xi_Z m_Z^2 \right)^2 \left(q^2\left(B_{11}\left(q^2;0,0\right)+B_1\left(q^2;0,0\right)\right)+2
   B_{00}\left(q^2;0,0\right)\right)\right).
\end{align}
Using the explicit form of the Passarino-Veltman functions, we find the result is pure transverse and the gauge parameter $\xi_Z$ drops out:
\begin{equation}
    \label{eq:Imunu-SE}
   I_{\mu \nu}(q) = i \left( \frac{g}{4 c_W}\right)^4 \frac{(q^2 g_{\mu \nu} - q_\mu q_\nu)}{18 \pi^2(q^2 - m_Z^2)^2} \left[5 + 3 \Delta_{\rm E} + 3 \log\left( \frac{\mu^2}{-q^2} \right) \right],
\end{equation}
where $\Delta_{\rm E}\equiv 1/\epsilon -\gamma_{\rm E} + \log(4\pi)$ with $\epsilon\to 0^+$. Therefore, the discontinuity part of the $g_{\mu \nu}$ term in Eq.~(\ref{eq:Imunu-SE}) is given by (with the replacement $q^2 \to t$):
\begin{equation}
    \text{Disc}[I_0(t)] = -\left(\frac{g}{4 c_W}\right)^4\frac{t}{3 \pi (t - m_Z^2)^2}\;.\label{eq:discI0SE}
\end{equation}
As one can see, the infinity in $\Delta_{\rm E}$ drops out after taking the discontinuity.
Using Eq. (\ref{eq:V0}), we can then find the neutrino force:
\begin{align}
    \label{eq:V0SEapp}
    V_0^{\rm SE}(r) &= -\left(\frac{g}{4 c_W}\right)^4 \frac{1}{24 \pi^3 r} \int_0^ \infty  {\rm d}t \frac{t e^{- \sqrt{t} r}}{(t-m_Z^2)^2}\;, \nonumber \\
    &= \left(\frac{g}{4 c_W}\right)^4 \frac{1}{48 \pi^3 r} \left[e^x\left(2+x\right) {\rm Ei}\left(-x\right)+e^{-x}\left(2-x\right) {\rm Ei}\left(x\right)+2\right],
\end{align}
where $x\equiv m_Z r$ and ${\rm Ei} (x) \equiv -\int_{-x}^\infty {\rm d}t \  t^{-1} e^{-t}$ is the exponential integral function. In the first line of Eq.~(\ref{eq:V0SEapp}), there is a double pole caused by the $Z$ propagators. In order to get the finite result in the second line, the denominator needs to be shifted by an  infinitesimally small imaginary part, which corresponds to the prescription from the Feynman propagator (see Eq.~(\ref{eq:V0SEapp2})) --- a careful treatment of the integral in the presence of additional poles is provided in Appendix~\ref{app:dispersion}.

Using the asymptotic behavior of ${\rm Ei} (x)$ at long distances ($x\gg 1$):
\begin{align}
    e^x\left(2+x\right) {\rm Ei}\left(-x\right)+e^{-x}\left(2-x\right) {\rm Ei}\left(x\right)+2 = -\frac{24}{x^4}+{\cal O}\left(\frac{1}{x^6}\right),
\end{align}
we recover the known $1/r^5$ form:
\begin{align}
    V_0^{\rm SE}\left(r\gg m_Z^{-1}\right) = \left(\frac{g}{4c_W}\right)^4 \frac{1}{48\pi^3 r}\left(-\frac{24}{m_Z^4 r^4}\right) = -\frac{G_F^2}{16\pi^3 r^5}\;.
\end{align}

\subsection{The penguin diagrams}
For the penguin diagram in Fig.~\ref{subfig:penguin1}, where the vertex correction is connected to $\chi_1$, the amplitude is given by:
\begin{align}
    i\mathcal{M} &= -\frac{g^4}{128 c_W^2} \int \frac{{\rm d^4} k}{(2 \pi)^4}\left[
    \frac{\Bar{u}(p'_1) \gamma^\rho (1 - \gamma^5) (\slashed{p'_1} +\slashed{k}) \gamma^\mu (1 - \gamma^5) (\slashed{p_1} + \slashed{k})\gamma^\sigma (1 - \gamma_5) u(p_1)}{(p'_1+k)^2 (p_1 +k)^2} \right. \nonumber \\
    &\left.\quad\times\Bar{v}(p_2)\gamma^\nu (c_V^{\chi_2} - c_A^{\chi_2} \gamma_5) v(p'_2) G^Z_{\mu \nu}(q)G^W_{\rho \sigma}(k)\right]\nonumber\\  
    &\equiv -\frac{g^4}{128 c_W^2}  \overline{u}(p'_1) \Gamma^\mu u(p_1) \Bar{v}(p_2)\gamma^\nu (c_V^{\chi_2} - c_A^{\chi_2} \gamma^5) v(p'_2) G^Z_{\mu \nu}(q)\;,        
\end{align}
where the minus sign comes from the anti-commutativity between fermion and anti-fermion fields, and the vertex correction part at $\chi_1$ reads:
\begin{align}
     \Gamma^\mu  =   \int \frac{{\rm d}^4 k}{(2\pi)^4} \frac{\gamma^\rho(1-\gamma^5)(\slashed{p'_1} +\slashed{k}) \gamma^\mu (1 - \gamma^5) (\slashed{p_1} + \slashed{k})\gamma^\sigma (1 - \gamma^5)}{(p'_1+k)^2 (p_1 +k)^2}G^W_{\rho \sigma}(k)\;.
\end{align}
The vertex function $\Gamma^\mu$ can be generally decomposed into the following form:
\begin{align}
    \Gamma^\mu (q) &= \gamma^\mu F_1(q^2) + \frac{i \sigma^{\mu \nu}q_\nu}{2 m_1} \left[ F_2(q^2) + \gamma^5 G_2(q^2)\right] + \frac{q^\mu}{m_1} \left[F_3(q^2)+ \gamma^5 G_3(q^2)\right]\nonumber \\
    & + \left(\gamma^\mu- \frac{\slashed{q} q^\mu}{q^2} \right) \gamma^5 G_1(q^2)\;.
\end{align}
At the leading order of $m^2_{1}/m_W^2$, the form factors $F_1= - G_1$, while $F_2, G_2, F_3, G_3$ all vanish.
Moreover, under the leading order of the NR approximation, we have
\begin{align}
    \overline{u}(p'_1) \frac{\slashed{q} q^\mu}{q^2} \gamma^5 u(p_1) \Bar{v}(p_2)\gamma^\nu (c_V^{\chi_2} - c_A^{\chi_2} \gamma^5) v(p'_2) G^Z_{\mu \nu }   &= 0 \;.
\end{align}
Then the vertex function simply reduces to:
\begin{align}
    \Gamma^\mu = F_1(q^2) \gamma^\mu \left( 1 - \gamma^5\right). \label{eq:vertexfunction}
\end{align}
Therefore, at the leading order, the amplitude can be simplified to:
\begin{align}
    i \mathcal{M} = \overline{u}(p'_1)\gamma^\mu (1-\gamma^5) u(p_1) \Bar{v}(p_2)\gamma^\nu (c_V^{\chi_2} - c_A^{\chi_2} \gamma_5) v(p'_2) \left[g_{\mu\nu}I_0(q^2)+ q_\mu q_\nu I_1(q^2)\right],
\end{align}
where $I_1$ is not relevant to our purpose while $I_0$ is given by
\begin{align}
    I_0 (q^2) =  \frac{g^4}{128 c_W^2}\frac{F_1(q^2)}{(q^2 - m_Z^2)}\;.\label{eq:I0PG}
\end{align}
The only relevant form factor $F_1$ reads:
\begin{align}
F_1(q^2) &= \frac{i}{12 \pi^2 (q^2-4 m_1^2)^2} \Bigg\{\pi^2 m_W^4 + 2 \left(3 + \pi^2\right) m_W^2 q^2 + \left(15 + \pi^2\right) q^4 \nonumber \\
&\quad - 3 q^4 \xi_W \left[\Delta_{\rm E}+ \log\left(\frac{\mu^2}{\xi_W m_W^2}\right) + 1\right] + 6 \left(m_W^2 + q^2\right)^2 \text{Li}_2\left(\frac{m_W^2 + q^2}{q^2}\right) \nonumber \\
&\quad + 3 \log\left(-\frac{m_W^2}{q^2}\right) \left[q^2 \left(2 m_W^2 + 3q^2\right) + \left(m_W^2 + q^2\right)^2 \log\left(-\frac{m_W^2}{q^2}\right)\right] \Bigg\}\;,\label{eq:F1}
\end{align}
where ${\rm Li}_2$ is the dilogarithm  function defined as 
\begin{align*}
{\rm Li}_2(z) \equiv -\int_0^z {\rm d}u\, \frac{\log\left(1-u\right)}{u}\;.    
\end{align*}

Although $F_1$ is divergent and gauge dependent, it should be noticed that the discontinuity part is finite and the gauge parameter drops out. In fact, the terms in Eq.~(\ref{eq:F1}) that contribute to the nonzero discontinuity are found to be (with $t>0$):
\begin{align}
{\rm Disc}\left[\log\left(-\frac{m_W^2}{t}\right)\right] &\equiv \log\left(-\frac{m_W^2}{t+i\epsilon}\right) -\log\left(-\frac{m_W^2}{t-i\epsilon}\right) = 2\pi i \;,\label{eq:disclog}\\
{\rm Disc}\left[{\rm Li}_2\left(1+\frac{m_W^2}{t}\right)\right] &\equiv {\rm Li}_2\left(1+\frac{m_W^2}{t+i\epsilon}\right) - {\rm Li}_2\left(1+\frac{m_W^2}{t-i\epsilon}\right) = -2\pi i \log\left(1+\frac{m_W^2}{t}\right),\label{eq:discLi2}\\
{\rm Disc}\left[\log^2\left(-\frac{m_W^2}{t}\right)\right] &\equiv\log^2\left(-\frac{m_W^2}{t+i\epsilon}\right) -\log^2\left(-\frac{m_W^2}{t-i\epsilon}\right) = 4\pi i \log\left(\frac{m_W^2}{t}\right).\label{eq:disclog2}
\end{align}
Combining Eqs.~(\ref{eq:disclog})-(\ref{eq:disclog2}) together with (\ref{eq:I0PG}) and (\ref{eq:F1}), we obtain
\begin{equation}
    \text{Disc}\left[I_0(t)\right] = -\frac{g^4}{256 c_W^2} \frac{t(2m_W^2 + 3t) - 2 (m_W^2 + t)^2 \log \left(1+\frac{t}{m_W^2}\right)}{\pi (t- 4 m_1^2)^2(t-m_Z^2)}\;.
\end{equation}
To get the result for the other penguin diagram in Fig.~\ref{subfig:penguin2} with the vertex correction at $\overline{\chi_2}$, one only needs to replace $m_1$ with $m_2$ in the above calculation. 
At the leading order of $m_i^2/m_Z^2$ (for $i=1,2$), using the dispersion technique in Eq.~(\ref{eq:V0}), we obtain:
\begin{align}
\label{eq:V0PGapp}
V_0^{\rm PG}(r) = -\frac{g^4}{2048 \pi^3  c_W^2 r} \int_0^\infty {\rm d}t\,e^{-\sqrt{t}\,r}\,\frac{t\left(2m_W^2 + 3t\right) - 2 \left(m_W^2 + t\right)^2 \log \left(1+\frac{t}{m_W^2}\right)} {t^2\left(t-m_Z^2\right)} \;.
\end{align} 
Note that the above integral contains a pole at $t=m_Z^2$, which comes from the $Z$ propagator in the penguin diagram. Such a divergence is superficial and can be removed by introducing the $i \epsilon$ prescription (see Appendix~\ref{app:dispersion} for details).

At long distances $r\gg m_Z^{-1}$, the result can be worked out analytically:
\begin{align}
     V_0^{\rm PG}\left(r\gg m_Z^{-1}\right) = -\frac{g^4}{2048\pi^3 c_W^2 r}\int_0^\infty {\rm d}t\, e^{-\sqrt{t}\,r} \frac{2t}{3m_W^2 m_Z^2} = -\frac{G_F^2}{8\pi^3 r^5}\;.
\end{align}

\subsection{The box diagram}
For the box diagram in Fig.~\ref{subfig:box}, the amplitude reads
\begin{align}
   i \mathcal{M}  = & -\left(\frac{g}{2\sqrt{2}}\right)^4 \int \frac{{\rm d}^4 k}{(2 \pi)^4}  
    \left[\frac{\overline{v}(p_2)  \gamma^\mu (1-\gamma^5) (\slashed{p}_1 - \slashed{k}) \gamma^\nu (1- \gamma^5) u(p_1)}{(p_1-k)^2} G^W_{\nu \sigma} (k) \right.  \; \nonumber \\
    &\left. \times G^W_{\mu \rho} (p_2 +p_1-k) \frac{\overline{u}(p'_1) \gamma^\sigma( 1- \gamma^5) (\slashed{p'_1} - \slashed{k}) \gamma^\rho (1- \gamma^5) v(p'_2)}{(p'_1-k)^2} \right].
\end{align}
At leading order in $m_i^2/m_W^2$ (for $i=1,2$) and non-relativistic approximation, the amplitude is reduced to
\begin{align}
     i\mathcal{M} = \overline{v}(p_2)  \gamma^\mu (1-\gamma_5)u(p_1) \overline{u}(p'_1)  \gamma^\nu (1-\gamma_5)v(p'_2) g_{\mu \nu} I_0\;,
\end{align}
where
\begin{align}
I_0(t) = & - \frac{i g^4}{3072 \pi^2\, t^2 m_W^4 \left(2 m_W^2 + t\right)^2} \Bigg\{- 24 m_W^6 t \left(2 m_W^2 + t\right)\, 
    \log\left(-\frac{m_W^2}{t} \right)\nonumber\\ 
& + 24  m_W^4 t\left(m_W^2 + t\right)\left(t^2 - 2 m_W^4\right)\, C_0\left(0, 0, t; 0, m_W, 0\right) \nonumber \\
&+24 m_W^6 t \left(2 m_W^4 + 6 m_W^2 t + 3 t^2\right)\, C_0\left(0, 0, -t; m_W, 0, m_W\right) \nonumber \\
& - 2 t \left(2 m_W^2 + t\right)^2 
    \left[t^2+2m_W^2 t \left(\xi_W-5\right)+m_W^4\left(\xi_W-1\right)^2 \right]
    \Lambda\left(-t, m_W, m_W \sqrt{\xi_W}\right) \nonumber \\
& + t^2 \left(2 m_W^2 + t\right)^2 \left(4 m_W^2 \xi_W + t\right)\, 
    \Lambda\left(-t, m_W \sqrt{\xi_W}, m_W \sqrt{\xi_W}\right) \nonumber \\
& + t\left(2 m_W^2 + t\right)\left(-24 m_W^6 - 64 m_W^4 t - 18 m_W^2 t^2 + t^3\right)\,
    \Lambda\left(-t, m_W, m_W\right) \nonumber \\
& - m_W^2 \left(2 m_W^2 + t\right)^2 \log(\xi_W)
   \left[3t^2\left(\xi_W+3\right)+3m_W^2 t\left(\xi_W^2-4\xi_W+3\right)+m_W^4\left(\xi_W-1\right)^3\right] \nonumber \\
& - 2 t m_W^4 \left(2 m_W^2 + t\right)^2 \left(\xi_W^2-2\xi_W+13\right)
\Bigg\}\;,\label{eq:I0box}
\end{align}
in which the function
\begin{align*}
    \Lambda\left(t, m_0,m_1\right) \equiv \frac{\sqrt{\lambda\left(t,m_0^2,m_1^2\right)}}{t}\log\left(\frac{-t+m_0^2+m_1^2+\sqrt{\lambda\left(t,m_0^2,m_1^2\right)}}{2m_0 m_1}\right)
\end{align*}
comes from the branch cut of Passarino-Veltman $B_0$ function in the $t$-plane and $\lambda(a,b,c)\equiv a^2+b^2+c^2-2ab-2ac-2bc$ is the K\"all\'en kinematic function. The $C_0$ in Eq.~(\ref{eq:I0box}) are the three-point Passarino-Veltman functions, and their explicit expressions are given by
\begin{align}
&C_0\left(0,0,t;0,m_W,0\right) =  \frac{1}{t}\left[\frac{\pi^2}{6} + \frac{1}{2} \log^2\left(-\frac{m_W^2}{t}\right)+{\rm Li}_2\left(1+\frac{m_W^2}{t}\right)\right],\\
&C_0\left(0,0,-t;m_W,0,m_W\right)\nonumber\\ &=\frac{1}{t}\left[\frac{\pi^2}{6}+\frac{1}{2}\log^2\left(\frac{t-\sqrt{t\left(t+4m_W^2\right)}}{2t}\right)-\frac{1}{2}\log^2\left(\frac{t+2m_W^2+\sqrt{t\left(t+4m_W^2\right)}}{-t+\sqrt{t\left(t+4m_W^2\right)}}\right)\right.\nonumber\\
&\left. \quad\quad- {\rm Li}_2\left(1+\frac{t}{m_W^2}\right)- {\rm Li}_2\left(\frac{2\left(t+m_W^2\right)}{t-\sqrt{t\left(t+4m_W^2\right)}}\right) + {\rm Li}_2\left(\frac{2m_W^2}{-t+\sqrt{t\left(t+4m_W^2\right)}}\right)\right.\nonumber\\
&\left. \quad\quad +{\rm Li}_2\left(\frac{2\left(t+m_W^2\right)}{t+2m_W^2-\sqrt{t\left(t+4m_W^2\right)}}\right)-{\rm Li}_2\left(\frac{2m_W^2}{t+2m_W^2+\sqrt{t\left(t+4m_W^2\right)}}\right)\right].
\end{align}

Then we calculate the discontinuity of $I_0$. First, we note that $\Lambda(-t,m_0,m_1)$ has no branch cut for positive $t$, so it does not contribute to the discontinuity for positive $t$. Then only the first three lines in Eq.~(\ref{eq:I0box}) can contribute to the discontinuity. The contribution in the first line from the logarithm has been calculated in Eq.~(\ref{eq:disclog}). For the second line in Eq.~(\ref{eq:I0box}), using Eqs.~(\ref{eq:discLi2}) and (\ref{eq:disclog2}) we obtain
\begin{align}
{\rm Disc}\left[C_0\left(0,0,t;0,m_W,0\right)\right]  &\equiv C_0\left(0,0,t+i\epsilon;0,m_W,0\right)-C_0\left(0,0,t-i\epsilon;0,m_W,0\right)\nonumber\\
&= -\frac{2\pi i}{t}\log\left(1+\frac{t}{m_W^2}\right).
\end{align}
Finally, for the third line in Eq.~(\ref{eq:I0box}), an explicit calculation gives
\begin{align}
&{\rm Disc}\left[C_0\left(0,0,-t;m_W,0,m_W\right)\right]\nonumber\\
&\equiv C_0\left(0,0,-t-i\epsilon;m_W,0,m_W\right)-C_0\left(0,0,-t+i\epsilon;m_W,0,m_W\right)=0\;.    
\end{align}
Therefore, we obtain the discontinuity of Eq.~(\ref{eq:I0box}) as follows:
\begin{align}
    {\rm Disc}[I_0(t)] &= -\frac{ig^4}{3072 \pi^2} \frac{-24 m_W^6 t\left(2m_W^2+t\right)(2\pi i) + 24 m_W^4 \left(m_W^2+t\right)\left(t^2-2m_W^4\right)(-2\pi i)\log\left(1+\frac{t}{m_W^2}\right)}{m_W^4 t^2 \left(t+2m_W^2\right)^2}
    \nonumber\\
    &= -\frac{g^4}{64 \pi}\frac{m_W^2 t\left(t+2 m_W^2\right) + \left(t^2-2m_W^4\right)\left(t+m_W^2\right)\log\left(1+\frac{t}{m_W^2}\right)}{t^2 \left(t+2m_W^2\right)^2}\;.
\end{align}

Now using the Fierz transformation, the amplitude can be recasted into the standard form of Eq.~(\ref{eq:form}) with effective couplings, $G_V^{\chi_1} = G_V^{\chi_2} =  G_A^{\chi_1} = G_A^{\chi_2} = 1$. Finally, using Eq.~(\ref{eq:V0}), we obtain:
\begin{align}
    \label{eq:V0boxapp}
    V_0^{\rm box}(r) = -\frac{g^4}{512\pi^3 r} \int_0^\infty {\rm d}t\, e^{-\sqrt{t}\,r} \frac{m_W^2 t\left(t+2 m_W^2\right) + \left(t^2-2m_W^4\right)\left(t+m_W^2\right)\log\left(1+\frac{t}{m_W^2}\right)}{t^2 \left(t+2m_W^2\right)^2}\;.
\end{align}
In the long-range limit, again it is reduced to the familiar $1/r^5$ form:
\begin{align}
     V_0^{\rm box}\left(r\gg m_Z^{-1}\right) &= -\frac{g^4}{512 \pi^3 r}\int_0^\infty {\rm d}t\, e^{-\sqrt{t}\,r} \frac{t}{3m_W^4} = -\frac{G_F^2}{4\pi^3 r^5}\;.
\end{align}

\section{Quantum forces from other Standard Model fermions}
\label{app:fermion-force}

In this appendix, we calculate the quantum forces mediated by other SM fermions $\chi$ (with mass $m_\chi$) beyond neutrinos, corresponding to the ``photon penguin'' diagrams in Figs.~\ref{fig:4-fermi-photon} and \ref{fig:photon_penguins}. The calculation can be split into two regions: (i) $r\ll m_\chi^{-1}$ and (ii) $r \gg m_Z^{-1}$. In region (i) we do the calculation in the full electroweak theory (see fig.~\ref{fig:photon_penguins}) and neglect the fermion mass $m_\chi$, while in region (ii) we keep $m_\chi$ and calculate the force in the four-Fermi effective theory (see fig.~\ref{fig:4-fermi-photon}). Since $m_\chi \ll m_Z$, the combination of the calculations in the two regions can cover the whole range of the quantum force mediated by $\chi$. Moreover, as a cross check, the results in region (i) and (ii) should match with each other in the middle range $ m_Z^{-1} \ll r \ll m_\chi^{-1}$. 

\subsection{Region $r \ll m_{\chi}^{-1}$}
In the full theory, there are two types of diagrams: the self-energy (Figs.~\ref{subfig:SE-gamma1} and \ref{subfig:SE-gamma2}) and the penguin (figs.~\ref{subfig:penguin-gamma1} and \ref{subfig:penguin-gamma2}). Notice that, for the self-energy diagram, $\chi$ in the loop can be any charged fermion, whereas for the penguin diagram, $\chi$ in the loop must be the same as either $\chi_1$ or $\chi_2$.
The calculation is similar to that of the neutrino force, where the final result is both finite and gauge invariant.

\renewcommand\arraystretch{1.2}
\begin{table}[h]
\begin{center}
\begin{tabular}{|c | c | c | c | c|} 
 \hline
 Effective Couplings & ${\gamma\rm SE}_1$ & ${\gamma\rm PG}_1$ & ${\gamma\rm SE}_2$ & ${\gamma\rm PG}_2$\\ [0.5ex] 
 \hline
 $G_V^{\chi_1}$ & $ c_V^{\chi_1} c_V^\chi$ & $  \left(c_V^{\chi_1}\right)^2+ \left(c_A^{\chi_1}\right)^2$ & $Q_\chi Q_{\chi_1}$ & $Q_{\chi_1} Q_{\chi_2}$  \\ 
 \hline
 $G_V^{\chi_2}$ & $Q_{\chi} Q_{\chi_2}$  &  $Q_{\chi_1} Q_{\chi_2}$  &$c_V^{\chi_2} c_V^{\chi}$ & $ \left(c_V^{\chi_2}\right)^2+ \left(c_A^{\chi_2}\right)^2 $ \\
 \hline
 $G_A^{\chi_1}$ & $ c_A^{\chi_1}c_V^{\chi}$ & $2 c_V^{\chi_1}c_A^{\chi_1}$ & 0 & 0 \\
 \hline
 $G_A^{\chi_2}$ & 0 & 0 & $ c_A^{\chi_2}c_V^{\chi}$ & $2 c_V^{\chi_2}c_A^{\chi_2}$ \\
 \hline
\end{tabular}
\caption{\label{table:photon-couplings}Effective couplings $G_V^\chi$ and $G_A^\chi$ for the self-energy diagram ($\rm{SE}_1$, $\rm{SE}_2$) and penguin diagram ($\rm{PG}_1$, $\rm{PG}_2$) in Fig.~\ref{fig:photon_penguins}, where the fundamental couplings $c_V^\chi$ and $c_A^\chi$ are given in Eq.~(\ref{eq:fundcoupling}), and $Q_\chi$ denotes the electric charge of $\chi$ (in unit of $e$). The lower index 1 (or 2) of SE and PG refers to the diagram with $\chi_1$ (or $\overline{\chi_2}$) connected to $Z$ boson.}
\end{center}
\end{table}
\renewcommand\arraystretch{1}
As the case of neutrino forces, we only need to calculate $V_0$ for each diagram, while the spin-independent and the spin-dependent parity-violating forces can be calculated from $V_0$ using Eqs.~(\ref{eq:SI}) and (\ref{eq:PV}), with the effective couplings $G_V$ and $G_A$ given in Tab.~\ref{table:photon-couplings}.
For $V_0$, we obtain (defining $x\equiv m_Z r$):
\begin{align}
    V^{\gamma\rm SE}_0(r) &= \left(\frac{eg}{4 c_W}\right)^2 \frac{1}{48 \pi^3 r}\left[e^x {\rm Ei}\left(-x\right)+e^{-x} {\rm Ei}\left(x\right)\right],\label{eq:V0gammaSE} \\
    V_0^{\gamma\rm PG}(r) &= -\left(\frac{eg}{4 c_W} \right)^2 \frac{1}{64 \pi^3 r} \int_0^\infty {\rm d}t\,e^{-\sqrt{t}\,r}\, \frac{t(2m_Z^2 + 3t) - 2 (m_Z^2 + t)^2 \log \left(1+\frac{t}{m_Z^2}\right)} {t\left(t-4 m_\chi^2\right)^2} \;.\label{eq:V0gammaPG}
\end{align} 

One can easily find the asymptotic behavior of the force in the short- and long-range limit. For the self-energy diagram, we obtain:
\begin{align}
    &r \ll m_Z^{-1}, \quad V_0^{\gamma\text{SE}}(r) = \frac{e^2 g^2}{384 \pi^3 c_W^2 r} \left[\gamma_{\rm E} + \log(m_Z r) \right],  \\ 
    &m_Z^{-1}\ll r \ll m_\chi^{-1}, \quad  V_0^{\gamma\text{SE}
    }(r) = \frac{e^2 g^2}{384 \pi^3 m_W^2 r^3}=\frac{\alpha G_F}{12\sqrt{2}\pi^2r^3}\;. 
\end{align}
For the penguin diagram, it turns out to be:
\begin{align}
&r \ll m_Z^{-1},\quad
    V_0^{\gamma\rm PG}(r) = \frac{e^2 g^2}{1024 \pi^3 c_W^2}\frac{1}{r}\left\{4\log\left(m_Z r\right)\left[2\gamma_{\rm E}+\log\left(m_Z r\right)\right]+6\log\left(m_\chi r\right)\right.\nonumber\\
&\left.\qquad\qquad\qquad\qquad\qquad\qquad\qquad\qquad\quad+\pi^2+2\gamma_{\rm E}\left(3+2\gamma_{\rm E}\right)+3\left(1+2\log 2\right)
\right\},\\ 
&m_Z^{-1} \ll r \ll m_\chi^{-1},\quad
V_0^{\gamma\rm PG}(r) = \frac{e^2 g^2}{768 \pi^3 m_W^2 r^3} = \frac{\alpha G_F}{24\sqrt{2}\pi^2r^3} \;. 
\end{align}

\subsection{Region $r \gg m_Z^{-1}$}
In this region, we can calculate the quantum force in the four-Fermi theory, as shown in Fig.~\ref{fig:4-fermi-photon}, where the $Z$ boson is integrated out. Moreover, as $r\gtrsim m_\chi^{-1}$, the fermion mass in the loop cannot be neglected. Therefore, we will keep $m_\chi$ in the following calculation.

\renewcommand\arraystretch{1.2}
\begin{table}[h]
\begin{center}
\begin{tabular}{|c | c | c | c | c|} 
 \hline
 Effective Couplings & $\chi =\chi_1$ & $\chi \neq \chi_1$ & $\chi =\chi_2$ & $\chi \neq \chi_2$\\ [0.5ex] 
 \hline
 $G_V^{\chi_1}$ & $3\left(c_V^{\chi_1}\right)^2+ \left(c_A^{\chi_1}\right)^2$ &  $2c_V^{\chi_1} c_V^\chi$ & $Q_{\chi_1} Q_{\chi_2}$  & $Q_{\chi} Q_{\chi_1}$  \\ 
 \hline
 $G_V^{\chi_2}$ &  $Q_{\chi_1} Q_{\chi_2}$ &  $Q_{\chi} Q_{\chi_2}$ & $3\left(c_V^{\chi_2}\right)^2+ \left(c_A^{\chi_2}\right)^2$ & $2 c_V^{\chi_2} c_V^\chi$\\
 \hline
 $G_A^{\chi_1}$ &  $4 c_V^{\chi_1} c_A^{\chi_1}$  & $2 c_A^{\chi_1} c_V^{\chi}$ & 0 & 0 \\
 \hline
 $G_A^{\chi_2}$ &  0 & 0 & $4 c_V^{\chi_2} c_A^{\chi_2}$  & $2 c_A^{\chi_2} c_V^{\chi}$ \\
 \hline
\end{tabular}
\caption{\label{table:photon-couplings-effective}Effective couplings $G_V^\chi$ and $G_A^\chi$ for the photon penguin diagram in Fig.~\ref{fig:4-fermi-photon}, where the fundamental couplings $c_V^\chi$ and $c_A^\chi$ are given in Eq.~(\ref{eq:fundcoupling}), and $Q_\chi$ denotes the electric charge of $\chi$ (in unit of $e$). The first two and last two columns correspond to the case of $\chi_1\chi_1\chi\chi$ and $\chi_2\chi_2\chi\chi$ effective vertex, respectively.}
\end{center}
\end{table}
\renewcommand\arraystretch{1}
The spin-independent part and the spin-dependent parity-violating part of the force can be computed from $V_0$ using Eqs.~(\ref{eq:SI}) and (\ref{eq:PV}), where the effective couplings $G_V$ and $G_A$ are given in Tab.~\ref{table:photon-couplings-effective} (obtained by matching the electroweak theory onto the four-Fermi effective theory at the tree level). The $V_0$ is straightforward to calculate using the dispersion technique in Eq.~(\ref{eq:V0}):
\begin{align}
    V_0^{\gamma \chi} (r) &= \frac{\alpha G_F}{48\sqrt{2} \pi^2 r} \int_{4m_\chi^2}^\infty\frac{ {\rm d}t}{t^2}e^{-\sqrt{t} r}\left(2m_\chi^2 +t\right)\sqrt{t\left(t-4m_\chi^2\right)}  \nonumber  \\ 
    &= \frac{\alpha G_F}{24\sqrt{2} \pi^2 r} e^{-2m_\chi r} \int_{0}^\infty {\rm d}x\,\frac{ e^{-xr}}{\left(x+2m_\chi\right)^2}\left(x^2 + 4m_\chi x + 6 m_\chi^2\right) \sqrt{x^2 + 4m_\chi x}\;, \label{eq:V0gammachi}
\end{align}
where in the second line we have shifted the variable to $x\equiv\sqrt{t}-2m_\chi$. The asymptotic behavior turns out to be:
\begin{align}
     &m_Z^{-1} \ll r \ll m_\chi^{-1}, \quad V_0^{\gamma \chi} (r) =  \frac{\alpha G_F}{24\sqrt{2} \pi^2 r^3}\;,  \\
    & r \gg m_\chi^{-1}, \quad V_0^{\gamma \chi} (r) = \frac{\alpha G_F}{16\pi^2} \sqrt{\frac{\pi m_\chi}{2}} \frac{e^{-2 m_\chi r}}{r^{5/2}}\;.
\end{align}

\section{Dispersion formalism in the presence of
additional poles}
\label{app:dispersion}

In this appendix, we generalize the dispersion formalism~\cite{Feinberg:1989ps} for calculating quantum potentials in the presence of additional poles on the
real axis apart from the branch cut, which is relevant to the calculation of the neutrino force from self-energy and penguin diagrams. 

\begin{figure}[t!]
\centering
\includegraphics[width=0.45\linewidth]{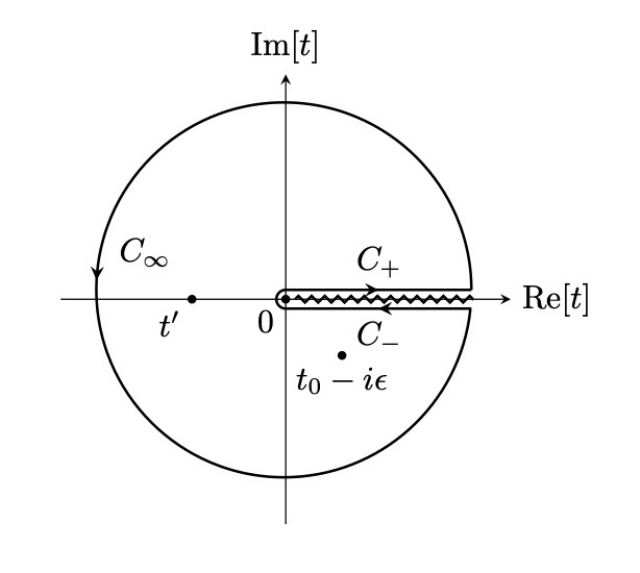}
\vspace{-1cm}
\caption{\label{fig:contour}The complex plane of $t$ and the contour $C$ used in Eq.~(\ref{eq:Cauchy}) to compute the one-loop amplitude, where the branch cut is on the positive real axis.}
\end{figure}

According to the Born approximation, the potential is given by the Fourier transform of the $t$-channel scattering amplitude in the non-relativistic limit:
\begin{align}
V(r) = -\int \frac{{\rm d}^3\vecq}{\left(2\pi\right)^3}e^{i\vecq \cdot \vecr} {\cal M}(\vecq)\;, \label{eq:Fourier}   
\end{align}
where $q^\mu\approx (0,\vecq)$ is the momentum transfer of the $t$-channel scattering process. The dispersion formalism developed by Feinberg, Sucher, and Au~\cite{Feinberg:1989ps} provides an alternative way to compute the potential from the absorptive part of the amplitude:
\begin{align}
V(r) = -\frac{1}{4\pi^2 r} \int_{0}^\infty {\rm d}t\,e^{-\sqrt{t}\,r}\,{\rm Im}\left[{\cal M}(t)\right], \label{eq:Laplace}  
\end{align}
where the lower limit of the integral is zero because we have neglected the mass of the mediator. (For our purpose, this is a good approximation since the neutrino is effectively massless at the atomic scale.) Here, $t$ is the value of the momentum transfer squared after analytic continuation (see Fig.~\ref{fig:contour}). The method in Eq.~(\ref{eq:Laplace}) is particularly useful to calculate the potential generated at the one-loop level (known as the quantum force), because the loop amplitude is typically divergent, while its imaginary part is finite according to the optical theorem.

However, for Eq.~(\ref{eq:Laplace}) to hold, two crucial assumptions have been made~\cite{Feinberg:1989ps}: 
\begin{enumerate}
    \item ${\cal M}(t)$ vanishes as $t\to \infty$;
    \item ${\cal M}(t)$ does not contain additional poles other than the branch cut on the real axis of $t$.
\end{enumerate}
For our purpose of calculating the full-range neutrino force from three kinds of diagrams (self-energy, penguin, box), the first assumption is satisfied, while the second assumption is not satisfied for self-energy and penguin diagrams because
their amplitudes contain additional poles from the $Z$ propagator.
Now, our goal here is to generalize Eq.~(\ref{eq:Laplace}) in the presence of additional poles. 

\vspace{0.2cm}
In general, the relevant amplitude in this work can be factorized into
\begin{align}
{\cal M}(t) = \frac{{\cal M}_0(t)}{\left(t-t_0 + i\epsilon\right)^n}\;,    
\end{align}
where ${\cal M}_0$ denotes the part that does not contain additional poles on the real axis, $t_0\equiv m_Z^2$ is the pole caused by the $Z$ propagator, $\epsilon \to 0^+$ comes from the use of the Feynman propagator, and $n=0, 1, 2$ for box, penguin, and self-energy, respectively.

The Mandelstam variable for $t$-channel scattering is defined as $t'\equiv q^2 = -\vecq^2$, which is always negative in the physical domain. As a result, the $t$-channel amplitude ${\cal M}(t')$ is always real when the argument is in the physical domain. Since the potential in Eq.~(\ref{eq:Fourier}) is obtained by Fourier transforming this real amplitude (and also note that the amplitude is an even function of $|\vecq|$), it too must be real.

One can extend the value of $t'$ from the negative real axis to the whole complex plane using Cauchy's theorem: 
\begin{align}
     \frac{1}{2\pi i} \int_C {\rm d}t\, \frac{ {\cal M}(t)}{t-t'}&={\cal M}(t') + {\rm Res}\left(\frac{{\cal M}(t)}{t-t'}; t= t_0-i\epsilon\right)\equiv {\cal M}(t') + {\rm Res}(n)\;,\label{eq:Cauchy}
\end{align}
where $C$ is a closed contour that does not contain additional poles for ${\cal M}(t)$ apart from $t=t_0-i\epsilon$. The first and second terms on the right-hand side of Eq.~(\ref{eq:Cauchy}) come from the residue of the function ${\cal M}(t)/(t-t')$ at $t=t'$ and at $t=t_0-i \epsilon$, respectively. We split the contour as in Fig.~\ref{fig:contour} to avoid the branch cut on the positive real axis:
\begin{align}
  C=C_+ + C_\infty + C_-\;.  
\end{align}
Then the original amplitude in the physical domain can be expressed as
\begin{align}
{\cal M}(t') &= \frac{1}{2\pi i} \left[\int_0^\infty {\rm d}t \frac{{\cal M}_0(t+i 0^+)}{\left(t-t_0+i\epsilon\right)^n \left(t-t'\right)}+\int_{C_\infty}+\int_\infty^0 {\rm d}t \frac{{\cal M}_0(t-i 0^+)}{\left(t-t_0+i\epsilon\right)^n \left(t-t'\right)}\right]-{\rm Res}(n)\nonumber\\
&=\frac{1}{2\pi i} \left[\int_0^\infty {\rm d}t \frac{{\cal M}_0(t+i 0^+)-{\cal M}_0(t-i 0^+)}{\left(t-t_0+i\epsilon\right)^n \left(t-t'\right)}\right]-{\rm Res}(n)\nonumber\\
&=\frac{1}{\pi} \left[\int_0^\infty {\rm d}t \frac{{\rm Im} \left[{\cal M}_0(t)\right]}{\left(t-t_0+i\epsilon\right)^n \left(t-t'\right)}\right]-{\rm Res}(n)\;,\label{eq:contourintegral}
\end{align}
where the integral along the large circle vanishes as $t\to \infty$, and $\mathcal{M}_0(t)$ obeys the Schwartz reflection principle ($\mathcal{M}^*_0(z) =\mathcal{M}_0(z^*) $) which leads to:
\begin{align}
{\cal M}_0(t+i 0^+)-{\cal M}_0(t-i 0^+)  \equiv {\rm Disc}\left[{\cal M}_0(t)\right] \equiv 2i {\rm Im}\left[{\cal M}_0(t)\right].\label{eq:disc}    
\end{align}

The key observation here is that since ${\cal M}(t')$ is real, the imaginary part of the two terms on the right-hand side of Eq.~(\ref{eq:contourintegral}) \emph{must} cancel each other. In the following, we explicitly show that this is indeed the case. We discuss separately the cases of $n=1$ and $n=2$, which correspond to the cases of penguin and self-energy diagrams, respectively.

\subsection{Simple pole}
For the case of the simple pole, the residue is given by
\begin{align}
{\rm Res}(1) &= \lim_{t\to t_0-i\epsilon}\left(t-t_0+i\epsilon\right) \frac{{\cal M}_0(t)}{\left(t-t_0+i\epsilon\right)\left(t-t'\right)}= \frac{{\cal M}_0(t_0-i\epsilon)}{t_0 -t'}\;,\label{eq:Res1}
\end{align}
where, in the second step, we have set $\epsilon \to 0$ in the denominator because there is no pole (recall that $t_0=m_Z^2 >0$ while $t'<0$). Note that $i\epsilon$ in the numerator of Eq.~(\ref{eq:Res1}) is crucial for the sign of the imaginary part because it determines the branch of the logarithm and dilogarithm functions. Using the relation
\begin{align}
{\rm Im}\left[{\cal M}_0(t_0-i\epsilon)\right] = - {\rm Im}\left[{\cal M}_0(t_0+i\epsilon)\right]\equiv -{\rm Im}\left[{\cal M}_0(t_0)\right],    
\end{align}
where the imaginary part of the amplitude is defined by Eq.~(\ref{eq:disc}) and we used the notation ${\rm Im}\left[{\cal M}_0(t_0)\right]\equiv {\rm Im}\left[{\cal M}_0(t)\right]|_{t=t_0}$, we have
\begin{align}
{\rm Im}\left[{\rm Res}(1)\right] = -  \frac{{\rm Im}\left[{\cal M}_0(t_0)\right]}{t_0-t'}\;.\label{eq:ImRes1} 
\end{align}
On the other hand, the imaginary part of the first term on the right-hand side of Eq.~(\ref{eq:contourintegral}) reads:
\begin{align}
&\frac{1}{\pi}  \int_0^\infty {\rm d}t \frac{{\rm Im} \left[{\cal M}_0(t)\right]}{\left(t-t'\right)}{\rm Im}\left[\frac{1}{t-t_0+i\epsilon}\right] = \frac{1}{\pi}\int_0^\infty {\rm d}t \frac{{\rm Im} \left[{\cal M}_0(t)\right]}{\left(t-t'\right)} (-\pi)\delta\left(t-t_0\right) = -\frac{{\rm Im} \left[{\cal M}_0(t_0)\right]}{t_0 - t'}\;,\label{eq:ImPG}
\end{align}
which exactly cancels Eq.~(\ref{eq:ImRes1}). 

Therefore, we have verified that the imaginary part of the two terms on the right-hand side of Eq.~(\ref{eq:contourintegral}) cancels each other for $n=1$. Then Eq.~(\ref{eq:contourintegral}) can be written as
\begin{align}
{\cal M}(t') = \frac{1}{\pi}{\rm Re} \left[\int_0^\infty {\rm d}t \frac{{\rm Im} \left[{\cal M}_0(t)\right]}{\left(t-t_0+i\epsilon\right) \left(t-t'\right)}\right] - \frac{{\rm Re}\left[{\cal M}_0(t_0)\right]}{t_0-t'}\;.\label{eq:MtRe1} 
\end{align}
Substituting Eq.~(\ref{eq:MtRe1}) into Eq.~(\ref{eq:Fourier}), we obtain
\begin{align}
V(r) = & -\frac{1}{\pi} {\rm Re} \left[ \int_0^\infty {\rm d}t \frac{{\rm Im} \left[{\cal M}_0(t)\right]}{\left(t-t_0+i\epsilon\right)}
\int \frac{{\rm d}^3\vecq}{\left(2\pi\right)^3}e^{i\vecq \cdot \vecr} \frac{1}{t-t'}\right]\nonumber\\
&  +{\rm Re}\left[{\cal M}_0(t_0)\right]
\int \frac{{\rm d}^3\vecq}{\left(2\pi\right)^3}e^{i\vecq \cdot \vecr} \frac{1}{t_0 -t'}\;.\label{eq:Vsimplepole}
\end{align}
The integral of $\vecq$ can easily be worked out as follows by recalling that $t'=-\vecq^2<0$ while both $t$ and $t_0$ are positive in the integration region:
\begin{align}
    \int\frac{{\rm d}^3\vecq}{\left(2\pi\right)^3}e^{i\vecq\cdot\vecr} \frac{1}{t-t'} &= -\frac{i}{4\pi^2 r} \int_{-\infty}^\infty {\rm d}|\vecq|\,\frac{|\vecq|}{t+|\vecq|^2 }e^{i|\vecq|r} = \frac{1}{4\pi r} e^{-\sqrt{t}r}\;,\label{eq:integraloverq}\\
    \int\frac{{\rm d}^3\vecq}{\left(2\pi\right)^3}e^{i\vecq\cdot\vecr} \frac{1}{t_0-t'} &= -\frac{i}{4\pi^2 r} \int_{-\infty}^\infty {\rm d}|\vecq|\,\frac{|\vecq|}{t_0+|\vecq|^2 }e^{i|\vecq|r} = \frac{1}{4\pi r} e^{-\sqrt{t_0}r}\;,
\end{align}
Substituting them back to Eq.~(\ref{eq:Vsimplepole}) leads to
\begin{align}
\boxed{
V(r) = -\frac{1}{4\pi^2 r}   {\rm Re} \left[ \int_0^\infty {\rm d}t \frac{{\rm Im} \left[{\cal M}_0(t)\right]}{\left(t-t_0+i\epsilon\right)} e^{-\sqrt{t} r}\right] + \frac{1}{4\pi r} {\rm Re}\left[{\cal M}_0(t_0)\right] e^{-\sqrt{t_0}r}\;.\label{eq:VRe1}
}
\end{align}
\vspace{0.2cm}

Next, we apply the general expression in Eq.~(\ref{eq:VRe1}) to compute the potential from the penguin diagram. For the penguin diagram, the amplitude ${\cal M}_0$ is given by
\begin{align}
{\cal M}_0 (q^2) =  -\frac{g^4}{128 c_W^2}iF_1(q^2)\;,   
\end{align}
where the function $F_1(q^2)$ is given in Eq.~(\ref{eq:F1}).

\begin{figure}[t]
    \centering
    \subfigure[Cut at the neutrino loop]{
        \includegraphics[scale = 1]{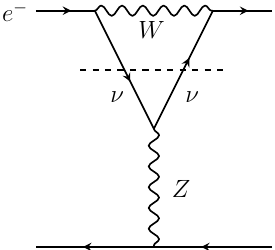}
        \label{subfig:cutNu}
    }
    \hspace{1cm}
    \subfigure[Cut at the $Z$ propagator]{
        \includegraphics[scale = 1]{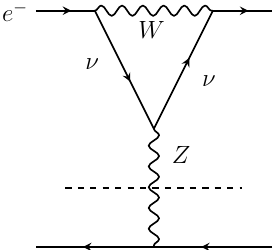}
        \label{subfig:cutZ}
    }
    \caption{ \label{fig:cut} Different cuts in the penguin diagram, corresponding to the two real terms in Eq.~(\ref{eq:VRe1}). The cut at the neutrino loop (a) contributes to the long-range neutrino force $V_{2\nu}^{\rm PG}(r)$. The cut at the $Z$ propagator (b) gives an additional term $V_\text{$Z$-pole}^{\rm PG}(r)\sim e^{-m_Zr}/r$, which is a contact correction to the tree-level $Z$ force.}
\end{figure}

There are two real terms in Eq.~(\ref{eq:VRe1}), where the first term corresponds to the cut at the neutrino loop in Fig.~(\ref{subfig:cutNu}), giving
\begin{align}
V_{2\nu}^{\rm PG}(r) &=  \frac{g^4}{512\pi^2 c_W^2 r}{\rm Re}\left[\int_0^\infty {\rm d}te^{-\sqrt{t}r} \frac{{\rm Im}\left[i F_1(t)\right]}{\left(t-m_Z^2+i\epsilon\right)}\right]\nonumber\\
&=-\frac{g^4}{2048 \pi^3  c_W^2 r} {\rm Re}\left[\int_0^\infty {\rm d}t\,e^{-\sqrt{t}\,r}\,\frac{t\left(2m_W^2 + 3t\right) - 2 \left(m_W^2 + t\right)^2 \log \left(1+\frac{t}{m_W^2}\right)} {t^2\left(t-m_Z^2+i\epsilon\right)}\right],\label{eq:VPG2Nu}
\end{align}
where ${\rm Im}\left[i F_1(t)\right]$ is calculated using Eqs.~(\ref{eq:F1})-(\ref{eq:disclog2}). Note that Eq.~(\ref{eq:VPG2Nu}) is exactly the equation we used to calculate the penguin neutrino force $V_0^{\rm PG}$ in the main text  (see Eq.~(\ref{eq:V0PGmod})).

On the other hand, the second term in Eq.~(\ref{eq:VRe1}) corresponds to the cut at $Z$ propagator in Fig.~(\ref{subfig:cutZ}), which gives
\begin{align}
V_\text{$Z$-pole}^{\rm PG}(r) = -\frac{g^4}{512 \pi c_W^2 } {\rm Re}\left[iF_1(m_Z^2)\right]  \frac{e^{-m_Z r}}{r}\;,\label{eq:Z-pole}
\end{align}
 
Note that the real part of $iF_1(q^2)$ contains a UV divergent part, and one needs to define a subtraction scheme to remove the divergence. The exact value of ${\rm Re}\left[iF_1(m_Z^2)\right]$ after renormalization depends on which specific subtraction scheme is used. It is not surprising that the real part of the amplitude from a single diagram is UV sensitive, because there also exist other one-loop diagrams (including those without two neutrinos in the loop) that contribute to the $eeZ$ vertex. In principle, one should add the amplitudes from all of these diagrams together, find a counterterm to remove the divergence in the real part, and fix the finite part. In this work, however, we are not aiming to perform a full renormalization in the electroweak theory, which is already a known result in the literature~\cite{Denner:1991kt}. Instead, we are interested in the long-range neutrino force\footnote{To be more precise, by ``neutrino force'', what we exactly mean is the long-range potential mediated by two neutrinos that has a nontrivial and \emph{different} scaling behavior from the $Z$ force, i.e., beyond the contact correction of the $Z$ force.} and its effect in APV, which is precisely the part missed in the early work \cite{Marciano:1978ed,Marciano:1982mm,Marciano:1983ss}.

Based on the above consideration, we adopt the following renormalization scheme for the $eeZ$ vertex so that the contribution from Eq.~(\ref{eq:Z-pole}) can be absorbed into the $eeZ$ coupling. First, we note that $F_1$ only couples to the left-handed current at the leading order of ${\cal O}(m_e^2/m_W^2)$ (see Eq.~(\ref{eq:vertexfunction})). At the tree level, the coupling between the left-handed electron and the $Z$ boson reads
\begin{align}
{\cal L}\supset \frac{g_0}{2c_W}\left(-\frac{1}{2}+s_W^2\right)  \bar{e}\gamma^\mu\left(1-\gamma^5\right)e\,Z_\mu\;,
\end{align}
where $g_0$ is the bare ${\rm SU(2)_L}$ gauge coupling.
To remove the divergence in the vertex function, we introduce the counterterm: 
\begin{align}
g_0\equiv g_R \left(1+\delta_g\right),    
\end{align}
where $g_R$ is the renormalized gauge coupling, while $\delta_g$ is the counterterm starting at the order $g_R^2$. At the one-loop level, the renormalized vertex of $\bar{e}\Gamma^\mu eZ_\mu$ is given by
\begin{align}
    \Gamma^\mu(q^2) &= \frac{g_0}{2 c_W}\left(-\frac{1}{2}+s_W^2\right)\gamma^\mu \left(1-\gamma^5\right)+\frac{g_0^3}{32c_W}iF_1(q^2)\gamma^\mu \left(1-\gamma^5\right)
    \nonumber\\
    &= \frac{g_R}{2c_W}\left(-\frac{1}{2}+s_W^2\right)\left[1+\widetilde{F}_1(q^2)+\delta_g\right]\gamma^\mu \left(1-\gamma^5\right) + {\cal O}\left(g_R^5\right),\label{eq:Gammamu}
\end{align}
where 
\begin{align}
\widetilde{F}_1(q^2) \equiv   \frac{g_R^2}{16\left(-\frac{1}{2}+s_W^2\right)} iF_1(q^2)\;.
\end{align}
Note that if other one-loop diagrams are included, then $\widetilde{F}_1$ in Eq.~(\ref{eq:Gammamu}) should be understood as the sum of all relevant one-loop vertex functions. To determine the counterterm, we choose the reference point at $Z$ pole and define the renormalization condition as follows:
\begin{align}
{\rm Re}~\Gamma^\mu\left(q^2=m_Z^2\right) = \frac{g_R}{2c_W}\left(-\frac{1}{2}+s_W^2\right) \gamma^\mu \left(1-\gamma^5\right), \label{eq:renormalizationcondition} 
\end{align}
from which we obtain the counterterm:
\begin{align}
\delta_g = -{\rm Re}\left[\widetilde{F}_1(m_Z^2)\right].  
\end{align}
Substituting the counterterm back to Eq.~(\ref{eq:Gammamu}) we obtain
\begin{align}
    \Gamma^\mu (q^2) &=\frac{g_R}{2c_W}\left(-\frac{1}{2}+s_W^2\right) \left[1 + \left(\widetilde{F}_1(q^2) - {\rm Re}\left[\widetilde{F}_1(m_Z^2)\right] \right)\right]\gamma^\mu \left(1-\gamma^5\right).\label{eq:Gammamu2}
\end{align}
In this renormalization scheme, $g_R$ is an input parameter, which is determined by the measurement of the gauge coupling at the energy scale of the $Z$ mass. The result in Eq.~(\ref{eq:Gammamu2}) is finite and can be used to predict the coupling strength of $\bar{e}\gamma^\mu\left(1-\gamma^5\right)e$ at arbitrary scales.

Since the potential in Eq.~(\ref{eq:Z-pole}) only depends on the real value of $\widetilde{F}_1(q^2)$ evaluated at $q^2=m_Z^2$, it \emph{vanishes} after adding the counterterm according to the renormalization condition in Eq.~(\ref{eq:renormalizationcondition}). Therefore, in this renormalization scheme, the expressions for both the neutrino forces and the tree-level $Z$ force are not affected by Eq.~(\ref{eq:Z-pole}); the only difference is that the bare coupling $g_0$ should be replaced with the renormalized coupling $g_R$ everywhere. Note that in this work, the physical observable relevant to APV is the neutrino force contribution \emph{relative to} the $Z$ force contribution, defined as $\eta_\nu \equiv C_{AB}^\nu/C_{AB}^Z$, which is of order $g_R^2$. However, the difference between $g_0$ and $g_R$ starts at ${\cal O}(g_R^3)$ because the counterterm $\delta_g$ starts at ${\cal O}(g_R^2)$. Therefore, we conclude that all results of $\eta_\nu$ in the main text are not affected by Eq.~(\ref{eq:Z-pole}) in this renormalization scheme.

To sum up, for the case of the simple pole (penguin diagram), the correct formula to calculate the potential is given by Eq.~(\ref{eq:VRe1}).
We have shown that the imaginary part of the contribution from the branch cut caused by $i\epsilon$ is exactly canceled by the imaginary part of the contribution from the $Z$ pole, making the final result real. In addition, the real part of the contribution from the $Z$ pole only contributes a contact correction to the tree-level $Z$ force, which only has a short-distance effect and is fundamentally different from the neutrino force we considered throughout this work. In particular, we explicitly demonstrated above that this term can be absorbed into the $eeZ$ coupling through a proper renormalization scheme and therefore does not affect the neutrino force contribution to APV calculated in the main text.

\subsection{Double pole}
Next, we turn to the case of the double pole, which is relevant to the neutrino force from the self-energy diagram. The residue is found to be
\begin{align}
{\rm Res}(2)= \lim_{t\to t_0-i\epsilon}\frac{{\rm d}}{{\rm d}t}\left[\left(t-t_0+i\epsilon\right)^2
\frac{{\cal M}_0(t)}{\left(t-t_0+i\epsilon\right)^2\left(t-t'\right)}\right] = \frac{{\cal M}'_0(t_0-i\epsilon)}{t_0-t'} - \frac{{\cal M}_0(t_0-i\epsilon)}{\left(t_0-t'\right)^2}\;,   
\end{align}
where we have denoted ${\cal M}_0'(t_0)\equiv [{\rm d}{\cal M}_0(t)/{\rm d}t]|_{t=t_0}$. 
The imaginary part of the residue reads
\begin{align}
    {\rm Im}\left[{\rm Res}(2)\right]= - \frac{{\rm Im}\left[{\cal M}'_0(t_0)\right]}{t_0 - t'} + \frac{{\rm Im}\left[{\cal M}_0(t_0)\right]}{\left(t_0-t'\right)^2}\;.\label{eq:ImRes2}
\end{align}
On the other hand, the imaginary part of the first term on the right-hand side of Eq.~(\ref{eq:contourintegral}) reads
\begin{align}
\frac{1}{\pi}  \int_0^\infty {\rm d}t \frac{{\rm Im} \left[{\cal M}_0(t)\right]}{\left(t-t'\right)}\,{\rm Im}\left[\frac{1}{\left(t-t_0+i\epsilon\right)^2}\right]\;.\label{eq:Imsquare} 
\end{align}
Using the identity
\begin{align*}
{\rm Im}\left[\frac{1}{\left(t-t_0+i\epsilon\right)^2}\right] = \pi \delta'\left(t-t_0\right) \;,    
\end{align*}
and integration by parts, Eq.~(\ref{eq:Imsquare}) is reduced to
\begin{align}
-\int_0^\infty {\rm d}t \frac{{\rm d}}{{\rm d}t}\left(\frac{{\rm Im} \left[{\cal M}_0(t)\right]}{t-t'}\right)\delta\left(t-t_0\right) &=  -\int_0^\infty {\rm d}t\left(\frac{{\rm Im} \left[{\cal M}'_0(t)\right]}{t-t'} - \frac{{\rm Im} \left[{\cal M}_0(t)\right]}{\left(t-t'\right)^2}\right)\delta\left(t-t_0\right)\nonumber\\
&= - \frac{{\rm Im} \left[{\cal M}'_0(t_0)\right]}{t_0-t'} + \frac{{\rm Im} \left[{\cal M}_0(t_0)\right]}{\left(t_0-t'\right)^2}\;,
\end{align}
which again cancels the imaginary part in Eq.~(\ref{eq:ImRes2}). Therefore, we have also proved that for $n=2$, the imaginary part of the two terms on the right-hand side of Eq.~(\ref{eq:contourintegral}) cancel each other. Then the original amplitude in Eq.~(\ref{eq:contourintegral}) can be written as
\begin{align}
{\cal M}(t)=    \frac{1}{\pi}{\rm Re} \left[\int_0^\infty {\rm d}t \frac{{\rm Im} \left[{\cal M}_0(t)\right]}{\left(t-t_0+i\epsilon\right)^2 \left(t-t'\right)}\right]  - {\rm Re}\left[{\rm Res}(2)\right],\label{eq:MtRe2} 
\end{align}
where
\begin{align}
{\rm Re}\left[{\rm Res}(2)\right] = \frac{1}{\left(t_0-t'\right)}{\rm Re}\left[\frac{{\rm d}}{{\rm d}t} {\cal M}_0(t)\right]_{t=t_0} - \frac{1}{\left(t_0-t'\right)^2}   {\rm Re}\left[{\cal M}_0(t_0)\right].
\end{align}
In our case, ${\cal M}_0$ is the self-energy of the $Z$ boson. (Note that the spin-independent neutrino force depends only on the transverse part of the self-energy amplitude.) Recall that in the standard on-shell renormalization scheme, the renormalization conditions require both ${\rm Re}\left[{\cal M}_0(t)\right]$ and its derivative to vanish at the mass shell $t=t_0\equiv m_Z^2$ (see e.g., Section 3.2 in \cite{Denner:1991kt}). So we have ${\rm Re}\left[{\rm Res}(2)\right]=0$ after on-shell renormalization. Also note that the renormalization procedure affects only the real part of the amplitude and leaves the imaginary part untouched.
Substituting Eq.~(\ref{eq:MtRe2}) into Eq.~(\ref{eq:Fourier}) and working out the integral over $\vecq$ using Eq.~(\ref{eq:integraloverq}), we arrive at
\begin{align}
\boxed{
V(r) = -\frac{1}{4\pi^2 r}   {\rm Re} \left[ \int_0^\infty {\rm d}t \frac{{\rm Im} \left[{\cal M}_0(t)\right]}{\left(t-t_0+i\epsilon\right)^2} e^{-\sqrt{t} r}\right].\label{eq:VRe2}
}
\end{align}

\vspace{0.2cm}

Then, we apply Eq.~(\ref{eq:VRe2}) to computing the neutrino force from the self-energy diagram. The imaginary part of ${\cal M}_0$ corresponds to the cut at the neutrino loop, which gives (see Eq.~(\ref{eq:discI0SE}))
\begin{align}
 {\rm Im} \left[{\cal M}_0(t)\right] =  \left(\frac{g}{4c_W}\right)^4\frac{t}{6\pi}\;.  \label{eq:ImM0SE} 
\end{align}
Substituting it into Eq.~(\ref{eq:VRe2}) one obtains (with $t_0\equiv m_Z^2$ and $x\equiv m_Zr$)
\begin{align}
V_0^{\rm SE}(r) &= -\left(\frac{g}{4c_W}\right)^4\frac{1}{24\pi^3 r}  {\rm Re} \left[ \int_0^\infty {\rm d}t \frac{t}{\left(t-t_0+i\epsilon\right)^2} e^{-\sqrt{t} r}\right]\nonumber\\
&= \left(\frac{g}{4c_W}\right)^4\frac{1}{48\pi^3 r} {\rm Re}\left[e^{x}\left(2+x\right){\rm Ei}(-x)+e^{-x}\left(2-x\right){\rm Ei}(x)+2 + i \pi\left(2-x\right)e^{-x}\right]\nonumber\\
& = \left(\frac{g}{4c_W}\right)^4\frac{1}{48\pi^3 r}\left[e^{x}\left(2+x\right){\rm Ei}(-x)+e^{-x}\left(2-x\right){\rm Ei}(x)+2 \right],\label{eq:V0SEapp2}
\end{align}
which reproduces Eq.~(\ref{eq:V0SE}) in the main text.

Note that the integral in the first line of Eq.~(\ref{eq:V0SEapp2}) does lead to a finite imaginary part due to $i\epsilon$. But, as we have proved above, this imaginary part is canceled by the imaginary part from the double $Z$ pole. In fact, using Eqs.~(\ref{eq:ImRes2}) and (\ref{eq:ImM0SE}), we have
\begin{align}
&\frac{{\rm Im}\left[{\rm Res}(2)\right]}{\left(g/4c_W\right)^4} = \frac{t}{6\pi\left(t-m_Z^2\right)^2} = -\frac{\vecq^2}{6\pi\left(\vecq^2+m_Z^2\right)^2}\;\nonumber\\
\implies & -\int \frac{{\rm d}^3\vecq}{\left(2\pi\right)^3}e^{i\vecq \cdot \vecr}\,\frac{{\rm Im}\left[{\rm Res}(2)\right]}{\left(g/4c_W\right)^4} \overset{\rho\equiv |\vecq|}{=} \frac{1}{24 \pi^3 i r } \int_{-\infty}^{\infty}{\rm d} \rho \frac{\rho^3}{\left(\rho^2+m_Z^2\right)^2}e^{i\rho r} \nonumber\\
&\qquad\qquad\qquad\qquad\qquad\qquad\;\; = \frac{1}{48\pi^2 r}\left(2-m_Z r\right)e^{-m_Z r}\;,
\end{align}
which indeed cancels the imaginary part in the bracket in the second line of  Eq.~(\ref{eq:V0SEapp2}). This justifies the validity of only taking the real part in Eq.~(\ref{eq:V0SEapp2}).

\bibliography{biblio}{}

\providecommand{\href}[2]{#2}\begingroup\raggedright\begin{thebibliography}{10}

\bibitem{Feinberg}
G.~Feinberg and J.~Sucher, \emph{Long-range forces from neutrino-pair
  exchange}, \href{https://doi.org/10.1103/PhysRev.166.1638}{\emph{Phys. Rev.}
  {\bfseries 166} (1968) 1638}.

\bibitem{Feinberg:1989ps}
G.~Feinberg, J.~Sucher and C.K.~Au, \emph{{The Dispersion Theory of Dispersion
  Forces}}, \href{https://doi.org/10.1016/0370-1573(89)90111-7}{\emph{Phys.
  Rept.} {\bfseries 180} (1989) 83}.

\bibitem{Hsu:1992tg}
S.D.H.~Hsu and P.~Sikivie, \emph{{Long range forces from two neutrino exchange
  revisited}}, \href{https://doi.org/10.1103/PhysRevD.49.4951}{\emph{Phys. Rev.
  D} {\bfseries 49} (1994) 4951}
  [\href{https://arxiv.org/abs/hep-ph/9211301}{{\ttfamily hep-ph/9211301}}].

\bibitem{Grifols:1996fk}
J.A.~Grifols, E.~Masso and R.~Toldra, \emph{{Majorana neutrinos and long range
  forces}}, \href{https://doi.org/10.1016/S0370-2693(96)01304-4}{\emph{Phys.
  Lett. B} {\bfseries 389} (1996) 563}
  [\href{https://arxiv.org/abs/hep-ph/9606377}{{\ttfamily hep-ph/9606377}}].

\bibitem{Lusignoli:2010gw}
M.~Lusignoli and S.~Petrarca, \emph{{Remarks on the forces generated by
  two-neutrino exchange}},
  \href{https://doi.org/10.1140/epjc/s10052-011-1568-7}{\emph{Eur. Phys. J. C}
  {\bfseries 71} (2011) 1568}
  [\href{https://arxiv.org/abs/1010.3872}{{\ttfamily 1010.3872}}].

\bibitem{LeThien:2019lxh}
Q.~Le~Thien and D.E.~Krause, \emph{{Spin-Independent Two-Neutrino Exchange
  Potential with Mixing and $CP$-Violation}},
  \href{https://doi.org/10.1103/PhysRevD.99.116006}{\emph{Phys. Rev. D}
  {\bfseries 99} (2019) 116006}
  [\href{https://arxiv.org/abs/1901.05345}{{\ttfamily 1901.05345}}].

\bibitem{Costantino:2020bei}
A.~Costantino and S.~Fichet, \emph{{The Neutrino Casimir Force}},
  \href{https://doi.org/10.1007/JHEP09(2020)122}{\emph{JHEP} {\bfseries 09}
  (2020) 122} [\href{https://arxiv.org/abs/2003.11032}{{\ttfamily
  2003.11032}}].

\bibitem{Segarra:2020rah}
A.~Segarra and J.~Bernab\'eu, \emph{{Absolute neutrino mass and the
  Dirac/Majorana distinction from the weak interaction of aggregate matter}},
  \href{https://doi.org/10.1103/PhysRevD.101.093004}{\emph{Phys. Rev. D}
  {\bfseries 101} (2020) 093004}
  [\href{https://arxiv.org/abs/2001.05900}{{\ttfamily 2001.05900}}].

\bibitem{Horowitz:1993kw}
C.J.~Horowitz and J.T.~Pantaleone, \emph{{Long range forces from the
  cosmological neutrinos background}},
  \href{https://doi.org/10.1016/0370-2693(93)90800-W}{\emph{Phys. Lett. B}
  {\bfseries 319} (1993) 186}
  [\href{https://arxiv.org/abs/hep-ph/9306222}{{\ttfamily hep-ph/9306222}}].

\bibitem{Ferrer:1998ju}
F.~Ferrer, J.A.~Grifols and M.~Nowakowski, \emph{{Long range forces induced by
  neutrinos at finite temperature}},
  \href{https://doi.org/10.1016/S0370-2693(98)01489-0}{\emph{Phys. Lett. B}
  {\bfseries 446} (1999) 111}
  [\href{https://arxiv.org/abs/hep-ph/9806438}{{\ttfamily hep-ph/9806438}}].

\bibitem{Ferrer:1999ad}
F.~Ferrer, J.A.~Grifols and M.~Nowakowski, \emph{{Long range neutrino forces in
  the cosmic relic neutrino background}},
  \href{https://doi.org/10.1103/PhysRevD.61.057304}{\emph{Phys. Rev. D}
  {\bfseries 61} (2000) 057304}
  [\href{https://arxiv.org/abs/hep-ph/9906463}{{\ttfamily hep-ph/9906463}}].

\bibitem{Ghosh:2022nzo}
M.~Ghosh, Y.~Grossman, W.~Tangarife, X.-J.~Xu and B.~Yu, \emph{{Neutrino forces
  in neutrino backgrounds}},
  \href{https://doi.org/10.1007/JHEP02(2023)092}{\emph{JHEP} {\bfseries 02}
  (2023) 092} [\href{https://arxiv.org/abs/2209.07082}{{\ttfamily
  2209.07082}}].

\bibitem{Blas:2022ovz}
D.~Blas, I.~Esteban, M.C.~Gonzalez-Garcia and J.~Salvado, \emph{{On
  neutrino-mediated potentials in a neutrino background}},
  \href{https://doi.org/10.1007/JHEP04(2023)039}{\emph{JHEP} {\bfseries 04}
  (2023) 039} [\href{https://arxiv.org/abs/2212.03889}{{\ttfamily
  2212.03889}}].

\bibitem{VanTilburg:2024tst}
K.~Van~Tilburg, \emph{{Wake Forces}},
  \href{https://arxiv.org/abs/2401.08745}{{\ttfamily 2401.08745}}.

\bibitem{Ghosh:2024qai}
M.~Ghosh, Y.~Grossman, W.~Tangarife, X.-J.~Xu and B.~Yu, \emph{{The neutrino
  force in neutrino backgrounds: Spin dependence and parity-violating
  effects}}, \href{https://doi.org/10.1007/JHEP07(2024)107}{\emph{JHEP}
  {\bfseries 07} (2024) 107}
  [\href{https://arxiv.org/abs/2405.16801}{{\ttfamily 2405.16801}}].

\bibitem{Barbosa:2024tty}
S.~Barbosa and S.~Fichet, \emph{{Background-Induced Forces from Dark Relics}},
  \href{https://arxiv.org/abs/2403.13894}{{\ttfamily 2403.13894}}.

\bibitem{Fischbach:1996qf}
E.~Fischbach, \emph{Long range forces and neutrino mass},
  \href{https://doi.org/10.1006/aphy.1996.0044}{\emph{Annals Phys.} {\bfseries
  247} (1996) 213} [\href{https://arxiv.org/abs/hep-ph/9603396}{{\ttfamily
  hep-ph/9603396}}].

\bibitem{Smirnov:1996vj}
A.Y.~Smirnov and F.~Vissani, \emph{{Long range neutrino forces and the lower
  bound on neutrino mass}},
  \href{https://arxiv.org/abs/hep-ph/9604443}{{\ttfamily hep-ph/9604443}}.

\bibitem{Abada:1996nx}
A.~Abada, M.B.~Gavela and O.~Pene, \emph{To rescue a star},
  \href{https://doi.org/10.1016/0370-2693(96)01050-7}{\emph{Phys. Lett. B}
  {\bfseries 387} (1996) 315}
  [\href{https://arxiv.org/abs/hep-ph/9605423}{{\ttfamily hep-ph/9605423}}].

\bibitem{Kachelriess:1997cr}
M.~Kachelriess, \emph{Neutrino selfenergy and pair creation in neutron stars},
  \href{https://doi.org/10.1016/S0370-2693(98)00076-8}{\emph{Phys. Lett. B}
  {\bfseries 426} (1998) 89}
  [\href{https://arxiv.org/abs/hep-ph/9712363}{{\ttfamily hep-ph/9712363}}].

\bibitem{Kiers:1997ty}
K.~Kiers and M.H.G.~Tytgat, \emph{Neutrino ground state in a dense star},
  \href{https://doi.org/10.1103/PhysRevD.57.5970}{\emph{Phys. Rev. D}
  {\bfseries 57} (1998) 5970}
  [\href{https://arxiv.org/abs/hep-ph/9712463}{{\ttfamily hep-ph/9712463}}].

\bibitem{Abada:1998ti}
A.~Abada, O.~Pene and J.~Rodriguez-Quintero, \emph{Finite size effects on
  multibody neutrino exchange},
  \href{https://doi.org/10.1103/PhysRevD.58.073001}{\emph{Phys. Rev. D}
  {\bfseries 58} (1998) 073001}
  [\href{https://arxiv.org/abs/hep-ph/9802393}{{\ttfamily hep-ph/9802393}}].

\bibitem{Arafune:1998ft}
J.~Arafune and Y.~Mimura, \emph{Finiteness of multibody neutrino exchange
  potential energy in neutron stars},
  \href{https://doi.org/10.1143/PTP.100.1083}{\emph{Prog. Theor. Phys.}
  {\bfseries 100} (1998) 1083}
  [\href{https://arxiv.org/abs/hep-ph/9805395}{{\ttfamily hep-ph/9805395}}].

\bibitem{Orlofsky:2021mmy}
N.~Orlofsky and Y.~Zhang, \emph{{Neutrino as the dark force}},
  \href{https://doi.org/10.1103/PhysRevD.104.075010}{\emph{Phys. Rev. D}
  {\bfseries 104} (2021) 075010}
  [\href{https://arxiv.org/abs/2106.08339}{{\ttfamily 2106.08339}}].

\bibitem{Coy:2022cpt}
R.~Coy, X.-J.~Xu and B.~Yu, \emph{{Neutrino forces and the Sommerfeld
  enhancement}}, \href{https://doi.org/10.1007/JHEP06(2022)093}{\emph{JHEP}
  {\bfseries 06} (2022) 093}
  [\href{https://arxiv.org/abs/2203.05455}{{\ttfamily 2203.05455}}].

\bibitem{Xu:2021daf}
X.-j.~Xu and B.~Yu, \emph{{On the short-range behavior of neutrino forces
  beyond the Standard Model: from 1/r$^{5}$ to 1/r$^{4}$, 1/r$^{2}$, and 1/r}},
  \href{https://doi.org/10.1007/JHEP02(2022)008}{\emph{JHEP} {\bfseries 02}
  (2022) 008} [\href{https://arxiv.org/abs/2112.03060}{{\ttfamily
  2112.03060}}].

\bibitem{Dzuba:2022vrv}
V.A.~Dzuba, V.V.~Flambaum and P.~Munro-Laylim, \emph{{Long-range
  parity-nonconserving electron-nucleon interaction}},
  \href{https://doi.org/10.1103/PhysRevA.106.012817}{\emph{Phys. Rev. A}
  {\bfseries 106} (2022) 012817}
  [\href{https://arxiv.org/abs/2205.02569}{{\ttfamily 2205.02569}}].

\bibitem{Munro-Laylim:2022fsv}
P.~Munro-Laylim, V.A.~Dzuba and V.V.~Flambaum, \emph{{Effects of the long-range
  neutrino-mediated force in atomic phenomena}},
  \href{https://arxiv.org/abs/2207.07325}{{\ttfamily 2207.07325}}.

\bibitem{Kapner:2006si}
D.J.~Kapner, T.S.~Cook, E.G.~Adelberger, J.H.~Gundlach, B.R.~Heckel, C.D.~Hoyle
  et~al., \emph{{Tests of the gravitational inverse-square law below the
  dark-energy length scale}},
  \href{https://doi.org/10.1103/PhysRevLett.98.021101}{\emph{Phys. Rev. Lett.}
  {\bfseries 98} (2007) 021101}
  [\href{https://arxiv.org/abs/hep-ph/0611184}{{\ttfamily hep-ph/0611184}}].

\bibitem{Adelberger:2006dh}
E.G.~Adelberger, B.R.~Heckel, S.A.~Hoedl, C.D.~Hoyle, D.J.~Kapner and
  A.~Upadhye, \emph{{Particle Physics Implications of a Recent Test of the
  Gravitational Inverse Sqaure Law}},
  \href{https://doi.org/10.1103/PhysRevLett.98.131104}{\emph{Phys. Rev. Lett.}
  {\bfseries 98} (2007) 131104}
  [\href{https://arxiv.org/abs/hep-ph/0611223}{{\ttfamily hep-ph/0611223}}].

\bibitem{Chen:2014oda}
Y.J.~Chen, W.K.~Tham, D.E.~Krause, D.~Lopez, E.~Fischbach and R.S.~Decca,
  \emph{{Stronger Limits on Hypothetical Yukawa Interactions in the
  30\textendash{}8000 nm Range}},
  \href{https://doi.org/10.1103/PhysRevLett.116.221102}{\emph{Phys. Rev. Lett.}
  {\bfseries 116} (2016) 221102}
  [\href{https://arxiv.org/abs/1410.7267}{{\ttfamily 1410.7267}}].

\bibitem{Vasilakis:2008yn}
G.~Vasilakis, J.M.~Brown, T.W.~Kornack and M.V.~Romalis, \emph{{Limits on new
  long range nuclear spin-dependent forces set with a K - He-3
  co-magnetometer}},
  \href{https://doi.org/10.1103/PhysRevLett.103.261801}{\emph{Phys. Rev. Lett.}
  {\bfseries 103} (2009) 261801}
  [\href{https://arxiv.org/abs/0809.4700}{{\ttfamily 0809.4700}}].

\bibitem{Terrano:2015sna}
W.A.~Terrano, E.G.~Adelberger, J.G.~Lee and B.R.~Heckel, \emph{{Short-range
  spin-dependent interactions of electrons: a probe for exotic pseudo-Goldstone
  bosons}}, \href{https://doi.org/10.1103/PhysRevLett.115.201801}{\emph{Phys.
  Rev. Lett.} {\bfseries 115} (2015) 201801}
  [\href{https://arxiv.org/abs/1508.02463}{{\ttfamily 1508.02463}}].

\bibitem{Brax:2017xho}
P.~Brax, S.~Fichet and G.~Pignol, \emph{{Bounding Quantum Dark Forces}},
  \href{https://doi.org/10.1103/PhysRevD.97.115034}{\emph{Phys. Rev. D}
  {\bfseries 97} (2018) 115034}
  [\href{https://arxiv.org/abs/1710.00850}{{\ttfamily 1710.00850}}].

\bibitem{Fichet:2017bng}
S.~Fichet, \emph{{Quantum Forces from Dark Matter and Where to Find Them}},
  \href{https://doi.org/10.1103/PhysRevLett.120.131801}{\emph{Phys. Rev. Lett.}
  {\bfseries 120} (2018) 131801}
  [\href{https://arxiv.org/abs/1705.10331}{{\ttfamily 1705.10331}}].

\bibitem{Costantino:2019ixl}
A.~Costantino, S.~Fichet and P.~Tanedo, \emph{{Exotic Spin-Dependent Forces
  from a Hidden Sector}},
  \href{https://doi.org/10.1007/JHEP03(2020)148}{\emph{JHEP} {\bfseries 03}
  (2020) 148} [\href{https://arxiv.org/abs/1910.02972}{{\ttfamily
  1910.02972}}].

\bibitem{Banks:2020gpu}
H.~Banks and M.~Mccullough, \emph{{Charting the Fifth Force Landscape}},
  \href{https://doi.org/10.1103/PhysRevD.103.075018}{\emph{Phys. Rev. D}
  {\bfseries 103} (2021) 075018}
  [\href{https://arxiv.org/abs/2009.12399}{{\ttfamily 2009.12399}}].

\bibitem{Brax:2022wrt}
P.~Brax and S.~Fichet, \emph{{Scalar-mediated quantum forces between
  macroscopic bodies and interferometry}},
  \href{https://doi.org/10.1016/j.dark.2023.101294}{\emph{Phys. Dark Univ.}
  {\bfseries 42} (2023) 101294}
  [\href{https://arxiv.org/abs/2203.01342}{{\ttfamily 2203.01342}}].

\bibitem{Bauer:2023czj}
M.~Bauer and G.~Rostagni, \emph{{Fifth Forces from QCD Axions Scale
  Differently}},
  \href{https://doi.org/10.1103/PhysRevLett.132.101802}{\emph{Phys. Rev. Lett.}
  {\bfseries 132} (2024) 101802}
  [\href{https://arxiv.org/abs/2307.09516}{{\ttfamily 2307.09516}}].

\bibitem{Dzuba:2017cas}
V.A.~Dzuba, V.V.~Flambaum, P.~Munro-Laylim and Y.V.~Stadnik, \emph{{Probing
  Long-Range Neutrino-Mediated Forces with Atomic and Nuclear Spectroscopy}},
  \href{https://doi.org/10.1103/PhysRevLett.120.223202}{\emph{Phys. Rev. Lett.}
  {\bfseries 120} (2018) 223202}
  [\href{https://arxiv.org/abs/1711.03700}{{\ttfamily 1711.03700}}].

\bibitem{Ghosh:2019dmi}
M.~Ghosh, Y.~Grossman and W.~Tangarife, \emph{{Probing the two-neutrino
  exchange force using atomic parity violation}},
  \href{https://doi.org/10.1103/PhysRevD.101.116006}{\emph{Phys. Rev. D}
  {\bfseries 101} (2020) 116006}
  [\href{https://arxiv.org/abs/1912.09444}{{\ttfamily 1912.09444}}].

\bibitem{Safronova:2017xyt}
M.S.~Safronova, D.~Budker, D.~DeMille, D.F.J.~Kimball, A.~Derevianko and
  C.W.~Clark, \emph{{Search for New Physics with Atoms and Molecules}},
  \href{https://doi.org/10.1103/RevModPhys.90.025008}{\emph{Rev. Mod. Phys.}
  {\bfseries 90} (2018) 025008}
  [\href{https://arxiv.org/abs/1710.01833}{{\ttfamily 1710.01833}}].

\bibitem{Wieman:2019vik}
C.~Wieman and A.~Derevianko, \emph{{Atomic parity violation and the standard
  model}},  \href{https://arxiv.org/abs/1904.00281}{{\ttfamily 1904.00281}}.

\bibitem{Arcadi:2019uif}
G.~Arcadi, M.~Lindner, J.~Martins and F.S.~Queiroz, \emph{{New physics probes:
  Atomic parity violation, polarized electron scattering and neutrino-nucleus
  coherent scattering}},
  \href{https://doi.org/10.1016/j.nuclphysb.2020.115158}{\emph{Nucl. Phys. B}
  {\bfseries 959} (2020) 115158}
  [\href{https://arxiv.org/abs/1906.04755}{{\ttfamily 1906.04755}}].

\bibitem{Crivelli:2018vfe}
P.~Crivelli, \emph{{The Mu-MASS (MuoniuM lAser SpectroScopy) experiment}},
  \href{https://doi.org/10.1007/s10751-018-1525-z}{\emph{Hyperfine Interact.}
  {\bfseries 239} (2018) 49}
  [\href{https://arxiv.org/abs/1811.00310}{{\ttfamily 1811.00310}}].

\bibitem{Cassidy:2018tgq}
D.B.~Cassidy, \emph{{Experimental progress in positronium laser physics}},
  \href{https://doi.org/10.1140/epjd/e2018-80721-y}{\emph{Eur. Phys. J. D}
  {\bfseries 72} (2018) 53}.

\bibitem{Ohayon:2021dec}
B.~Ohayon, Z.~Burkley and P.~Crivelli, \emph{{Current status and prospects of
  muonium spectroscopy at PSI}},
  \href{https://doi.org/10.21468/SciPostPhysProc.5.029}{\emph{SciPost Phys.
  Proc.} {\bfseries 5} (2021) 029}.

\bibitem{Mu-MASS:2021uou}
{\scshape Mu-MASS} collaboration, \emph{{Precision Measurement of the Lamb
  Shift in Muonium}},
  \href{https://doi.org/10.1103/PhysRevLett.128.011802}{\emph{Phys. Rev. Lett.}
  {\bfseries 128} (2022) 011802}
  [\href{https://arxiv.org/abs/2108.12891}{{\ttfamily 2108.12891}}].

\bibitem{Janka:2021xxr}
G.~Janka, B.~Ohayon and P.~Crivelli, \emph{{Muonium Lamb shift: theory update
  and experimental prospects}},
  \href{https://doi.org/10.1051/epjconf/202226201001}{\emph{EPJ Web Conf.}
  {\bfseries 262} (2022) 01001}
  [\href{https://arxiv.org/abs/2111.13951}{{\ttfamily 2111.13951}}].

\bibitem{Adkins:2022omi}
G.S.~Adkins, D.B.~Cassidy and J.~P\'erez-R\'\i{}os, \emph{{Precision
  spectroscopy of positronium: Testing bound-state QED theory and the search
  for physics beyond the Standard Model}},
  \href{https://doi.org/10.1016/j.physrep.2022.05.002}{\emph{Phys. Rept.}
  {\bfseries 975} (2022) 1}.

\bibitem{Sheldon:2023eic}
R.E.~Sheldon, T.J.~Babij, S.H.~Reeder, S.D.~Hogan and D.B.~Cassidy,
  \emph{{Microwave spectroscopy of positronium atoms in free space}},
  \href{https://doi.org/10.1103/PhysRevA.107.042810}{\emph{Phys. Rev. A}
  {\bfseries 107} (2023) 042810}.

\bibitem{Fermi:1934sk}
E.~Fermi, \emph{{Trends to a Theory of beta Radiation. (In Italian)}},
  \href{https://doi.org/10.1007/BF02959820}{\emph{Nuovo Cim.} {\bfseries 11}
  (1934) 1}.

\bibitem{Frank:1971xx}
W.~Frank, D.J.~Land and R.M.~Spector, \emph{{Singular potentials}},
  \href{https://doi.org/10.1103/RevModPhys.43.36}{\emph{Rev. Mod. Phys.}
  {\bfseries 43} (1971) 36}.

\bibitem{Bedaque:2009ri}
P.F.~Bedaque, M.I.~Buchoff and R.K.~Mishra, \emph{{Sommerfeld enhancement from
  Goldstone pseudo-scalar exchange}},
  \href{https://doi.org/10.1088/1126-6708/2009/11/046}{\emph{JHEP} {\bfseries
  11} (2009) 046} [\href{https://arxiv.org/abs/0907.0235}{{\ttfamily
  0907.0235}}].

\bibitem{Chen:2009ch}
C.-H.~Chen and C.S.~Kim, \emph{{Sommerfeld Enhancement from Unparticle Exchange
  for Dark Matter Annihilation}},
  \href{https://doi.org/10.1016/j.physletb.2010.03.054}{\emph{Phys. Lett. B}
  {\bfseries 687} (2010) 232}
  [\href{https://arxiv.org/abs/0909.1878}{{\ttfamily 0909.1878}}].

\bibitem{Chaffey:2021tmj}
I.~Chaffey, S.~Fichet and P.~Tanedo, \emph{{Continuum-Mediated Self-Interacting
  Dark Matter}}, \href{https://doi.org/10.1007/JHEP06(2021)008}{\emph{JHEP}
  {\bfseries 06} (2021) 008}
  [\href{https://arxiv.org/abs/2102.05674}{{\ttfamily 2102.05674}}].

\bibitem{landau2013quantum}
L.D.~Landau and E.M.~Lifshitz, \emph{Quantum mechanics: non-relativistic
  theory}, vol.~3, Elsevier (2013).

\bibitem{Asaka:2018qfg}
T.~Asaka, M.~Tanaka, K.~Tsumura and M.~Yoshimura, \emph{{Precision electroweak
  shift of muonium hyperfine splitting}},
  \href{https://arxiv.org/abs/1810.05429}{{\ttfamily 1810.05429}}.

\bibitem{Bouchiat:1974kt}
M.A.~Bouchiat and C.C.~Bouchiat, \emph{{Weak Neutral Currents in Atomic
  Physics}}, \href{https://doi.org/10.1016/0370-2693(74)90656-X}{\emph{Phys.
  Lett. B} {\bfseries 48} (1974) 111}.

\bibitem{Bouchiat:1976vg}
M.A.~Bouchiat and L.~Pottier, \emph{{Search for a Parity Violation Induced by
  Neutral Currents in the 6S-7S Transition of Atomic Cesium}},
  \href{https://doi.org/10.1016/0370-2693(76)90087-3}{\emph{Phys. Lett. B}
  {\bfseries 62} (1976) 327}.

\bibitem{Bouchiat:1977zp}
C.~Bouchiat, \emph{{Parity Violation in Atomic Physics and Neutral Currents}},
  \href{https://doi.org/10.1007/978-1-4757-1565-1_18}{\emph{Stud. Nat. Sci.}
  {\bfseries 12} (1977) 435}.

\bibitem{Bouchiat:1980gp}
C.~Bouchiat, \emph{{Parity Violation Effects Induced by Neutral Currents in
  Atoms: Theory}},  in \emph{{7th International Conference on Atomic Physics}},
  9, 1980.

\bibitem{Bouchiat:1983uf}
C.~Bouchiat and C.A.~Piketty, \emph{{Parity Violation in Atomic Cesium and
  Alternatives to the Standard Model of Electroweak Interactions}},
  \href{https://doi.org/10.1016/0370-2693(83)90076-X}{\emph{Phys. Lett. B}
  {\bfseries 128} (1983) 73}.

\bibitem{Bouchiat:1984wz}
M.A.~Bouchiat, J.~Guena, L.~Pottier and L.~Hunter, \emph{{NEW OBSERVATION OF A
  PARITY VIOLATION IN CESIUM}},
  \href{https://doi.org/10.1016/0370-2693(84)91386-8}{\emph{Phys. Lett. B}
  {\bfseries 134} (1984) 463}.

\bibitem{Bouchiat:1984yh}
M.A.~Bouchiat and L.~Pottier, \emph{{Atomic parity violation experiments}},  in
  \emph{{9th International Conference on Atomic Physics}}, pp.~246--271, 1984.

\bibitem{Bouchiat:1986rm}
C.~Bouchiat and C.A.~Piketty, \emph{{A New Method for the Evaluation of the
  Parity Violating Electric Dipole 6s $\to$ 7s Amplitude in Atomic Cesium}},
  \href{https://doi.org/10.1209/0295-5075/2/7/004}{\emph{EPL} {\bfseries 2}
  (1986) 511}.

\bibitem{Bouchiat:1993ek}
M.-A.~Bouchiat, \emph{{Neutral currents in atoms, experiments in cesium and
  implications}},  in \emph{{International Conference on Neutral Currents:
  Twenty Years Later}}, pp.~247--268, 7, 1993.

\bibitem{Bouchiat:1997mj}
M.A.~Bouchiat and C.~Bouchiat, \emph{{Parity violation in atoms}},
  \href{https://doi.org/10.1088/0034-4885/60/11/004}{\emph{Rept. Prog. Phys.}
  {\bfseries 60} (1997) 1351}.

\bibitem{Bouchiat:2001tba}
M.-A.~Bouchiat, \emph{{Twenty Five Years Of Atomic Parity Violation In
  Cesium}},  in \emph{{35th Rencontres de Moriond: Electroweak Interactions and
  Unified Theories}}, (Hanoi), pp.~307--318, The Gioi, 2001.

\bibitem{Bouchiat:2011fs}
M.-A.~Bouchiat, \emph{{Atomic Parity Violation. Early days, present results,
  prospects}}, \href{https://doi.org/10.1393/ncc/i2012-11269-6}{\emph{Nuovo
  Cim. C} {\bfseries 35} (2012) 78}
  [\href{https://arxiv.org/abs/1111.2172}{{\ttfamily 1111.2172}}].

\bibitem{Guena:2004sq}
J.~Guena, M.~Lintz and M.A.~Bouchiat, \emph{{Measurement of the parity
  violating 6S-7S transition amplitude in cesium achieved within 2 x 10(-13)
  atomic-unit accuracy by stimulated-emission detection}},
  \href{https://doi.org/10.1103/PhysRevA.71.042108}{\emph{Phys. Rev. A}
  {\bfseries 71} (2005) 042108}
  [\href{https://arxiv.org/abs/physics/0412017}{{\ttfamily physics/0412017}}].

\bibitem{Guena:2005uj}
J.~Guena, M.~Lintz and M.-A.~Bouchiat, \emph{{Atomic parity violation:
  Principles, recent results, present motivations}},
  \href{https://doi.org/10.1142/S0217732305016853}{\emph{Mod. Phys. Lett. A}
  {\bfseries 20} (2005) 375}
  [\href{https://arxiv.org/abs/physics/0503143}{{\ttfamily physics/0503143}}].

\bibitem{Johnson:2003ba}
W.R.~Johnson, M.S.~Safronova and U.I.~Safronova, \emph{{Combined effect of
  coherent Z exchange and the hyperfine interaction in atomic PNC}},
  \href{https://doi.org/10.1103/PhysRevA.67.062106}{\emph{Phys. Rev. A}
  {\bfseries 67} (2003) 062106}
  [\href{https://arxiv.org/abs/hep-ph/0302029}{{\ttfamily hep-ph/0302029}}].

\bibitem{Wood:1997zq}
C.S.~Wood, S.C.~Bennett, D.~Cho, B.P.~Masterson, J.L.~Roberts, C.E.~Tanner
  et~al., \emph{{Measurement of parity nonconservation and an anapole moment in
  cesium}}, \href{https://doi.org/10.1126/science.275.5307.1759}{\emph{Science}
  {\bfseries 275} (1997) 1759}.

\bibitem{Marciano:1978ed}
W.J.~Marciano and A.I.~Sanda, \emph{{Parity Violation in Atoms Induced by
  Radiative Corrections}},
  \href{https://doi.org/10.1103/PhysRevD.17.3055}{\emph{Phys. Rev. D}
  {\bfseries 17} (1978) 3055}.

\bibitem{Marciano:1982mm}
W.J.~Marciano and A.~Sirlin, \emph{{RADIATIVE CORRECTIONS TO ATOMIC PARITY
  VIOLATION}}, \href{https://doi.org/10.1103/PhysRevD.27.552}{\emph{Phys. Rev.
  D} {\bfseries 27} (1983) 552}.

\bibitem{Marciano:1983ss}
W.J.~Marciano and A.~Sirlin, \emph{{On Some General Properties of the O(alpha)
  Corrections to Parity Violation in Atoms}},
  \href{https://doi.org/10.1103/PhysRevD.29.75}{\emph{Phys. Rev. D} {\bfseries
  29} (1984) 75}.

\bibitem{Porsev:2009pr}
S.G.~Porsev, K.~Beloy and A.~Derevianko, \emph{{Precision determination of
  electroweak coupling from atomic parity violation and implications for
  particle physics}},
  \href{https://doi.org/10.1103/PhysRevLett.102.181601}{\emph{Phys. Rev. Lett.}
  {\bfseries 102} (2009) 181601}
  [\href{https://arxiv.org/abs/0902.0335}{{\ttfamily 0902.0335}}].

\bibitem{Dzuba:2012kx}
V.A.~Dzuba, J.C.~Berengut, V.V.~Flambaum and B.~Roberts, \emph{{Revisiting
  parity non-conservation in cesium}},
  \href{https://doi.org/10.1103/PhysRevLett.109.203003}{\emph{Phys. Rev. Lett.}
  {\bfseries 109} (2012) 203003}
  [\href{https://arxiv.org/abs/1207.5864}{{\ttfamily 1207.5864}}].

\bibitem{ParticleDataGroup:2022pth}
{\scshape Particle Data Group} collaboration, \emph{{Review of Particle
  Physics}}, \href{https://doi.org/10.1093/ptep/ptac097}{\emph{PTEP} {\bfseries
  2022} (2022) 083C01}.

\bibitem{Dzuba:2002kx}
V.A.~Dzuba, V.V.~Flambaum and J.S.M.~Ginges, \emph{{Precise calculation of
  parity nonconservation in cesium and test of the standard model}},
  \href{https://doi.org/10.1103/PhysRevD.66.076013}{\emph{Phys. Rev. D}
  {\bfseries 66} (2002) 076013}
  [\href{https://arxiv.org/abs/hep-ph/0204134}{{\ttfamily hep-ph/0204134}}].

\bibitem{Cadeddu:2018izq}
M.~Cadeddu and F.~Dordei, \emph{{Reinterpreting the weak mixing angle from
  atomic parity violation in view of the Cs neutron rms radius measurement from
  COHERENT}}, \href{https://doi.org/10.1103/PhysRevD.99.033010}{\emph{Phys.
  Rev. D} {\bfseries 99} (2019) 033010}
  [\href{https://arxiv.org/abs/1808.10202}{{\ttfamily 1808.10202}}].

\bibitem{Patel:2015tea}
H.H.~Patel, \emph{{Package-X: A Mathematica package for the analytic
  calculation of one-loop integrals}},
  \href{https://doi.org/10.1016/j.cpc.2015.08.017}{\emph{Comput. Phys. Commun.}
  {\bfseries 197} (2015) 276}
  [\href{https://arxiv.org/abs/1503.01469}{{\ttfamily 1503.01469}}].

\bibitem{Patel:2016fam}
H.H.~Patel, \emph{{Package-X 2.0: A Mathematica package for the analytic
  calculation of one-loop integrals}},
  \href{https://doi.org/10.1016/j.cpc.2017.04.015}{\emph{Comput. Phys. Commun.}
  {\bfseries 218} (2017) 66}
  [\href{https://arxiv.org/abs/1612.00009}{{\ttfamily 1612.00009}}].

\bibitem{Passarino:1978jh}
G.~Passarino and M.J.G.~Veltman, \emph{{One Loop Corrections for e+ e-
  Annihilation Into mu+ mu- in the Weinberg Model}},
  \href{https://doi.org/10.1016/0550-3213(79)90234-7}{\emph{Nucl. Phys. B}
  {\bfseries 160} (1979) 151}.

\bibitem{Denner:1991kt}
A.~Denner, \emph{{Techniques for calculation of electroweak radiative
  corrections at the one loop level and results for W physics at LEP-200}},
  \href{https://doi.org/10.1002/prop.2190410402}{\emph{Fortsch. Phys.}
  {\bfseries 41} (1993) 307} [\href{https://arxiv.org/abs/0709.1075}{{\ttfamily
  0709.1075}}].

\end{thebibliography}\endgroup
\bibliographystyle{JHEP}
\end{document}